\pdfoutput=1
\documentclass[11pt,twoside,a4paper,cmspaper,final,collab]{cms-tdr}
\def\svnVersion{a0d25e5-D}\def\svnDate{2021/04/14}\def\cmsCernNoTag{CERN-EP-2020-182}\def\cmsCernDate{\today}\def\cmsMessage{"Published in the European Physical Journal C as \href{http://dx.doi.org/10.1140/epjc/s10052-021-08949-5}{\doi{10.1140/epjc/s10052-021-08949-5}.}"}
\begin{document}\cmsNoteHeader{GEN-19-001}

\newcommand{\HerwigS} {{$\HERWIG\,7$}\xspace}
\newcommand{\HerwigSTitle} {{\textsc{HERWIG\,7}}\xspace}
\newcommand{\Matchbox} {\textsc{matchbox}\xspace}
\newcommand{\SoftTune} {\ensuremath{\text{SoftTune}}\xspace}
\newcommand{\CH} {{\textsc{CH}}\xspace}
\newcommand{\CHo} {{\textsc{CH1}}\xspace}
\newcommand{\CHt} {{\textsc{CH2}}\xspace}
\newcommand{\CHth} {{\textsc{CH3}}\xspace}
\newcommand{\HerwigSix} {\HERWIG}

\newcommand{\etaphi} {\ensuremath{\eta}-\ensuremath{\phi}\xspace}
\newcommand{\deltaPhi} {\ensuremath{\Delta\phi}\xspace}

\newcommand{\MZ}{\ensuremath{m_{\mathrm{\PZ}}}\xspace}
\newcommand{\alpSMZ}{\ensuremath{\alpS(\MZ)}\xspace}
\newcommand{\alpSPDFMZ}{\ensuremath{\alpS^{\mathrm{PDF}}(\MZ)}\xspace}

\newcommand{\PYTHIASIX} {{$\PYTHIA\,6$}\xspace}
\newcommand{\PYTHIAEIGHT} {{$\PYTHIA\,8$}\xspace}
\newcommand{\hdamp} {\ensuremath{h_{\text{damp}}}\xspace}
\newcommand{\NNPDF} {{{NNPDF\,3.1}}\xspace}
\newcommand{\MMHTLO} {{{MMHT\,2014 LO}}\xspace}

\newcommand{\POWHEGPYTHIA} {\POWHEG~+~\PYTHIA\xspace}
\newcommand{\POWHEGHERWIG} {\POWHEG~+~\HerwigS}

\newcommand{\Professor} {{\textsc{professor}}\xspace}
\newcommand{\chis} {\ensuremath{\chi^{2}}\xspace}
\newcommand{\deltachis} {\ensuremath{\Delta\chi^{2}}\xspace}
\newcommand{\nparam}{\ensuremath{N_{\text{param}}}\xspace}
\newcommand{\weight}{\ensuremath{w_{\mathcal{O}}}\xspace}
\newcommand{\Rivet} {{\textsc{rivet}}\xspace}
\newcommand{\bin}{\ensuremath{i}\xspace}
\newcommand{\nbins}{\ensuremath{N_{\text{bins}}}\xspace}
\newcommand{\ndof}{\ensuremath{N_{\text{dof}}}\xspace}

\newcommand{\ptsum} {\ensuremath{\pt^{\text{sum}}}\xspace}
\newcommand{\ptleadingtrack} {\ensuremath{\pt^{\text{max}}}\xspace}
\newcommand{\ptleadingtrackjet} {\ensuremath{\pt^{\text{jet}}}\xspace}
\newcommand{\SISCONE}{\textsc{SISCone}\xspace}
\newcommand{\antikt}{anti-\kt}
\newcommand{\Nch} {\ensuremath{N_{\text{ch}}}\xspace}
\newcommand{\dNdeta} {\ensuremath{\ddinline{N_{\text{ch}}}{\eta}}\xspace}

\newcommand{\ptmino} {\ensuremath{p_{\bot,\mathrm{0}}^\mathrm{min}}\xspace}
\newcommand{\ptmin} {\ensuremath{p_{\bot}^\text{min}}\xspace}
\newcommand{\bParameter} {\ensuremath{b}\xspace}
\newcommand{\musParameter} {\ensuremath{\mu^{2}}\xspace}
\newcommand{\musunits} {\ensuremath{\GeVns^{-2}}\xspace}
\newcommand{\PParameter} {\ensuremath{P}\xspace}
\newcommand{\No} {\ensuremath{N_\mathrm{0}}\xspace}
\newcommand{\preco} {\ensuremath{p_{\text{reco}}}\xspace}
\newcommand{\Eo} {\ensuremath{E_{\mathrm{0}}}\xspace}

\newcommand{\meanMPI}{\ensuremath{\langle n\rangle}\xspace}
\newcommand{\sigmaMPI}{\ensuremath{\sigma(s)}\xspace}
\newcommand{\impactParam}{\ensuremath{d}\xspace}
\newcommand{\Aoverlap}{\ensuremath{A(\impactParam)}\xspace}

\newcommand{\sqrts} {\ensuremath{\sqrt{s}}\xspace}

\newcommand{\mt} {\ensuremath{m_{\PQt}}\xspace}

\newcommand{\pthat}{\ensuremath{\hat{p}_{\mathrm{T}}}\xspace}

\newcommand{\ptjet}{\ensuremath{\pt^{\text{jet}}}\xspace}
\newcommand{\Fz}{\ensuremath{F(z)}\xspace}
\newcommand{\ptrel}{\ensuremath{\pt^{\text{rel}}}\xspace}
\newcommand{\fptrel}{\ensuremath{f(\pt^{\text{rel}}})\xspace}
\newcommand{\z}{\ensuremath{z}\xspace}

\newcommand{\rhor}{\ensuremath{\rho(\mathrm{r})}\xspace}
\newcommand{\jetmoment}{\ensuremath{\langle\delta R^2\rangle}\xspace}

\newcommand{\lambdazerozero}{\ensuremath{\lambda_0^0}\xspace}
\newcommand{\deltaRg}{\ensuremath{\Delta R_\mathrm{g}}\xspace}
\newcommand{\zg}{\ensuremath{z_\mathrm{g}}\xspace}
\newcommand{\eccentricity}{\ensuremath{\varepsilon}\xspace}

\newcommand{\thrust} {\ensuremath{T}\xspace}
\newcommand{\thrustmajor} {\ensuremath{T_{\mathrm{major}}}\xspace}
\newcommand{\oblateness} {\ensuremath{O}\xspace}
\newcommand{\sphericity} {\ensuremath{S}\xspace}

\newcommand{\deltaRDefn}{\ensuremath{\sqrt{\smash[b]{(\Delta\eta)^2+(\Delta\phi)^2}}}\xspace}

\newcommand{\MGaMC} {\MGvATNLO}
\newcommand{\FXFX} {FxFx\xspace}
\newcommand{\ptmumu}{\ensuremath{\pt(\mu\mu)}\xspace}
\newcommand{\transverseMass}{\ensuremath{\mT = \sqrt{\smash[b]{2\pt^\mu\ptmiss[1-\cos(\Delta\phi_{\mu,\ptvecmiss})]}}}\xspace}
\newcommand{\MGaMCHerwigS} {\MGaMC + \HerwigS}
\newcommand{\ZJets}{\PZ+jets\xspace}
\newcommand{\ptZ}{\ensuremath{\pt(\PZ)}\xspace}
\newcommand{\ptbal}{\ensuremath{\pt^{\text{bal}}}\xspace}
\newcommand{\ptvecZ}{\ensuremath{\ptvec(\PZ)}\xspace}
\newcommand{\JZB}{\ensuremath{\mathrm{JZB}}\xspace}

\newcommand{\captionColouredShadedBand}{The coloured band in the ratio plot represents the total experimental uncertainty in the data.  The vertical bars on the points for the different predictions represent the statistical uncertainties.\xspace}

\providecommand{\cmsTable}[1]{\resizebox{\textwidth}{!}{#1}}
\newlength\cmsTabSkip\setlength{\cmsTabSkip}{1ex}

\cmsNoteHeader{GEN-19-001}
\title{Development and validation of \texorpdfstring{\HerwigSTitle}{HERWIG 7} tunes from CMS underlying-event measurements}

\date{\today}

\abstract{
   This paper presents new sets of parameters (``tunes'') for the underlying-event model of the \HerwigS event generator. These parameters control the description of multiple-parton interactions (MPI) and colour reconnection in \HerwigS, and are obtained from a fit to minimum-bias data collected by the CMS experiment at $\sqrt{s}=0.9$, 7, and $13 \TeV$. The tunes are based on the \NNPDF next-to-next-to-leading-order parton distribution function (PDF) set for the parton shower, and either a leading-order or next-to-next-to-leading-order PDF set for the simulation of MPI and the beam remnants. Predictions utilizing the tunes are produced for event shape observables in electron-positron collisions, and for minimum-bias, inclusive jet, top quark pair, and \PZ and \PW boson events in proton-proton collisions, and are compared with data. Each of the new tunes describes the data at a reasonable level, and the tunes using a leading-order PDF for the simulation of MPI provide the best description of the data.
}

\hypersetup{%
pdfauthor={CMS Collaboration},%
pdftitle={Development and validation of HERWIG7 tunes from CMS underlying-event measurements},%
pdfsubject={CMS},%
pdfkeywords={CMS, HERWIG7, underlying event}}

\maketitle

\section{Introduction}
\label{sec:Introduction}

{\tolerance=1000 In hadron-hadron collisions, the hard scattering of partons is accompanied by additional activity from multiple-parton interactions (MPI) that take place within the same collision, and by interactions between the remnants of the hadrons.  To describe the underlying-event (UE) activity in a hard scattering process, and minimum-bias (MB) events, Monte Carlo (MC) event generators such as \HerwigS~\cite{Herwigpp,Herwig70,Herwig71} and \PYTHIAEIGHT~\cite{Pythia8} include a model of these additional interactions.  Because these processes are soft in nature, perturbative quantum chromodynamics (QCD) cannot be used to predict them, so they must be described by a phenomenological model.  The parameters of the models must be optimized to provide a reasonable description of measured observables that are sensitive to the UE and MB events.  An accurate description of the UE by MC event generators, along with an understanding of the uncertainties in the description, is of particular importance for precision measurements at hadron colliders, such as the extraction of the top quark mass.  This paper presents new sets of parameters (``tunes'') for the UE model of the \HerwigS event generator. \par}

{\tolerance=800 The \HerwigS event generator is a multipurpose event generator, which can perform matrix-element (ME) calculations beyond leading order (LO) in QCD, via the \Matchbox module~\cite{HerwigMatchboxAndDipole}, matched with parton shower (PS) calculations.  Both an angular-ordered and a dipole-based PS simulation are available in \HerwigS, and the former is used in this paper.  
The ME calculations can also be provided by an external ME generator, such as \POWHEG~\cite{Powheg_ref1,Powheg_ref2,Powheg_ref3} or \MGaMC~\cite{MG5aMCatNLO}.  The \HerwigS generator is built upon the development of the preceding \HerwigSix~\cite{Herwig6} and \HERWIGpp~\cite{Herwigpp} event generators.  
In addition to the simulation of hard scattering of partons in hadron-hadron collisions, a simulation of MPI, which is modelled by a combination of soft and hard interactions and by colour reconnection (CR)~\cite{Herwigpp,HerwigppMPI,Herwig7SoftModel,HerwigCR}, is included in \HerwigS.  As shown in Ref.~\cite{HerwigCR}, a model of CR is required in \HerwigS to describe the structure of colour connections between different MPI, and to obtain a good description of the mean charged-particle transverse momentum (\pt) as a function of the charged-particle multiplicity (\Nch).

The model describing the soft interactions, and also diffractive processes, was improved in version 7.1 of \HerwigS.  This resulted in a new tune of the MPI parameters, called \SoftTune, which improved the description of MB data~\cite{Herwig7SoftModel,Herwig71}. In particular, the description of final-state hadronic systems separated by a large rapidity gap~\cite{ATLASRapGap, CMSRapGap} is notably improved because a significant contribution is expected from diffractive events.
The tune \SoftTune is based on the \MMHTLO parton distribution function (PDF) set~\cite{MMHT}, and was derived by fitting MB data at $\sqrts=0.9,~7,$ and $13\TeV$ from the ATLAS experiment~\cite{ATLASdNdEta0p9And7TeV}.
The MB data used in the tuning include the pseudorapidity ($\eta$) and $\pt$ distributions of charged particles for various lower bounds on \Nch, namely $\Nch \geq 1,~2,~6,$ and 20. The mean charged-particle $\pt$ as a function of \Nch was also included in the tuning procedure.
Three models of CR are available in \HerwigS, and \SoftTune was derived with the plain colour reconnection (PCR) model implemented.  The same PCR model is considered in our studies. \par}

{\tolerance=1000 In this paper, we present new UE tunes for the \HerwigS (version 7.1.4) generator.
In contrast to \SoftTune, the tunes presented here are based on the \NNPDF PDF sets~\cite{NNPDF31}, and use the next-to-next-to-leading-order (NNLO) PDF set for the simulation of the PS, and either an LO or NNLO PDF set for the simulation of MPI and the beam remnants.  This choice of PDF sets is similar to that used to obtain tunes for the \PYTHIAEIGHT event generator in Ref.~\cite{GEN17001}, where it was shown that predictions from \PYTHIAEIGHT using LO, next-to-leading-order (NLO), and NNLO PDFs with their associated tunes can all give a reliable description of the UE.  Based on these findings and the wide use by the CMS Collaboration of the CP5 \PYTHIAEIGHT tune, we concentrate on deriving tunes for the \HerwigS generator that are also based on an NNLO PDF set for the simulation of the parton shower.  It is verified that using an NNLO PDF in the simulation of the PS in \HerwigS also provides a reliable description of MB data. A consistent choice of PDF in the \HerwigS and \PYTHIAEIGHT generators, as well as a similar method of the MPI model tuning, provides a better comparison of predictions from these two generators.
\par}

{\tolerance=800 The tunes are derived by fitting measurements from proton-proton collision data collected by the CMS experiment~\cite{TheCMSExperiment} at $\sqrts = 0.9$, $7$, and $13\TeV$.  The measurements used in the fitting procedure are chosen because of their sensitivity to the modelling of the UE in \HerwigS.  
Uncertainties in the parameters of one of the new tunes are also derived.  This quantifies the effect of the uncertainties in the fitted parameters for future analyses.
To validate the performance of the new tunes, the corresponding \HerwigS predictions are compared with a range of MB data from proton-proton and proton-antiproton collisions.  Comparisons are also made using event shape observables from electron-positron collisions collected at the CERN LEP accelerator, which are particularly sensitive to the choice of the strong coupling \alpS in the description of final-state radiation.  To further validate the new tunes, predictions of differential \ttbar, \PZ boson, and \PW boson cross sections are also obtained from matching ME calculations from \POWHEG and \MGaMC with the \HerwigS PS description.  The kinematics of the \ttbar system are studied, along with the multiplicity of additional jets, which are sensitive to the modelling by the PS simulation, in \ttbar, \PZ boson, and \PW boson events. 
The modelling of the UE in \PZ boson events, and the substructure of jets in \ttbar and in inclusive jet events are also investigated. Some of these comparisons are sensitive to the modelling by the ME calculations, and the purpose of those is to validate that the various predictions using the tunes do not differ from each other by a significant amount.  Other comparisons are more sensitive to the modelling of the PS and MPI simulation, allowing us to test the new tunes in data other than MB data. \par}

This paper is organized as follows.  In Section~\ref{sec:HerwigModel}, we summarize the UE model employed by \HerwigS, and describe the model parameters considered in the tuning.
The choice of PDF and the value of the strong coupling in the tunes is discussed in Section~\ref{sec:TuneProcedure} in addition to details of the fitting procedure.  The new tunes are presented in Section~\ref{sec:HerwigTunes}, and the corresponding predictions from \HerwigS are compared with MB data.  Uncertainties in one of the derived tunes are presented in Section~\ref{sec:HerwigTuneUncertainties}.  Further validation of the new tunes is performed in the following sections: their predictions are compared with event shape observables from the CERN LEP in Section~\ref{sec:LepComparisons}, and with top quark, inclusive jet, and \PZ and \PW boson production data in Sections~\ref{sec:TopComparisons}, \ref{sec:JetMetComparisons}, and~\ref{sec:VJetsComparisons}, respectively.  Finally, we present a summary in Section~\ref{sec:Summary}.

\section{The UE model in \texorpdfstring{\HerwigSTitle}{HERWIG 7}}
\label{sec:HerwigModel}

The UE in \HerwigS is modelled by a combination of soft and hard interactions~\cite{Herwigpp,HerwigppMPI,Herwig7SoftModel}.  The parameter \ptmin defines the transition between the soft and hard MPI.  The interactions with a pair of outgoing partons with \pt above \ptmin are treated as hard interactions, which are constructed from QCD two-to-two processes.  The \ptmin transition threshold depends on the centre-of-mass energy of the hadron-hadron collision and is given by:
\begin{equation}
    \ptmin = \ptmino \left(\frac{\sqrts}{\Eo}\right)^\bParameter,
\end{equation}
where \ptmino is the value of \ptmin at a reference energy scale \Eo, which is set to 7\TeV, \sqrts is the centre-of-mass energy of the hadron-hadron collision, and the parameter \bParameter controls the energy dependence of \ptmin.  
Decreasing the value of \ptmin increases the number of hard interactions whilst reducing the number of soft interactions, which typically increases the amount of activity in the UE.

The average number \meanMPI of these additional hard interactions per hadron-hadron collision is given by:
\begin{equation}
\label{eq:meanMPI}
\meanMPI=\Aoverlap\sigmaMPI,
\end{equation}
where \sigmaMPI is the production cross section of a pair of partons with $\pt>\ptmin$ and \Aoverlap describes the overlap between the two protons at a given impact parameter \impactParam.  The form of the overlap function is given by:
\begin{equation}
\Aoverlap = \frac{\musParameter}{96\pi}(\mu \impactParam)^{3} K_{3},
\end{equation}
where $\musParameter$ is the inverse proton radius squared, and $K_{3}\equiv{K_{3}(\mu \impactParam)}$ is the modified Bessel function of the third kind.  The overlap function is obtained by the convolution of the electromagnetic form factors of two protons. The number of additional hard interactions per hadron-hadron collision at a given $\impactParam$ is described by a Poissonian probability distribution with a mean given by Eq.~(\ref{eq:meanMPI}), which is then integrated over the impact parameter space.  Increasing \musParameter  increases the density of the partons in the hadrons, and results in a higher probability for additional hard scatterings to take place.

Additional soft interactions, which produce pairs of partons below \ptmin, are based on a model of multiperipheral particle production~\cite{Herwig7SoftModel}.  The number of additional soft interactions between the two hadron remnants is described in a similar way to the hard interactions above \ptmin.  In a soft interaction between the two hadron remnants, the mean number of particles produced is given by :
\begin{equation}
\langle N \rangle = \No \left( \frac{s}{1 \TeV^2} \right)^\PParameter \ln \frac{\left(p_{\mathrm{r1}} + p_{\mathrm{r2}}\right)^2}{m^{2}_{\mathrm{rem}}},
\end{equation}
where $p_{\mathrm{r1}}$ and $p_{\mathrm{r2}}$ are the four-momenta of the two remnants, and $m_{\mathrm{rem}}$ is the mass of a proton remnant, \ie the remaining valence quarks of a proton treated as a diquark system, and is set to $0.95\GeV$. The parameters \No and \PParameter control the energy dependence of the mean number of soft particles produced.  They were tuned to MB data, which resulted in the values $\PParameter=-0.08$ and $\No=0.95$~\cite{Herwig71}. In deriving the tune \SoftTune the values of \No and \PParameter were kept fixed at these values.

The cluster model~\cite{ClusterModel} is used to model the hadronization of quarks into hadrons.  After the PS calculation, gluons are split into quark-antiquark pairs, and a cluster is formed from each colour connected pair of quarks.  Before hadrons are produced from the clusters, CR can modify the configuration of the clusters.  With the PCR model, the quarks from two clusters can be reconfigured to form two alternative clusters.  The change of the cluster configuration takes place only if the sum of the masses of the new clusters is smaller than before.  If this condition is satisfied, the CR is accepted with a probability \preco, which is the only parameter of the PCR model.  The PCR model typically leads to clusters with smaller invariant mass compared with the clusters that would be obtained without CR, and will typically reduce the overall activity in the UE.

\section{Tuning procedure}
\label{sec:TuneProcedure}

We derive three tunes based on the \NNPDF PDF sets~\cite{NNPDF31}.  A different PDF set is chosen for each aspect of the \HerwigS simulation: hard scattering, parton showering, MPI, and beam remnant handling.  The value of \alpS at a scale equal to the \PZ boson mass \MZ in each tune is set to $\alpSMZ=0.118$ for all parts of the \HerwigS simulation, with a two-loop running of \alpS.

{\tolerance=800 The first tune, \CHo (``CMS \HERWIG{}''), uses an NNLO PDF set in all aspects of simulation in \HerwigS, where the PDF was derived with a value of $\alpSMZ=0.118$.  This is equivalent to the choice of PDF and \alpSMZ used in the CP5 \PYTHIAEIGHT tune~\cite{GEN17001}.
In the second tune, \CHt, an LO PDF set that was also derived with $\alpSMZ=0.118$, is used in the simulation of MPI and beam remnant handling, whereas an NNLO PDF set is used elsewhere.  The final tune, \CHth, is similar to \CHt, but uses an LO PDF set that was derived with $\alpSMZ=0.130$ for the simulation of MPI and remnant handling.
The choice of an LO PDF set for the simulation of MPI and beam remnant handling, regardless of the choice of PDF used in the PS and ME calculation, is motivated by ensuring that the gluon PDF is positive at the low energy scales involved, which is not necessarily the case with higher-order PDF sets.  However, as was shown in Ref.~\cite{GEN17001}, the gluon PDF in the NNLO \NNPDF set remains positive at low energy scales, and predictions from \PYTHIAEIGHT using LO and higher-order PDFs can both give a reliable description of MB data.
The configurations of PDF sets in the CH1, CH2, and CH3 tunes allow us to study whether using an NNLO PDF set consistently for all aspects of the \HerwigS simulation, or an LO PDF set for the simulation of MPI, can both give a reliable description of MB data.
For both of these choices the gluon PDF is positive at low energy scales. \par}

The names of the parameters being tuned in the \HerwigS configuration, and their allowed ranges in the fit, are shown in Table~\ref{tab:ParameterRanges}.
The values of $\No=0.95$ and $\PParameter=-0.08$ are fixed at the values that were used in the tune \SoftTune.  As shown later, no further tuning of these parameters is necessary, because of the good description of measured observables obtained with these values. 

\begin{table*}
\topcaption{Parameters considered in the tuning, and their allowed ranges in the fit.}
\centering
\renewcommand{\arraystretch}{1.2}
\cmsTable{
\begin{tabular}{ c  c  c }
Parameter & \HerwigS configuration parameter & Range  \\ 
\hline
\ptmino (\GeVns) & /Herwig/UnderlyingEvent/MPIHandler:pTmin0 & 1.0--5.0 \\
\bParameter & /Herwig/UnderlyingEvent/MPIHandler:Power & 0.1--0.5 \\
\musParameter (\musunits) & /Herwig/UnderlyingEvent/MPIHandler:InvRadius & 0.5--2.7 \\
\preco &  /Herwig/Hadronization/ColourReconnector:ReconnectionProbability & 0.05--0.90 \\
\end{tabular}
}
\label{tab:ParameterRanges}
\end{table*}

\begin{figure}[thbp]
  \centering
  \includegraphics[width=0.49\textwidth]{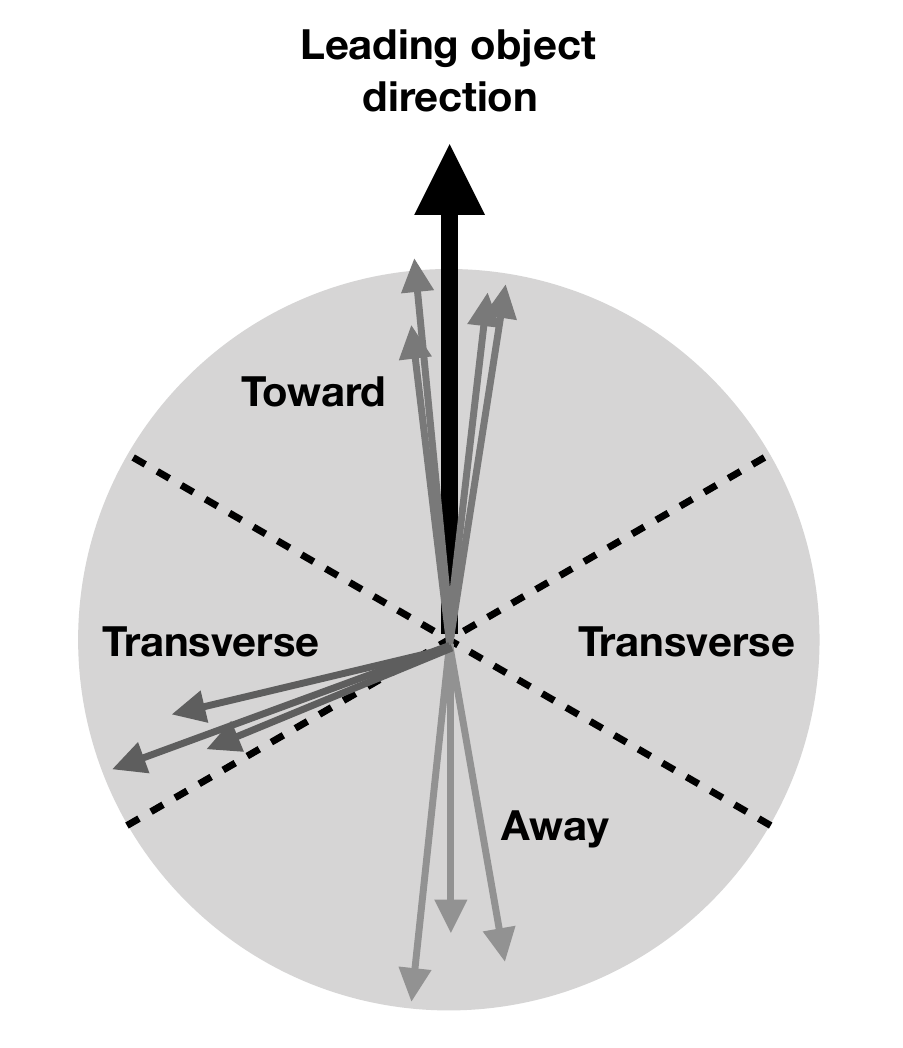}
  \caption{Illustration of the different $\phi$ regions, with respect to the leading object in an event, used to probe the properties of the UE in measurements. }
  \label{fig:UEDiagram}
\end{figure}

The tunes are derived by fitting unfolded MB data that are available in the \Rivet~\cite{RivetManual} toolkit.  The proton-proton collision data used in the fit were collected by the CMS experiment at $\sqrts=0.9$, 7, and 13\TeV.
In measurements probing the UE, charged particles in a particular event are typically categorized into different \etaphi regions with respect to a leading object in that event, such as the highest \pt track or jet, as illustrated in Fig.~\ref{fig:UEDiagram}.  The difference in azimuthal $\phi$ between each charged particle and the leading object (\deltaPhi) is used to assign each charged particle to a region, namely the toward ($\abs{\deltaPhi}\leq60^{\circ}$), away ($\abs{\deltaPhi}>120^{\circ}$), and transverse regions ($60<\abs{\deltaPhi}\leq120^{\circ}$).  The properties of the charged particles in the transverse regions are the most sensitive to the modelling of the UE.  The two transverse regions can be further divided into the transMin and transMax regions, which are the regions with the least and most charged-particle activity, respectively.  Data that have been categorized in this way are referred to as UE data in this paper.

At $\sqrts=7$ and 13\TeV, the \Nch and transverse momentum sum (\ptsum), with respect to the beam axis, as functions of the \pt of the leading track (\ptleadingtrack) in the transMin and transMax regions are used in the fit~\cite{CMS-PAS-FSQ-12-020,CMS-PAS-FSQ-15-007}.  At $\sqrts=0.9\TeV$, the observables used are the \Nch and \ptsum in the transverse region, as a function of the \pt of the leading jet (\ptleadingtrackjet)~\cite{CMSTradUE}.  The track jets are clustered using the \SISCONE algorithm~\cite{SISCone} with a distance parameter of 0.5.  The regions $\ptleadingtrack<3\GeV$ and $\ptleadingtrackjet<3\GeV$ are not included in the fit because the parameters of diffractive processes, which dominate this region, are not considered.  The charged-hadron multiplicity as a function of $\eta$, \dNdeta, as measured by CMS at $\sqrts=13\TeV$ with zero magnetic field strength ($\mathrm{B}=0\unit{T}$)~\cite{CMSdNdeta} is also used in the fitting procedure.  The charged-particle \pt and $\eta$ as measured by CMS in Ref.~\cite{CMS7TeVMB} are not considered here, since they are biased by predictions obtained with \PYTHIASIX~\cite{Pythia6}, as discussed in Ref.~\cite{Herwig7SoftModel}.

{\tolerance=800 The tuning is performed within the \Professor (v1.4.0) framework~\cite{Professor}.  Around 60 random choices of the parameters are made, and predictions for each of these choices are obtained using \Rivet.  Approximately 10 million MB events are generated for each choice of parameters, such that the uncertainty in the prediction in any bin is typically not larger than the uncertainty in the data in the same bin. \par}

The fit is performed by minimising the \chis function:
\begin{equation}
\label{eq:chis}
    \chis(p) = \sum_{\mathcal{O}}\weight\sum_{\bin\in\mathcal{O}}\frac{(f^{\bin}(p)-\mathcal{R}_\bin)^2}{\Delta^2_\bin},
\end{equation}
where $\mathcal{R}_\bin$ is the measured content of bin \bin of the distribution of observable $\mathcal{O}$, while $f^{\bin}(p)$ is the predicted content in bin \bin, which is obtained by \Professor from a parameterization of the dependence of the prediction on the tuning parameters $p$.  The total uncertainty in the data and the simulated prediction in bin \bin of a given observable is denoted by $\Delta^2_\bin$, and \weight is a weight that increases or decreases the importance of an observable $\mathcal{O}$ in the fit.  The weight is typically set to $\weight=1$.  However, for the \CHo tune, where the PDF set used in the simulation of MPI and beam remnants is an NNLO set instead of an LO set, the weight is set to $\weight=3$ for the \dNdeta distribution.  This is the smallest weight that ensures the distribution is well described after the tuning.
Beyond this, the parameters for the three tunes and their predictions are stable with respect to a change in the weight assigned to the \dNdeta distribution in the fit.
Correlations between the bins \bin are not taken into account when minimising Eq.~(\ref{eq:chis}), because these were not available for the used input distributions.  
A third-order polynomial is used to parameterize the dependence of the prediction on the tuning parameters. Using a fourth-order polynomial to perform this interpolation between the 60 choices of parameters has a negligible effect on the outcome of the fits.

The number of degrees of freedom (\ndof) in the fit is calculated as:
\begin{equation}
    \ndof = \frac{(\sum_{\mathcal{O}}\sum_{\bin\in\mathcal{O}}\weight)^2}{\sum_{\mathcal{O}}\sum_{\bin\in\mathcal{O}}\weight^2} - \nparam,
\end{equation}
where \nparam is the number of parameters being optimized in the fit.

\section{Results from the new \texorpdfstring{\HerwigSTitle}{HERWIG 7} tunes}
\label{sec:HerwigTunes}

The tuned values of the parameters and the \chis values from the fit, \ie the minimum values of Eq.~(\ref{eq:chis}), divided by the \ndof of the fit are shown in Table~\ref{tab:TunedValues}, along with the values of the parameters for the default tune \SoftTune. The \ndof in the fit is 118 for \CHo, and 152 for \CHt and \CHth.  To provide a comparison between the compatibilities of the \CH tunes and \SoftTune with the data, the \chis/\ndof corresponding to the prediction of SoftTune and the data is also shown with \ndof set to 152.

\begin{table*}
\topcaption{Value of the parameters for the \SoftTune~\cite{Herwig7SoftModel,Herwig71}, \CHo, \CHt, and \CHth tunes.}
\centering
\cmsTable{
\begin{tabular}{ c  c  c  c  c  c }
\multicolumn{2}{c}{}   & \SoftTune & CH1 & CH2 & CH3  \\ 
\hline
\multicolumn{2}{c}{\alpSMZ}   & 0.1262 & 0.118 & 0.118 & 0.118 \\[\cmsTabSkip]
\multirow{2}{*}{PS} & PDF set  & \MMHTLO & \NNPDF NNLO  & \NNPDF NNLO & \NNPDF NNLO \\
                    & \alpSPDFMZ  & 0.135 & 0.118  & 0.118 & 0.118 \\
MPI \& & PDF set  & \MMHTLO & \NNPDF NNLO  & \NNPDF LO & \NNPDF LO \\
remnants                    & \alpSPDFMZ & 0.135 & 0.118  & 0.118 & 0.130 \\[\cmsTabSkip]
\multicolumn{2}{c}{\ptmino (\GeVns)} & 3.502 & 2.322 & 3.138 & 3.040 \\
\multicolumn{2}{c}{\bParameter} & 0.416 & 0.157 & 0.120 & 0.136 \\
\multicolumn{2}{c}{\musParameter (\musunits)} & 1.402 & 1.532 & 1.174 & 1.284 \\
\multicolumn{2}{c}{\preco} & 0.5 & 0.400 & 0.479 & 0.471 \\[\cmsTabSkip]
\multicolumn{2}{c}{\chis/\ndof} & 12.8 & 6.75 & 1.54 & 1.71 \\
\end{tabular}
}
\label{tab:TunedValues}
\end{table*}

The values of the parameters of the MPI model are intertwined with each other since they are tuned simultaneously to reproduce the amount of UE activity observed in the data. Nonetheless, a general interpretation of the variations in the tuned parameters for each tune can be distinguished.
For example, the value of \ptmino is lower for all three \CH tunes than for \SoftTune, and significantly lower for \CHo, which increases the amount of MPI in an event compared to that with the tune \SoftTune.  

The lower value of \bParameter for all \CH tunes further increases the contribution of MPI in collisions at $\sqrts=13\TeV$.
Because of the lower values of \preco, the amount of CR in the \CH tunes is lower than in \SoftTune.  This also has the effect of increasing the overall amount of activity in the UE for the \CH tunes.  The value of \musParameter for \CHt and \CHth is lower than the corresponding value for \SoftTune.
Even though a lower value of \musParameter would lead to a lower amount of MPI in a given event, the combined effect of the parameters of the \CH tunes results in a larger amount of MPI compared with \SoftTune.

The tuned parameters of \CHt and \CHth are fairly similar, as are the values of \chis/\ndof of these two tunes, indicating that the choice of \alpSMZ used when deriving the LO PDF set in the simulation of MPI does not have a large effect.
The parameters for the tune \CHo differ from those for the tunes \CHt and \CHth, and the value of \chis/\ndof is larger, implying that using an LO PDF set is somewhat preferred over an NNLO PDF set for the simulation of MPI.  In the following, the predictions from the three \CH tunes are compared with the data used in the tuning procedure.  These predictions are obtained by generating events with the corresponding parameters shown in Table~\ref{tab:TunedValues} rather than from the parameterization of the tune parameters used in the fit.

\begin{figure}[htbp]
  \centering
  \includegraphics[width=0.49\textwidth]{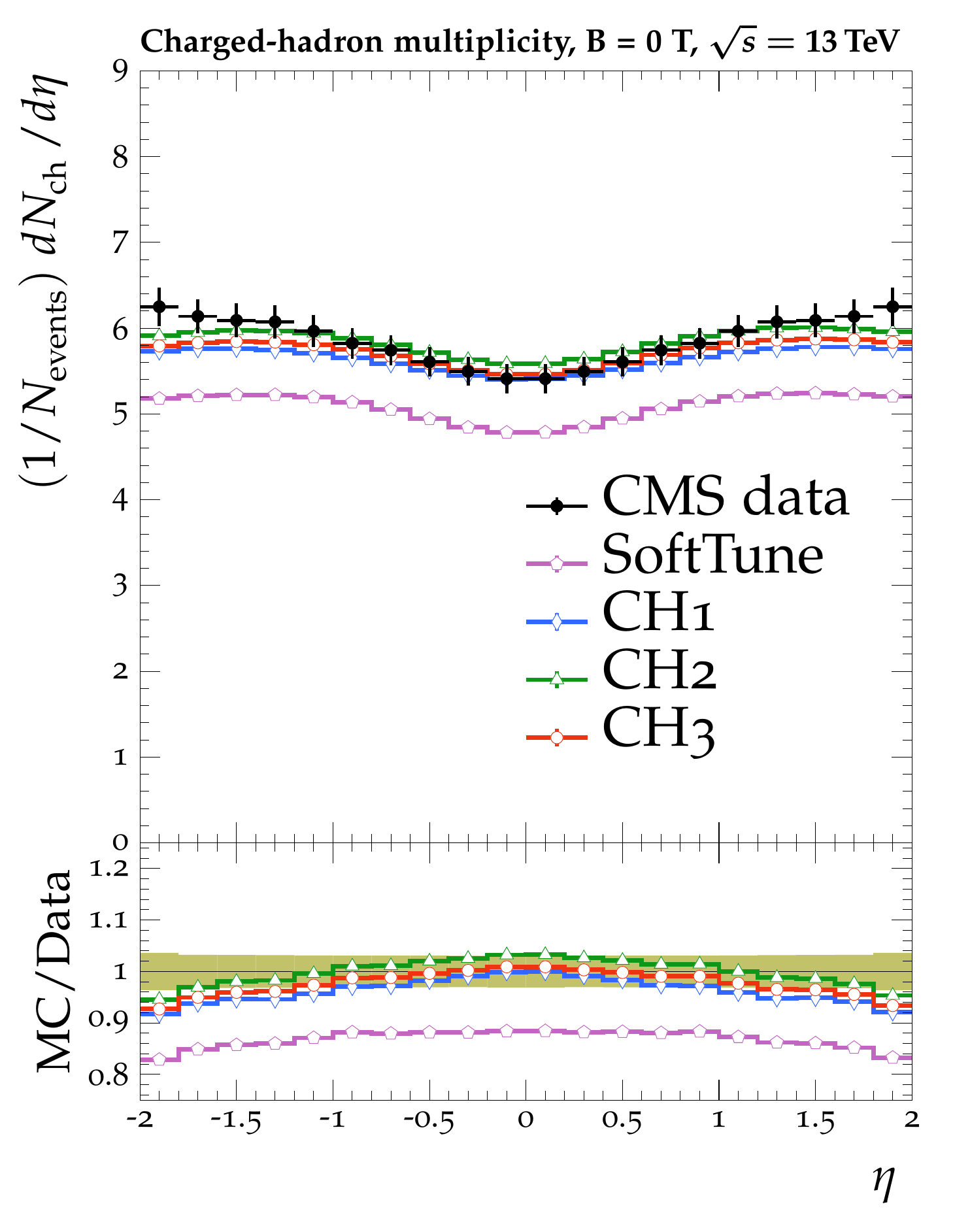}
  \caption{The normalized \dNdeta of charged hadrons as a function of $\eta$~\cite{CMSdNdeta}. CMS MB data are compared with \SoftTune and the \CH tunes.  \captionColouredShadedBand  }
  \label{fig:CMS_MB_13TeV}
\end{figure}

Figure~\ref{fig:CMS_MB_13TeV} shows the normalized \dNdeta of charged hadrons as a function of $\eta$ at 13\TeV in MB events.  Only the predictions for \SoftTune deviate significantly from the data, and underestimate the \dNdeta in data by 10--18\%.  The \CH tunes each provide a slightly different prediction, but all have a similar level of agreement with the data.  
The \CH tunes compared with SoftTune predict an increase in the UE activity, which is observed.

\begin{figure*}[htb]
  \centering
  \includegraphics[width=0.49\textwidth]{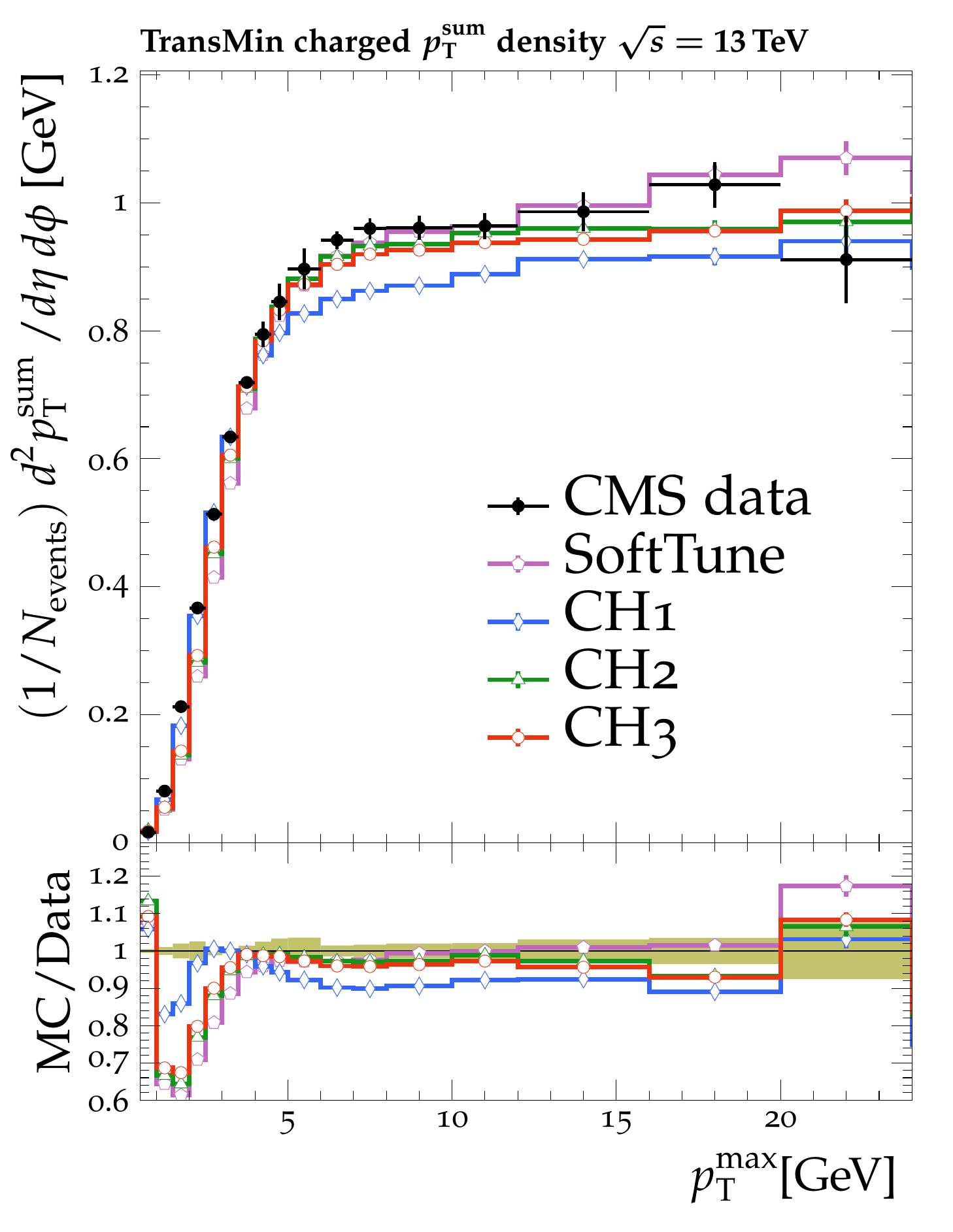}
  \includegraphics[width=0.49\textwidth]{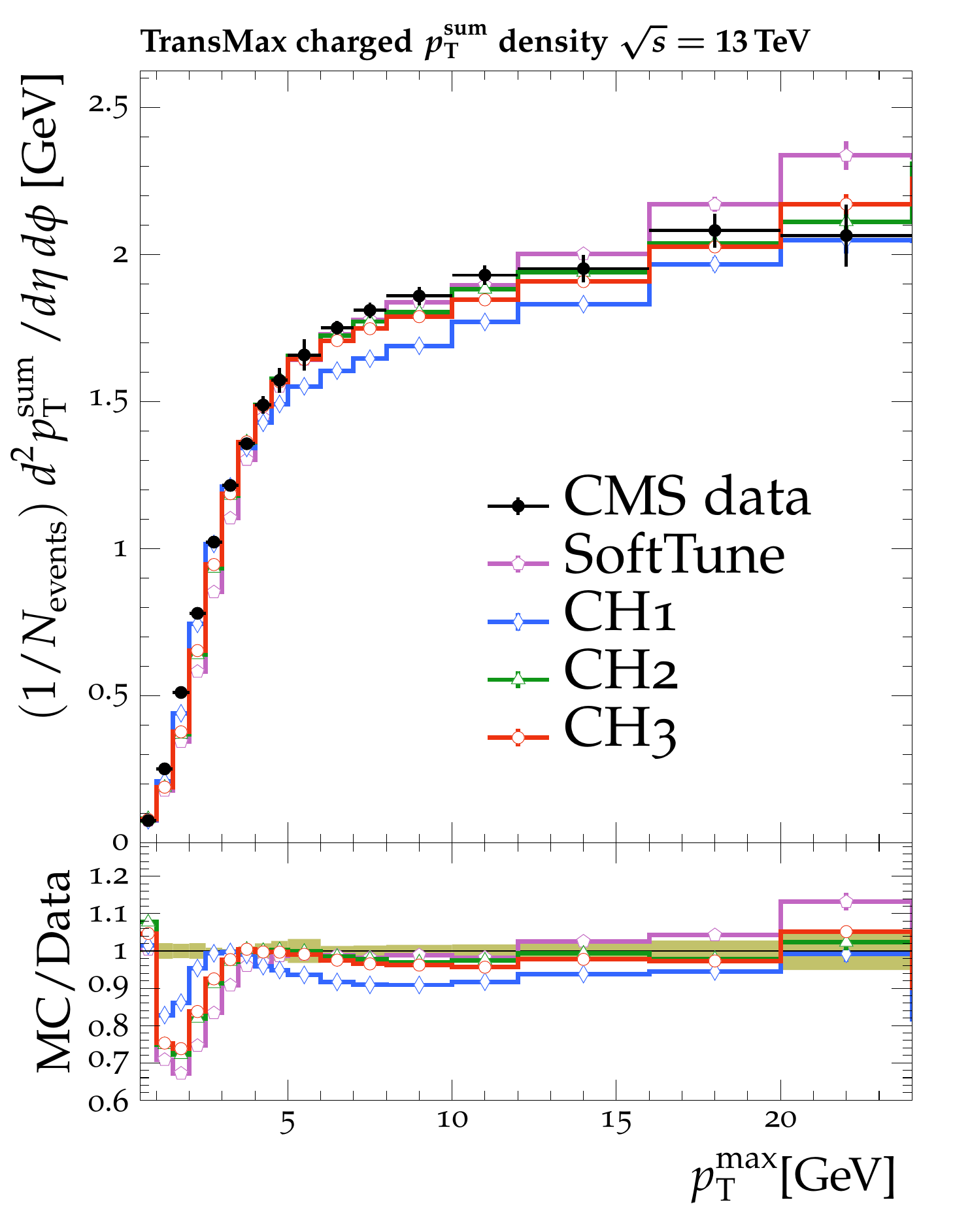} \\
  \includegraphics[width=0.49\textwidth]{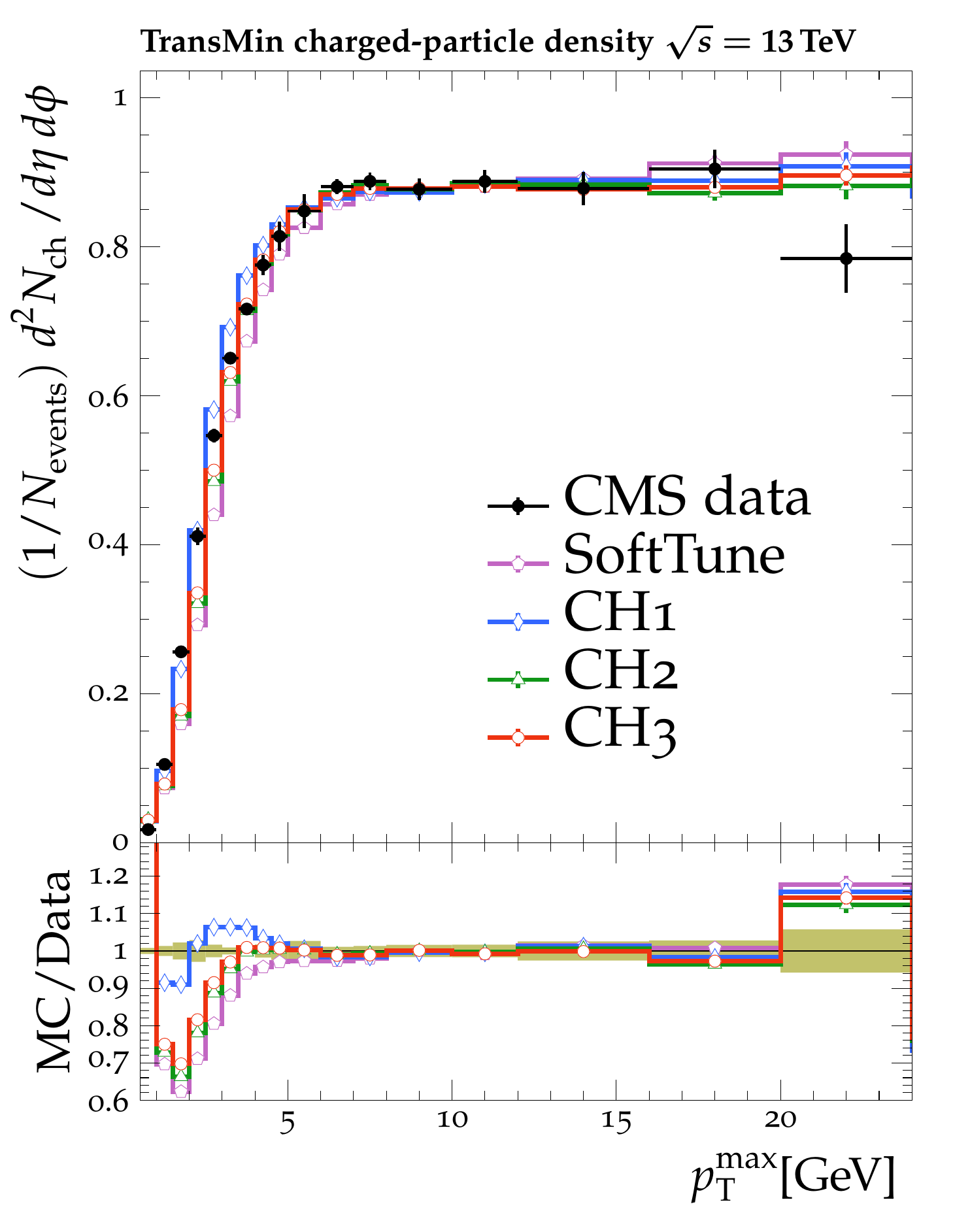}
  \includegraphics[width=0.49\textwidth]{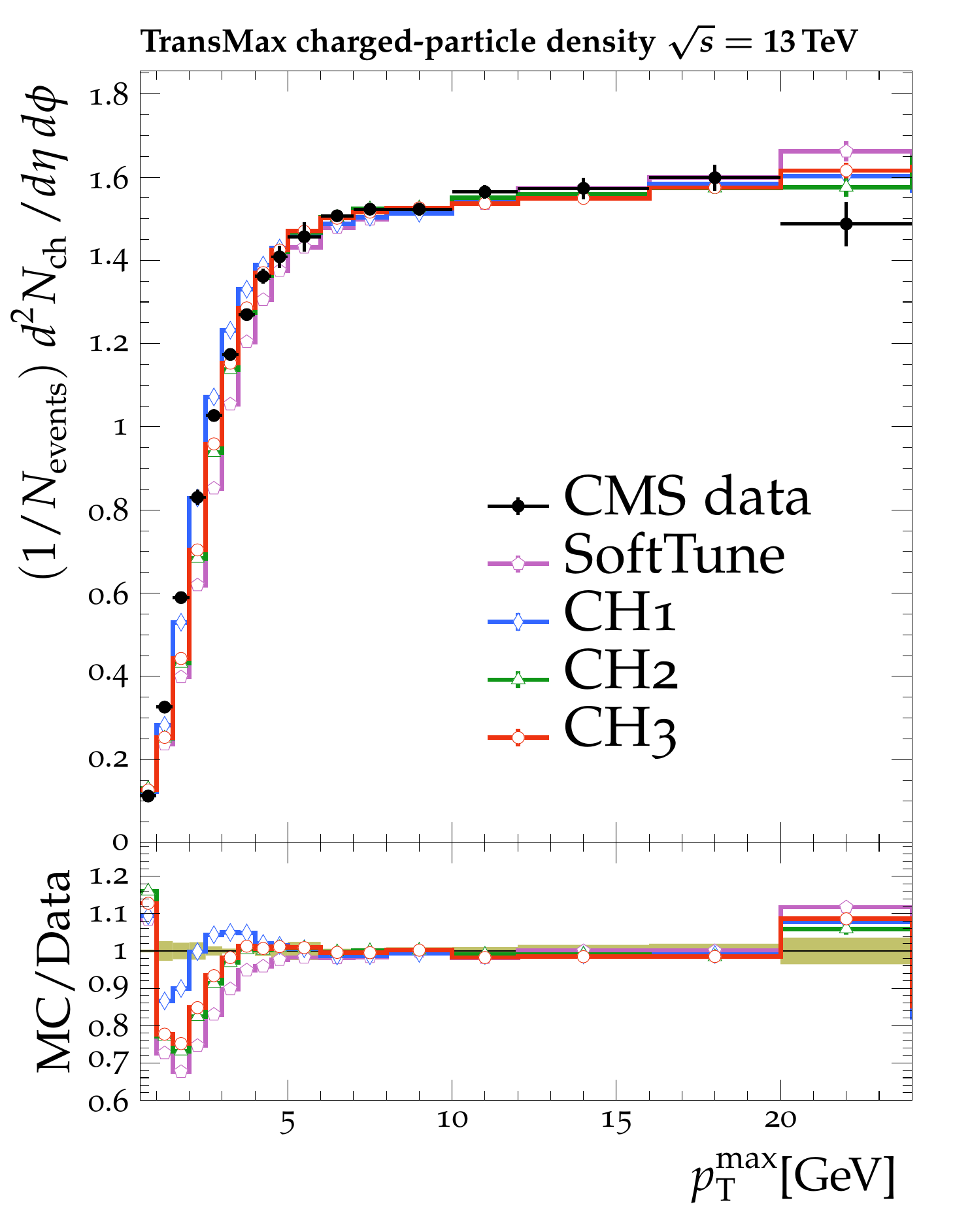}
  \caption{The normalized \ptsum (upper) and \Nch (lower) density distributions in the transMin (left) and transMax (right) regions, as a function of the \pt of the leading track, \ptleadingtrack~\cite{CMS-PAS-FSQ-15-007}.  CMS MB data are compared with the predictions from \HerwigS, with the \SoftTune and \CH tunes.  \captionColouredShadedBand}
  \label{fig:CMS_UE_13TeV}
\end{figure*}

{\tolerance=800 Figure~\ref{fig:CMS_UE_13TeV} shows the normalized \ptsum and \Nch densities as a function of \ptleadingtrack with comparisons from \SoftTune and the \CH tunes for both transMin and transMax.
The predictions of \SoftTune and the \CHt, \CHth tunes are broadly similar, and give a good description the data in the plateau region ($\ptleadingtrack\gtrsim4\GeV$).
In the rising part of the spectrum, the predictions from the tunes \CHt, \CHth, and \SoftTune deviate from the data in some bins by up to 40\%.
The \CHth tune provides the best predictions in the rising region of the spectrum.  However, only the region $\ptleadingtrack>3\GeV$ was included in the tuning procedure, because the region $\ptleadingtrack<3\GeV$ is dominated by diffractive processes whose model parameters are not used in the fit. \par}

The effect of using an NNLO PDF, instead of an LO PDF, in the simulation of MPI is seen from the predictions with the tune \CHo in Fig.~\ref{fig:CMS_UE_13TeV}.  This tune provides a good description of the \Nch distributions in both the transMin and transMax regions, and is typically within 10\% of the data.  However, the tune \CHo does not simultaneously provide a good description of the \ptsum distributions in either the transMin or transMax region, with a 10\% difference to the data in the plateau region of the corresponding transMax distribution.

\begin{figure*}[htb]
  \centering
  \includegraphics[width=0.49\textwidth]{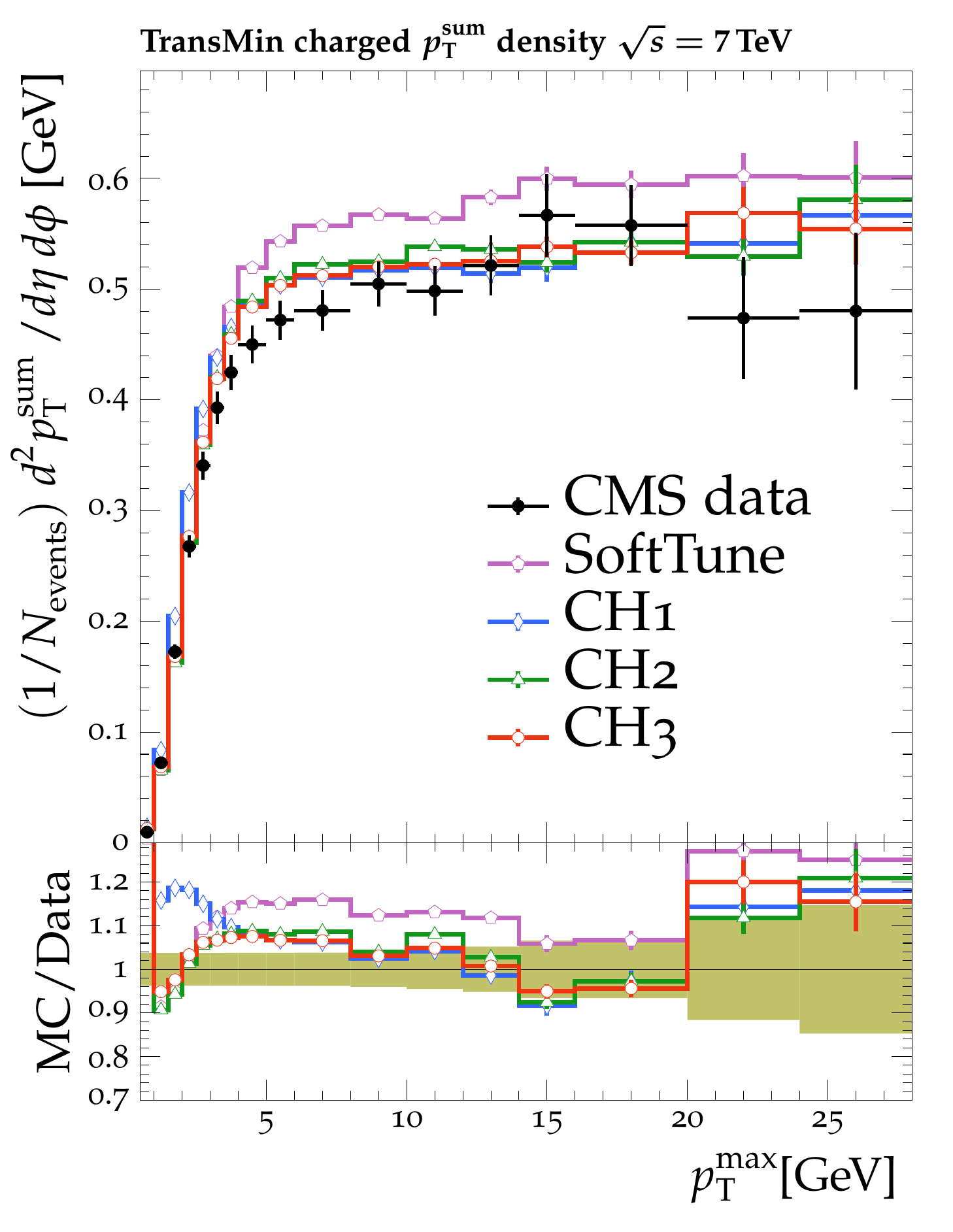}
  \includegraphics[width=0.49\textwidth]{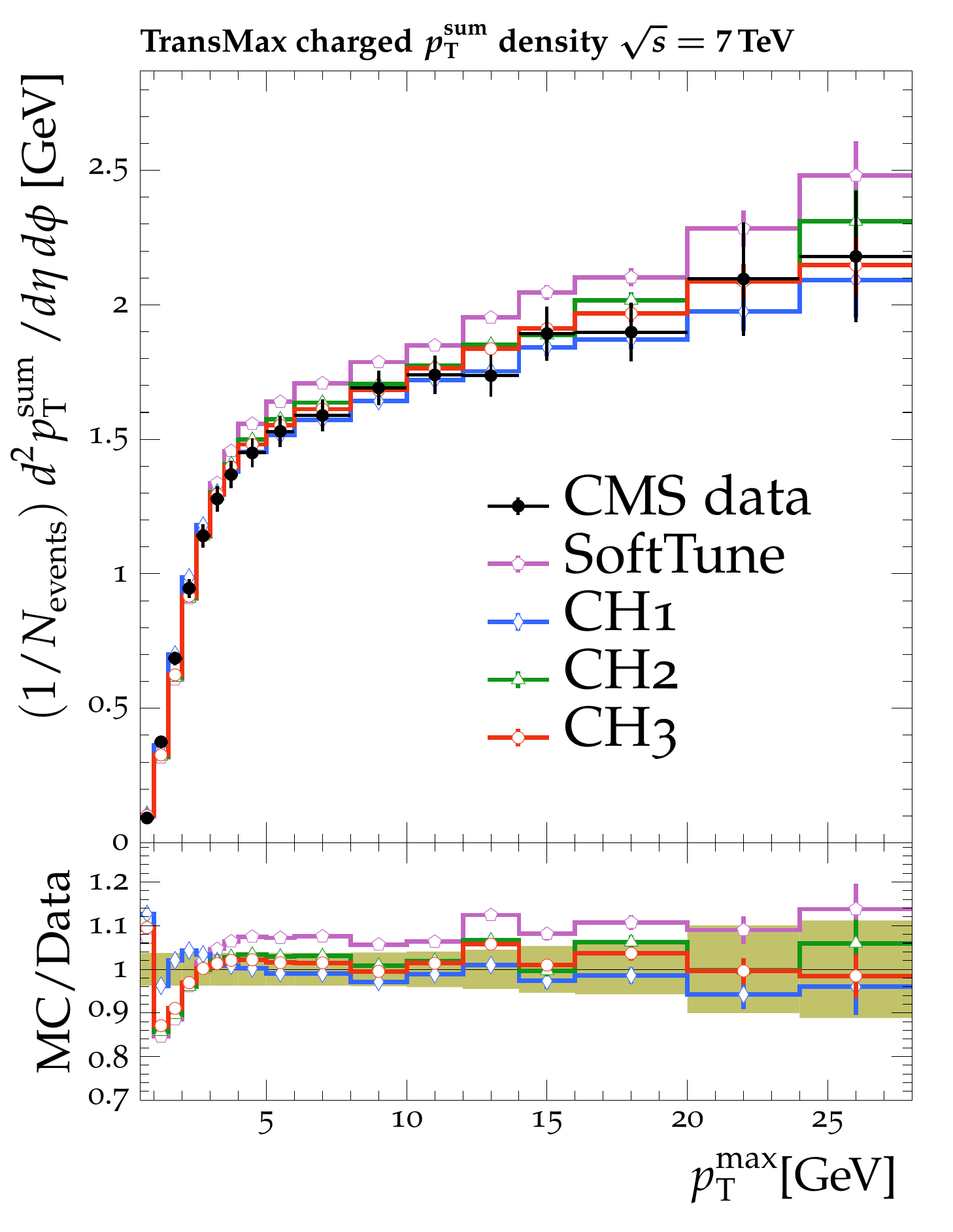} \\
  \includegraphics[width=0.49\textwidth]{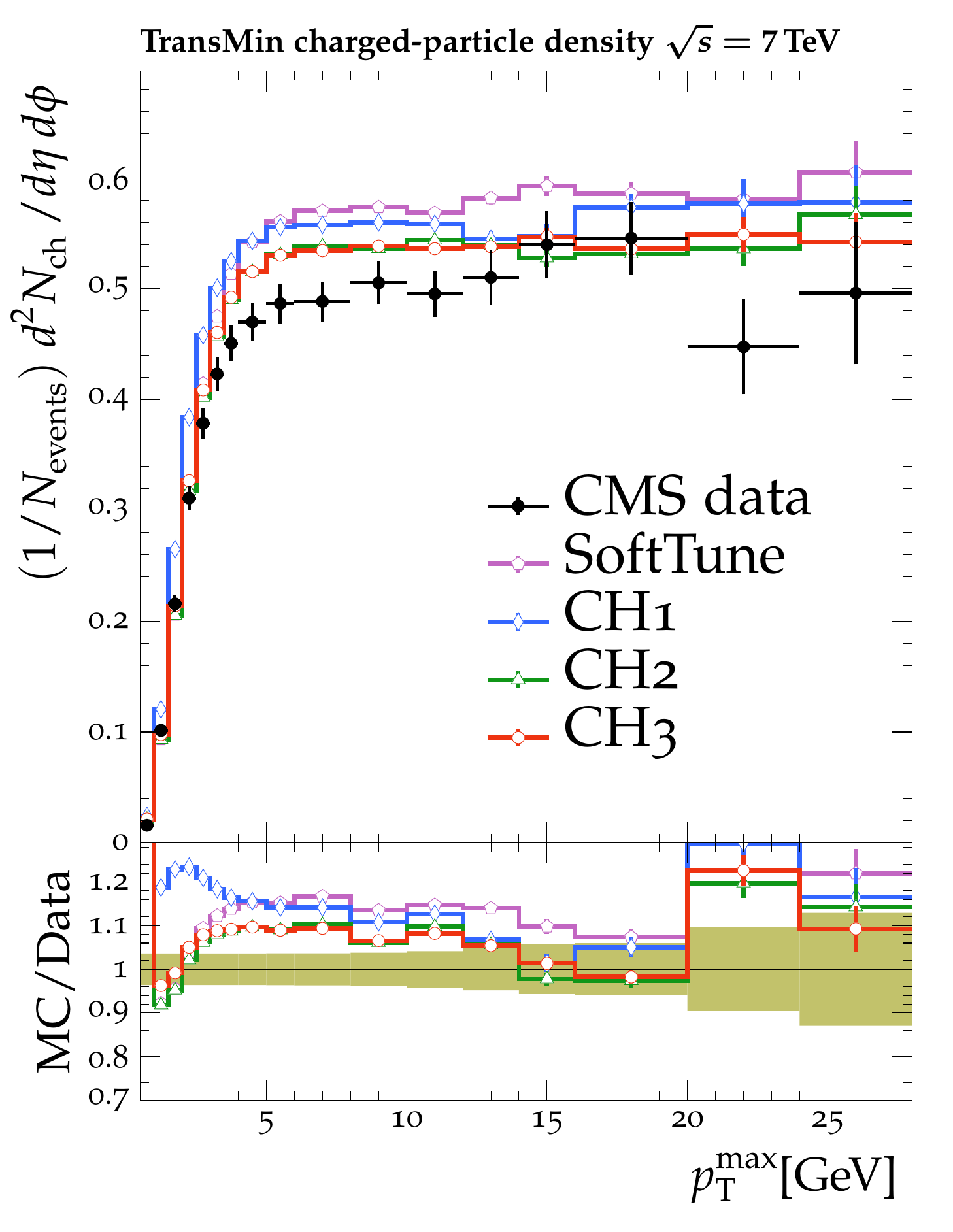}
  \includegraphics[width=0.49\textwidth]{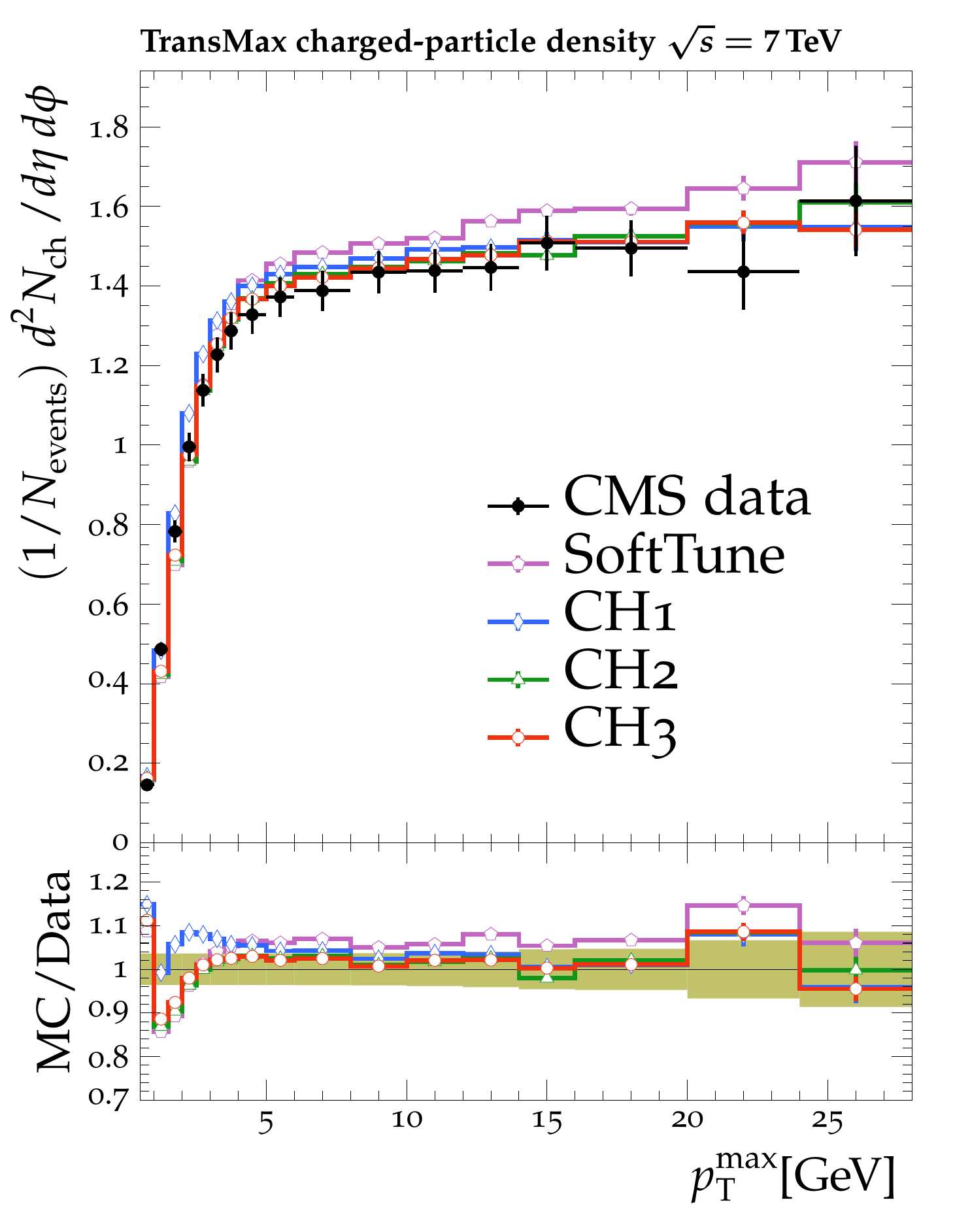}
  \caption{The \ptsum (upper) and \Nch (lower) density distributions in the transMin (left) and transMax (right) regions, as a function of the \pt of the leading track, \ptleadingtrack~\cite{CMS-PAS-FSQ-12-020}.  CMS MB data are compared with the predictions from \HerwigS, with the \SoftTune and \CH tunes.  \captionColouredShadedBand}
  \label{fig:CMS_UE_7TeV}
\end{figure*}

Figure~\ref{fig:CMS_UE_7TeV} shows the normalized \Nch and \ptsum densities as a function of \ptleadingtrack using UE data at 7 TeV and compared with various tunes. In the transMax region, the predictions from the \CH tunes describe the data well, with at most a 15\% discrepancy at low \ptleadingtrack.  In the transMin region, the predictions from all tunes
deviate from the data at intermediate values of $\ptleadingtrack\approx3\text{--}8\GeV$.  The deviation is up to $\approx$10\% for the \CHt and \CHth tunes, whereas the difference between data and the tunes \SoftTune and \CHo is larger than this.  The prediction of \CHo deviates further from the data at lower values of \ptleadingtrack.

\begin{figure*}[htb]
  \centering
  \includegraphics[width=0.49\textwidth]{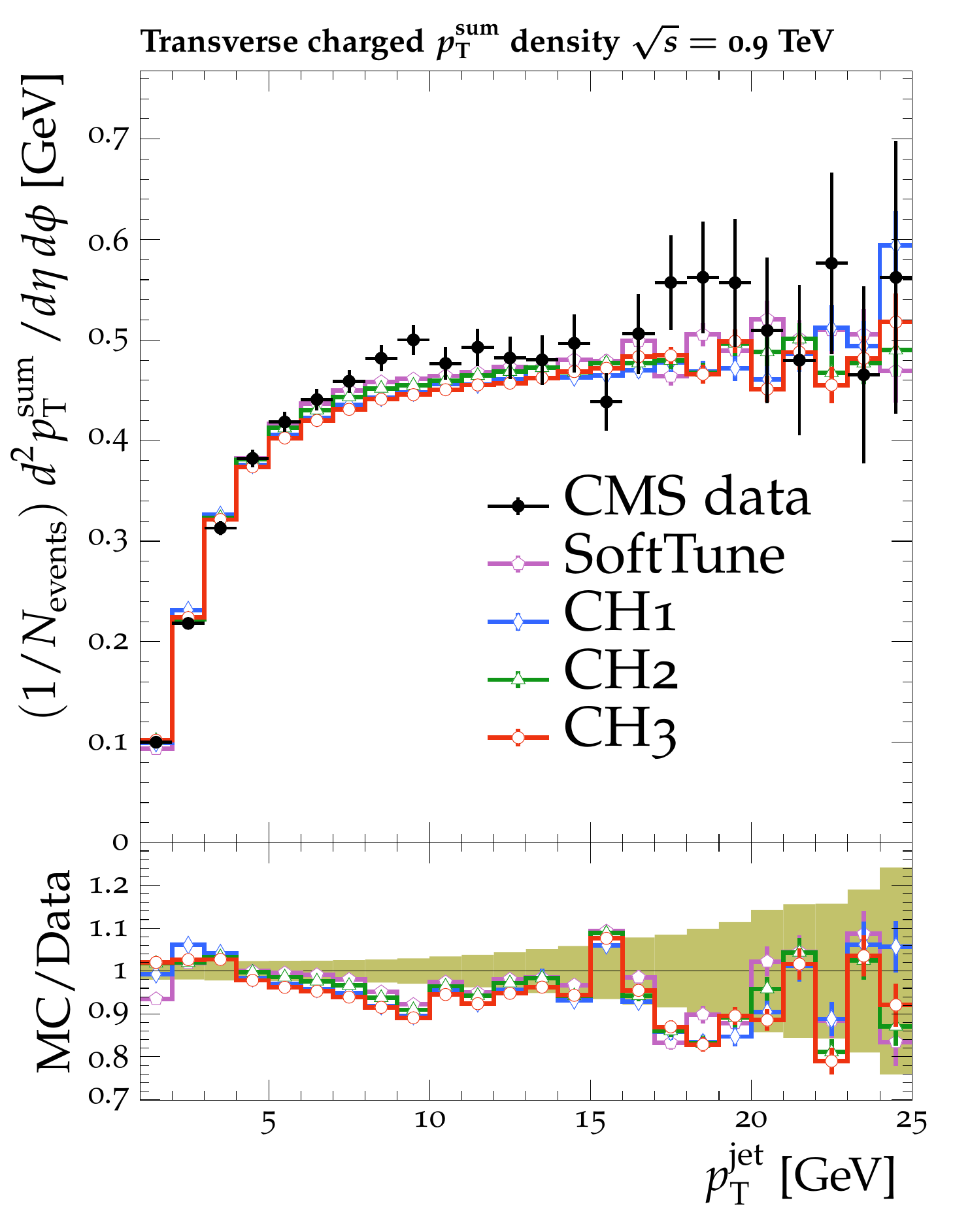}
  \includegraphics[width=0.49\textwidth]{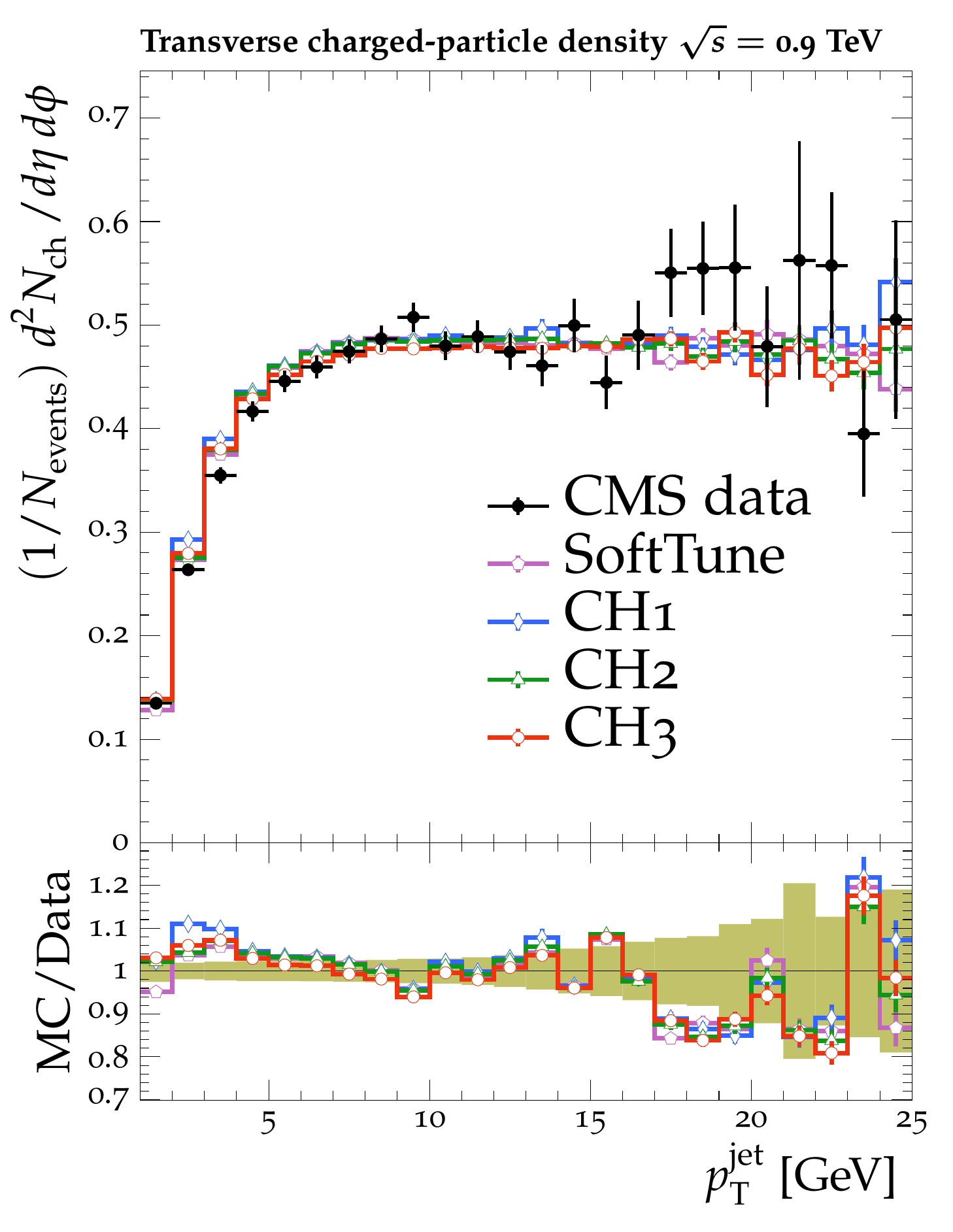} \\
  \caption{The \ptsum (left) and \Nch (right) density distributions in the transverse regions, as a function of the \pt of the leading track jet, \ptleadingtrackjet~\cite{CMSTradUE}.  CMS MB data are compared with the predictions from \HerwigS, with the \SoftTune and \CH tunes.  \captionColouredShadedBand}
  \label{fig:CMS_UE_0p9TeV}
\end{figure*}

The predictions are compared with UE data at $\sqrts=0.9\TeV$ to normalized \ptsum densities in the transverse regions in Fig.~\ref{fig:CMS_UE_0p9TeV}. All tunes provide a similar prediction of the observables above $\ptleadingtrackjet>4\GeV$, and agree with the data.  Some differences are apparent between the predictions at low \ptleadingtrackjet, with the tunes \CHt and \CHth providing a better description of the data compared to the tunes \CHo and \SoftTune.

\begin{figure*}[htb]
  \centering
  \includegraphics[width=0.49\textwidth]{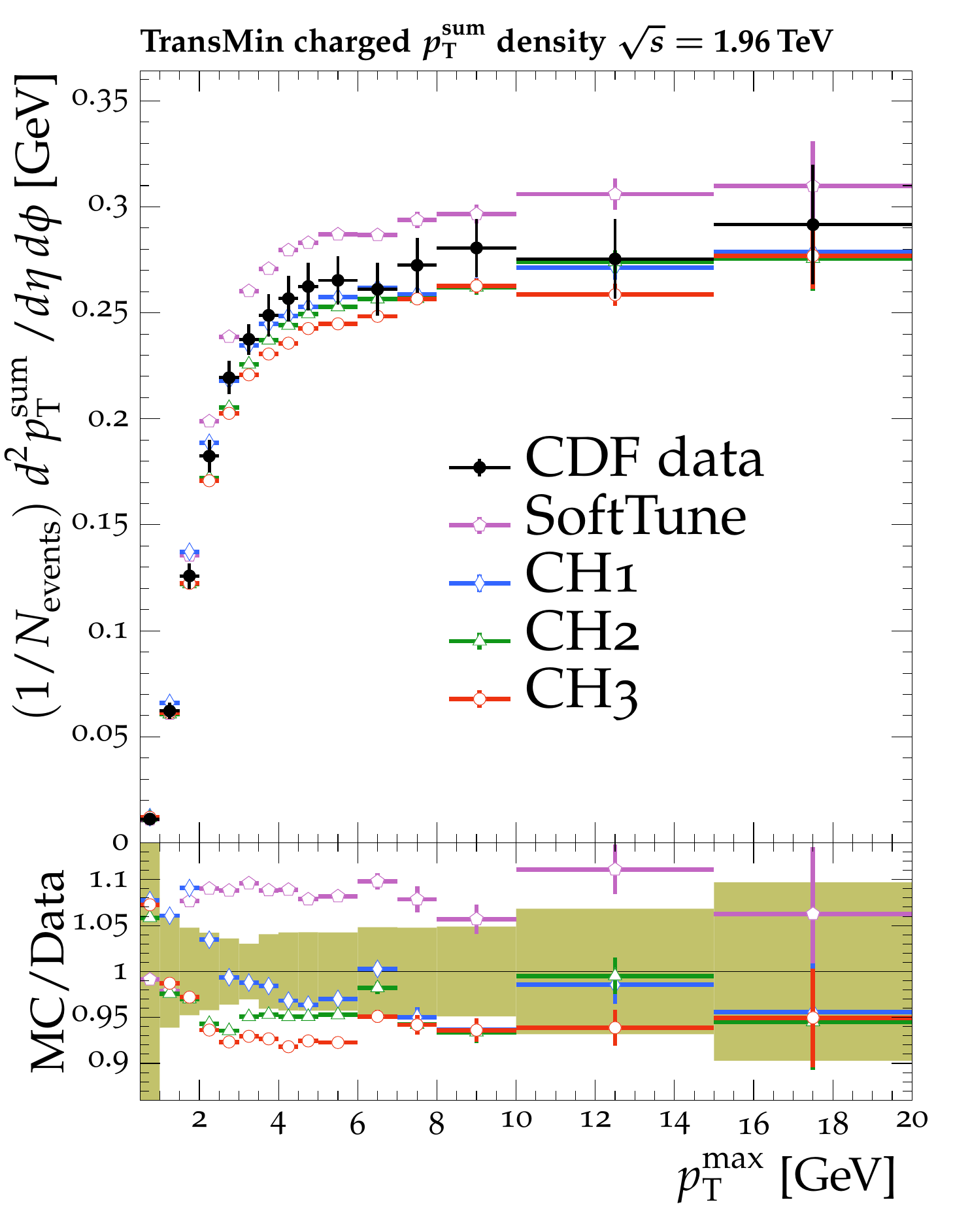}
  \includegraphics[width=0.49\textwidth]{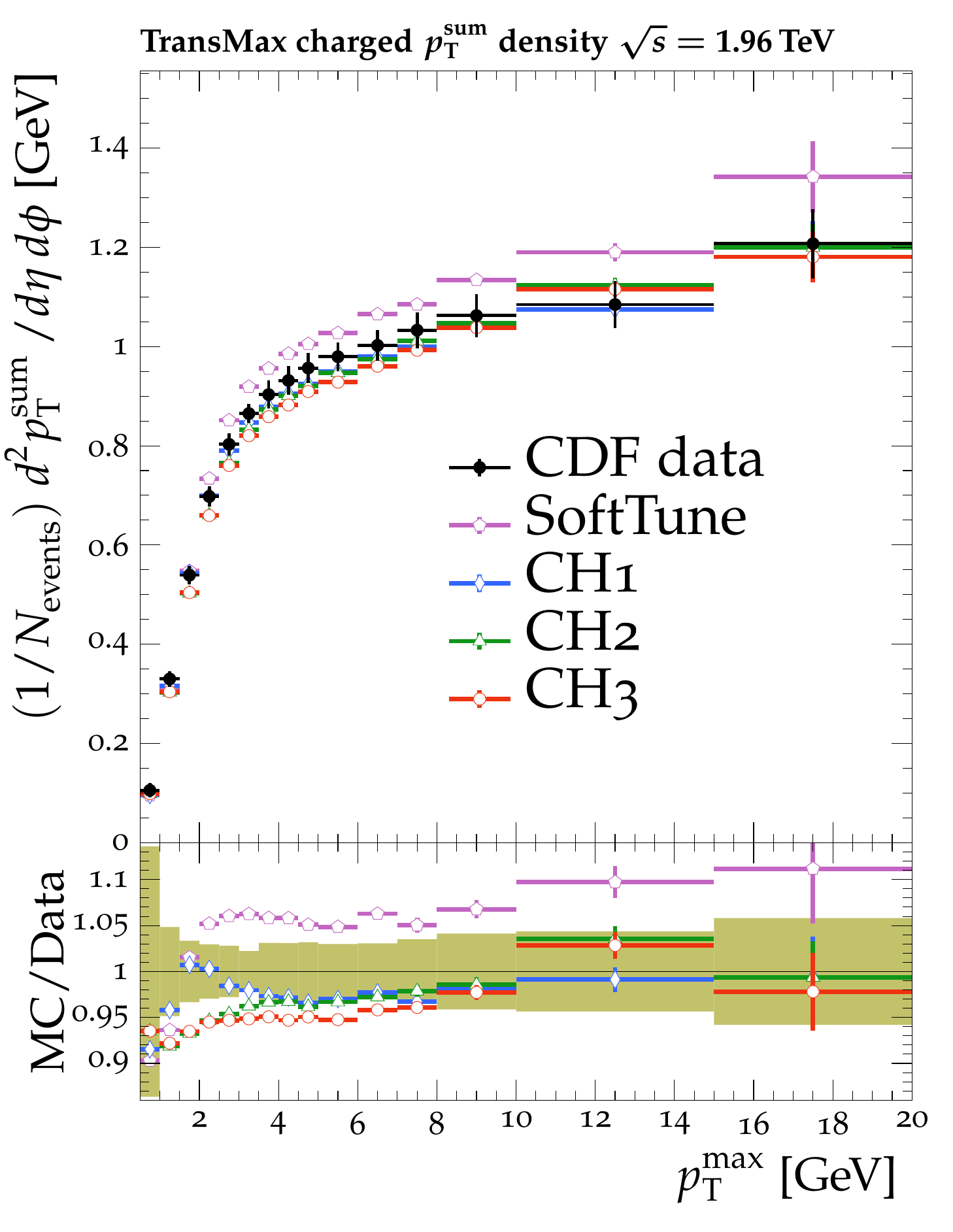} \\
  \includegraphics[width=0.49\textwidth]{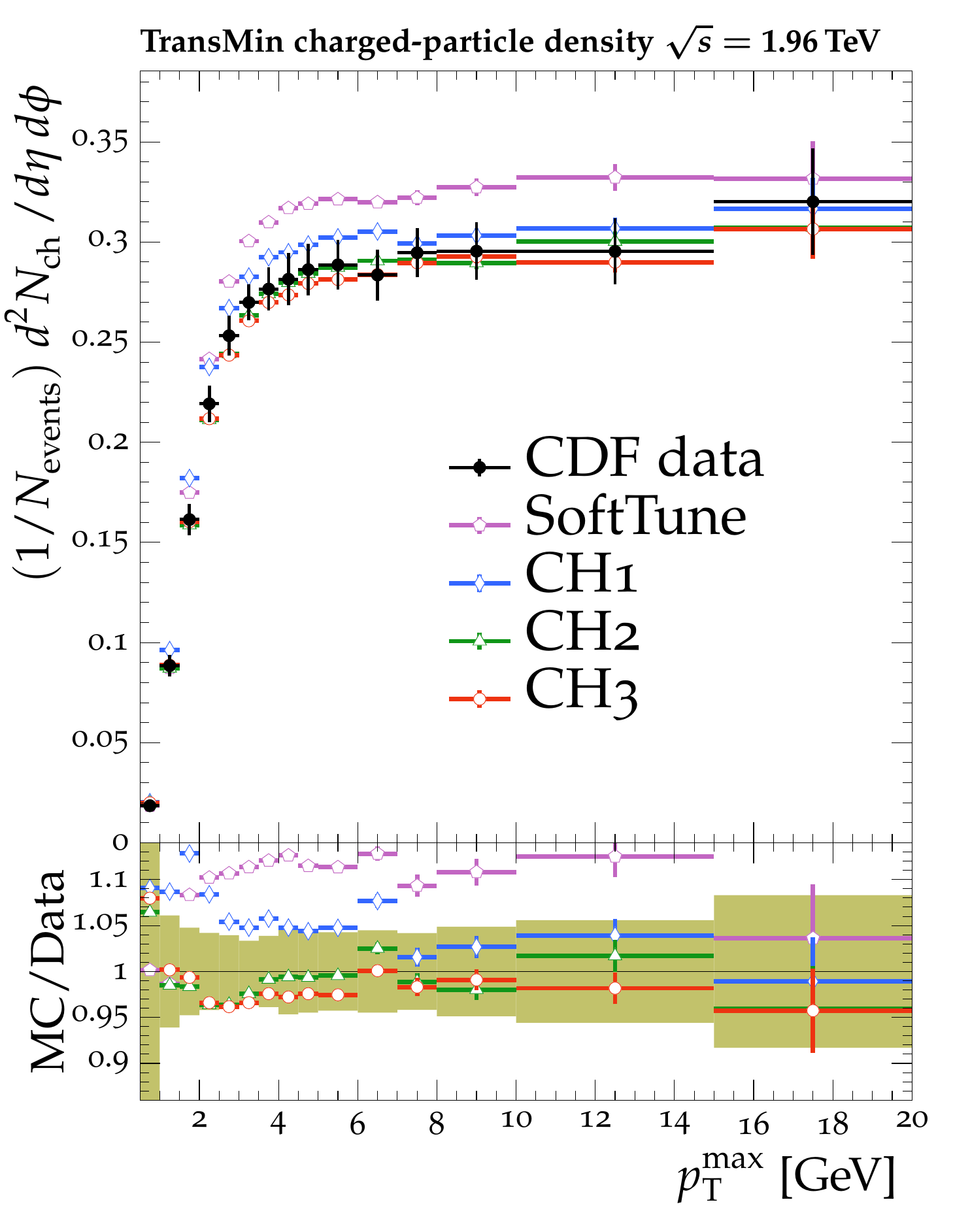}
  \includegraphics[width=0.49\textwidth]{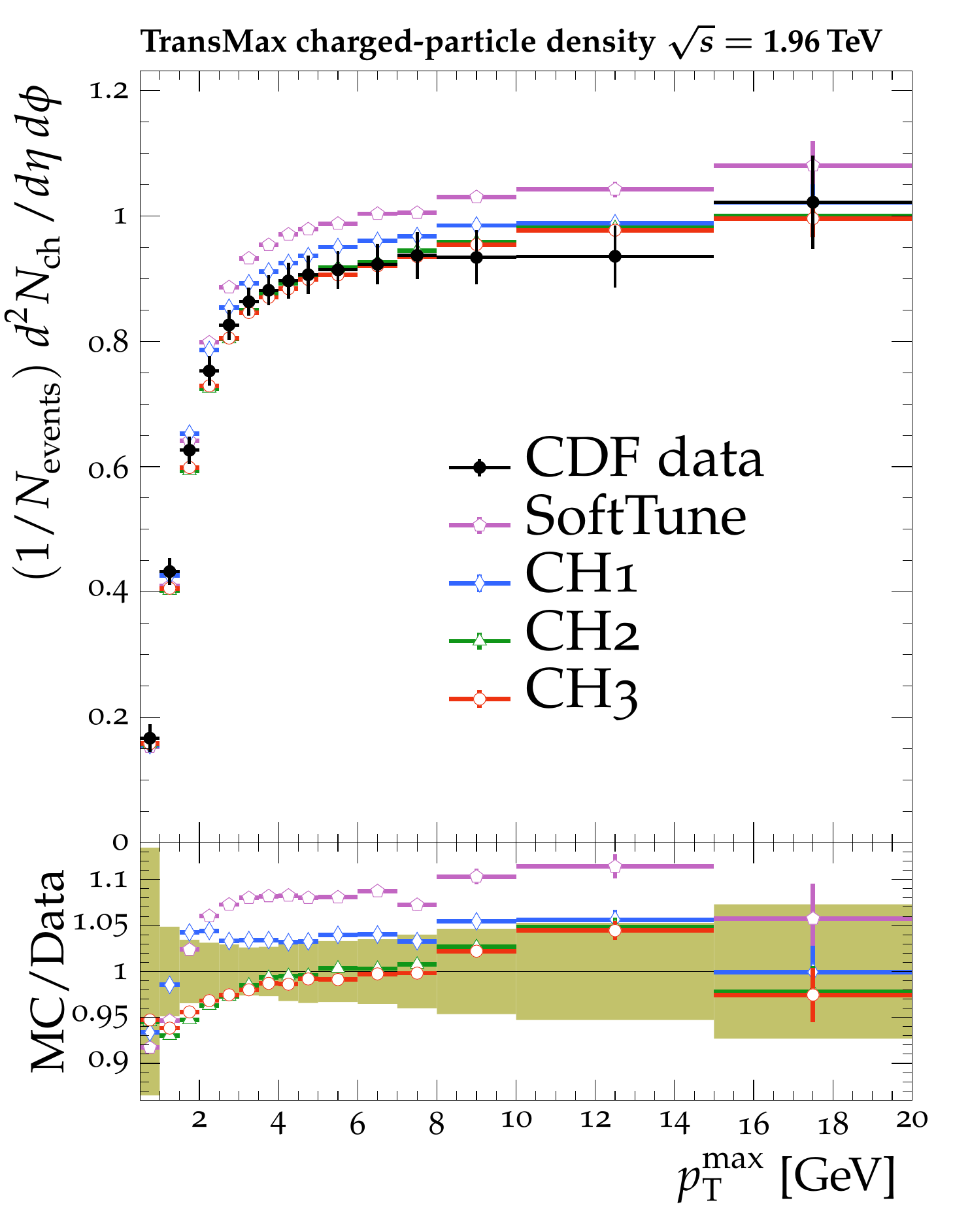}
  \caption{The \ptsum (upper) and \Nch (lower) density distributions in the transMin (left) and transMax (right) regions, as a function of the \pt of the leading track, \ptleadingtrack~\cite{CDFUE}.  CDF MB data are compared with the predictions from \HerwigS, with the \SoftTune and \CH tunes.  \captionColouredShadedBand}
  \label{fig:CDF_UE_1p96TeV}
\end{figure*}

Figure~\ref{fig:CDF_UE_1p96TeV} shows comparisons of the normalized \ptsum and \Nch densities using tune predictions with UE data collected by the CDF experiment at the Fermilab Tevatron at $\sqrts=1.96\TeV$~\cite{CDFUE}.  The \CH tunes describe the distributions in both transMin and transMax well, however the \CHth tune underestimates the \ptsum data somewhat at $\ptleadingtrack<10\GeV$, in both the transMin and transMax regions.  Although these data were not used in deriving any of the tunes considered here, they validate that the energy dependence of the new tunes is correctly modelled.  The tune \SoftTune overestimates the data by $\approx$5--15\% in all distributions.
Additional comparisons of the predictions of \HerwigS with the various tunes using MB data from the ATLAS experiment, which were used in deriving \SoftTune, are shown in Appendix~\ref{app:ATLASMB}.  One notable difference between the distribution of \dNdeta shown in Fig.~\ref{fig:CMS_MB_13TeV} and the one shown in Fig.~\ref{fig:ATLAS_13TeV_eta2p5} is that the former includes all charged particles, whereas the latter includes only charged particles with $\pt > 500\MeV$.

Based on the comparisons shown in this section, the tunes \CHt and \CHth both provide a similar description of the data, indicating that the choice between the two LO PDFs used for the simulation of MPI and remnant handling has little effect on the predictions.  These two PDFs are both LO PDFs, but a value of $\alpSMZ=0.118$ is used in deriving the PDF used with \CHt, and a value of $\alpSMZ=0.130$ is assumed for the PDF used with \CHth.  As stated in Section~\ref{sec:TuneProcedure}, $\alpSMZ=0.118$ is used in all parts of the \HerwigS simulation for the three \CH tunes.  From Table~\ref{tab:TunedValues}, the \chis/\ndof for the tune \CHt is slightly lower than that for the tune \CHth.  However, the use of the LO PDF in the tune \CHth, which was derived with $\alpSMZ=0.130$, is consistent with the value of \alpSMZ typically associated with LO PDFs and therefore is a preferred choice over the tune \CHt.
Both of the tunes \CHt and \CHth provide a better description of the data than the tune \CHo, where the NNLO NNPDF3.1 PDF was used for the simulation of MPI and remnant handling.  This suggests that the use of the LO NNPDF3.1 PDF is preferred in this aspect of the \HerwigS simulation, even though the gluon PDF in both the LO and NNLO PDF sets are positive at low energy scales, as discussed earlier.

\begin{figure*}[htb]
  \centering
  \includegraphics[width=0.49\textwidth]{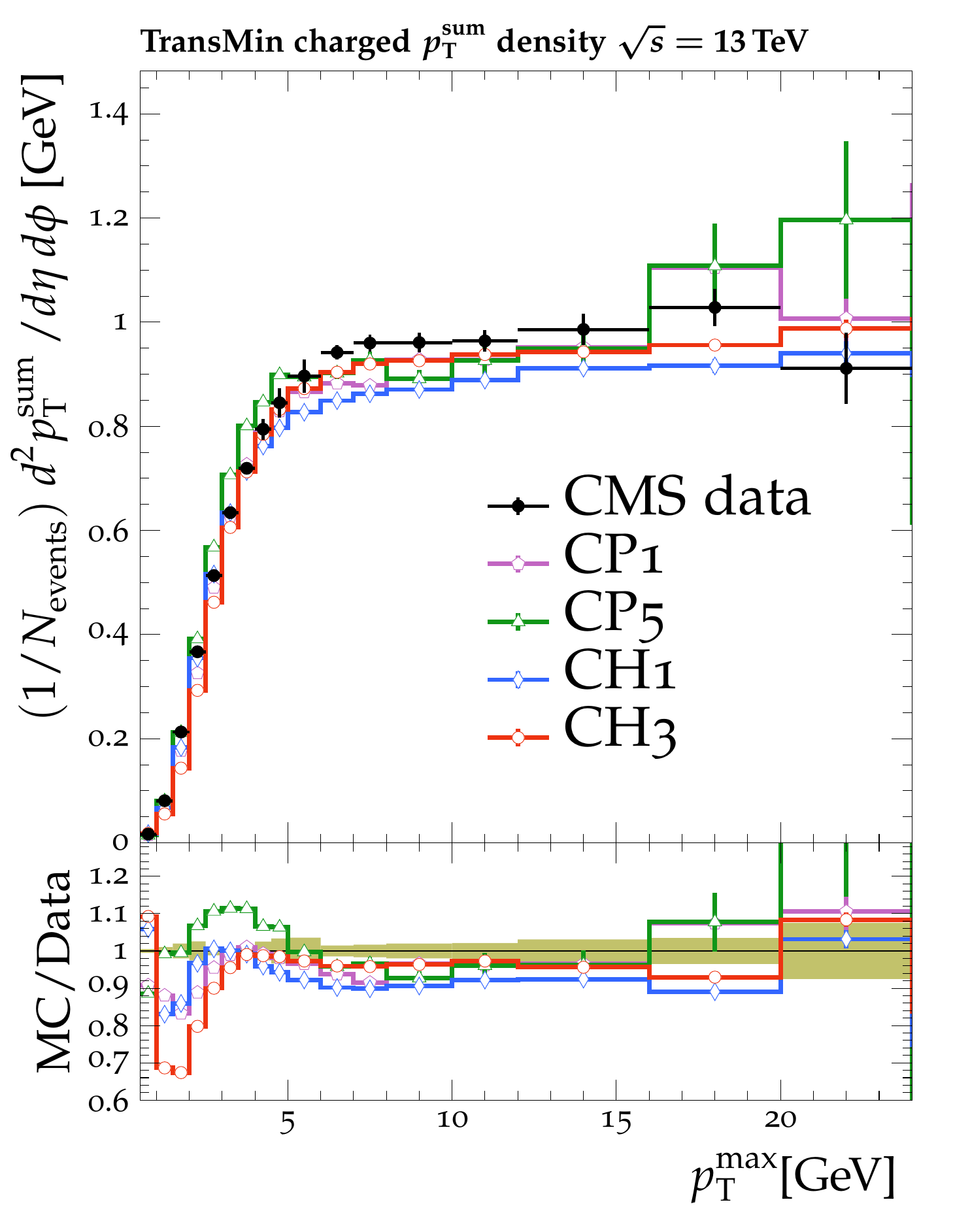}
  \includegraphics[width=0.49\textwidth]{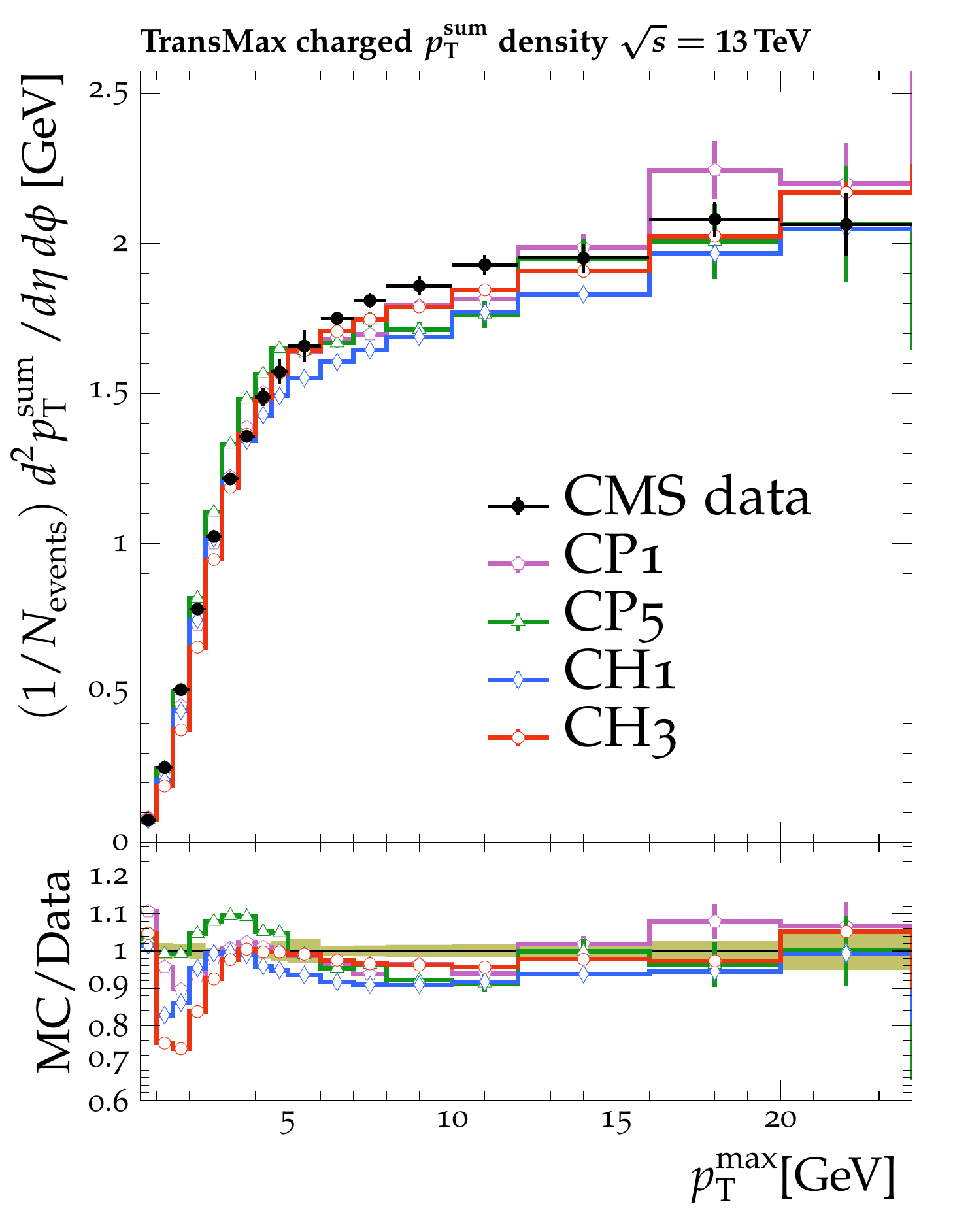} \\
  \includegraphics[width=0.49\textwidth]{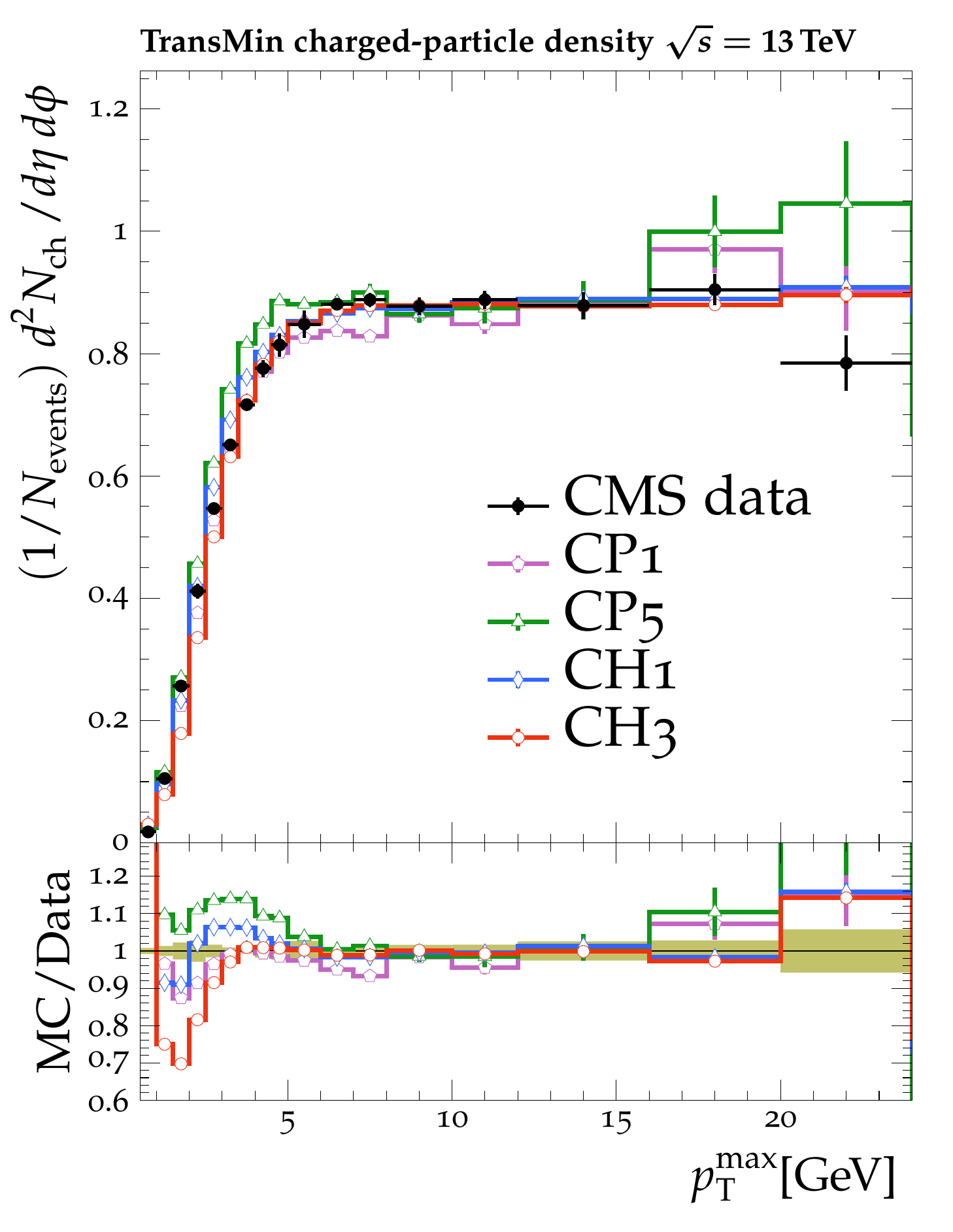}
  \includegraphics[width=0.49\textwidth]{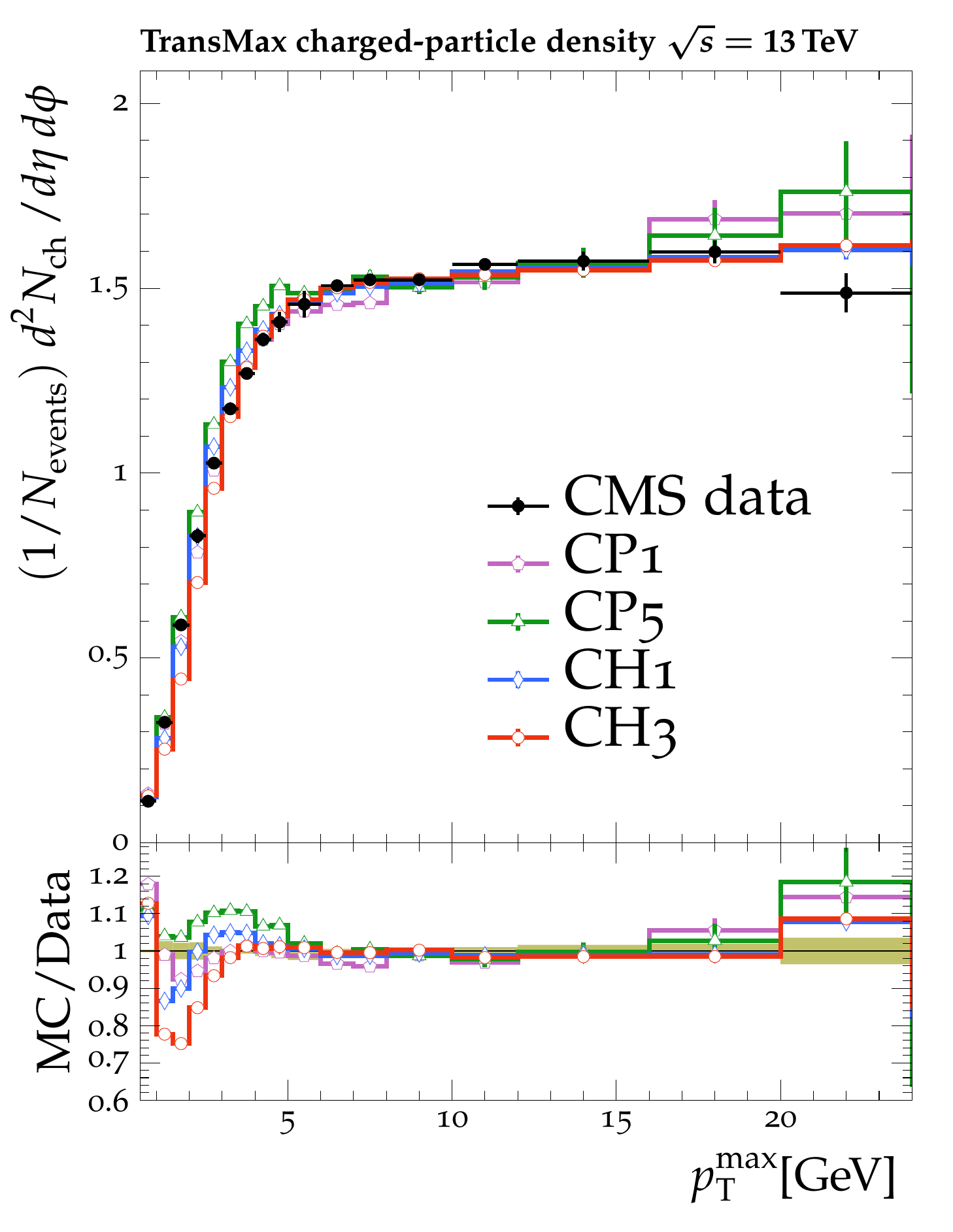}
  \caption{The \ptsum (upper) and \Nch (lower) density distributions in the transMin (left) and transMax (right) regions, as a function of the \pt of the leading track, \ptleadingtrack~\cite{CMS-PAS-FSQ-15-007}.  CMS MB data are compared with the predictions from \HerwigS, with the \CHo and \CHth tunes, and from \PYTHIAEIGHT, with the CP1 and CP5 tunes.  \captionColouredShadedBand}
  \label{fig:CMS_UE_13TeV_PythiaComp}
\end{figure*}

\begin{figure}[htb]
  \centering
  \includegraphics[width=0.49\textwidth]{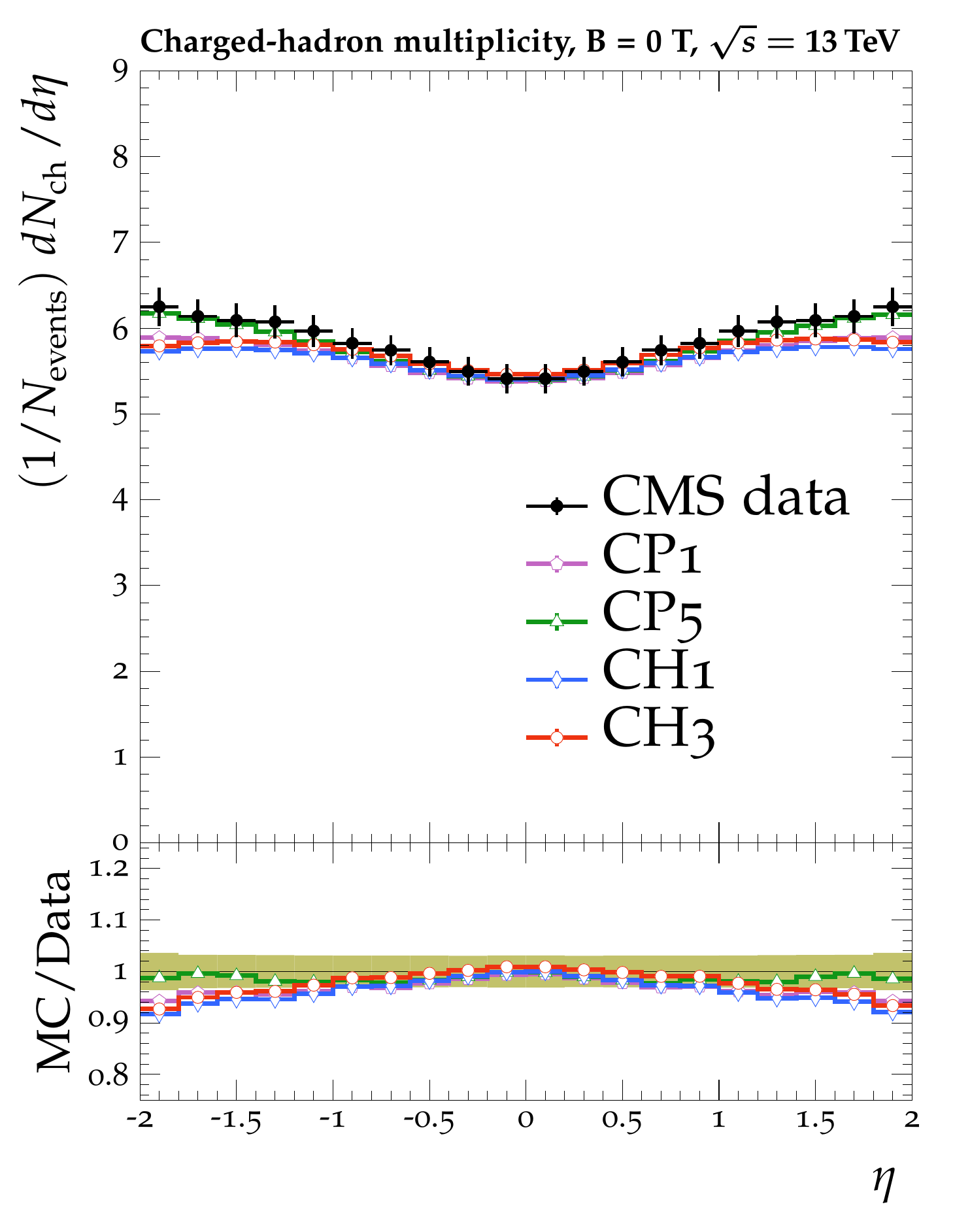}
  \caption{The normalized \dNdeta of charged hadrons as a function of $\eta$~\cite{CMSdNdeta}. CMS MB data are compared with the predictions from \HerwigS, with the \CHo and \CHth tunes, and from \PYTHIAEIGHT, with the CP1 and CP5 tunes.  \captionColouredShadedBand}
  \label{fig:CMS_MB_13TeV_PythiaComp}
\end{figure}

In Fig.~\ref{fig:CMS_UE_13TeV_PythiaComp} the normalized \Nch and \ptsum density predictions of the UE data at $\sqrts=13\TeV$ show a comparison of the \CHo and \CHth tunes with those obtained from the \PYTHIAEIGHT (version 8.230) using the tunes CP1 and CP5~\cite{GEN17001}.
The tune \CHt is not displayed, because its prediction is similar to the one of the tune \CHth.
The CP1 tune uses an LO NNPDF3.1 PDF set in all aspects of the \PYTHIAEIGHT simulation, an \alpSMZ value of 0.130 in the simulation of MPI and hard scattering, and an \alpSMZ value of 0.1365 for the simulation of initial- and final-state radiation.  The CP5 tune uses an NNLO PDF set with an \alpSMZ value of 0.118 in all aspects of simulation.  The choice of the PDF set and \alpSMZ value in the CP5 tune is the same as the \CHo~\HerwigS tune.  Although all the predictions show a reasonable agreement with the data in the plateau region of the UE distributions, the use of an LO PDF for MPI and remnant handling in \CHth provides a slightly improved description of the \ptsum data compared to using an NNLO PDF in \CHo.  This differs from the predictions of \PYTHIAEIGHT, where the use of an LO and NNLO PDF for simulating MPI give a similar description of the data in this region.
Each prediction exhibits different behaviour at low \ptleadingtrack.  None of the \HerwigS or \PYTHIAEIGHT tunes provides a perfect description of the data at low \ptleadingtrack, since they exhibit at least a 10\% difference between any one of the tunes and the data.  Figure~\ref{fig:CMS_MB_13TeV_PythiaComp} shows a similar comparison for the $\eta$ distribution of charged hadrons at 13\TeV.  The prediction from CP5 provides a better description of the data compared with the other tunes at larger values of $\abs{\eta}$.  The predictions from the \HerwigS tunes show a closer behaviour to the CP1 tune in this distribution.

\clearpage

\section{Uncertainties in the \texorpdfstring{\HerwigSTitle}{HERWIG 7} tunes}
\label{sec:HerwigTuneUncertainties}

Alternative tunes are derived in this section that provide an approximation to the uncertainties in the parameters of the tune CH3.  These are obtained from the eigentunes provided by \Professor.  These eigentunes are variations of the tuned parameters along the maximally independent directions in the parameter space by an amount corresponding to a change in the \chis (\deltachis) equal to the optimal \chis of the fit.
Because a change \deltachis in Eq.~(\ref{eq:chis}) does not result in a variation with a meaningful statistical interpretation, the value of \deltachis is chosen in an empirical way.  The change $\deltachis=\chis$, which is suggested by the \Professor Collaboration, results in variations that are similar in magnitude to the uncertainties in the fitted data points and judged to provide a reasonable set of variations that reflect the combined statistical and systematic uncertainty in the model parameters.
A consequence of this adopted procedure is that the uncertainty may not necessarily cover the data in every bin.
If the uncertainties in the fitted data points were uncorrelated between themselves, then the magnitude of the uncertainties in the data points depends on their bin widths.  For the data used in the fit, the uncertainties are typically dominated by uncertainties that are correlated between the bins.  
However, the uncertainties in the data points at high \ptleadingtrack and \ptleadingtrackjet,  \eg $\ptleadingtrack\gtrsim10\GeV$ for the UE observables at $\sqrts=13\TeV$, are dominated by statistical uncertainties, which are uncorrelated between bins.  This introduces some dependence of the eigentunes on the bin widths of the data used in the fit.

The variations of the tunes provided by the eight eigentunes are reduced to two variations, as explained below, one ``up'' and one ``down'' variation.  The ``up'' variation is obtained by considering the positive differences in each bin between each eigentune and the central prediction of the CH3 tune for the distributions used in the tuning procedure.  The difference for each eigentune is summed in quadrature.
Similarly, the ``down'' variation is obtained by considering the negative differences between the eigentunes and the central predictions.  The two variations are then fitted, using the same procedure described in Section~\ref{sec:TuneProcedure} to obtain a set of tune parameters that describe these two variations.  The parameters of the two variations are shown in Table~\ref{tab:TuneUncertainties}.
The values of each parameter of the variations do not necessarily encompass the corresponding values of the \CHth tune, as a result of the method of determining the variations from the differences between several eigentunes.
The two variations accurately replicate the combination of all eigentunes, \ie the sum in quadrature of all positive or negative differences with respect to the central prediction.  
By using these variations, the uncertainties in the tune \CHth are estimated by considering only two variations of the tune parameters, rather than eight variations.  However, the correlations between bins of an observable for each of the eight individual variations are not known when considering only the ``up'' and ``down'' variations.

\begin{table}[h]
\topcaption{Parameters of the central, ``up'', and ``down'' variations of the CH3 tune.}
\centering
\begin{tabular}{ c c c c }
& \multicolumn{3}{c}{CH3}  \\ 
& Down & Central & Up \\
\hline
\ptmino (\GeVns) & 2.349 & 3.040 & 3.382  \\
\bParameter & 0.298 & 0.136 & 0.328  \\
\musParameter (\musunits)& 1.160 & 1.284 & 1.539  \\
\preco & 0.641 & 0.471 & 0.191  \\
\end{tabular}
\label{tab:TuneUncertainties}
\end{table}

\begin{figure*}[tbp]
  \centering
  \includegraphics[width=0.49\textwidth]{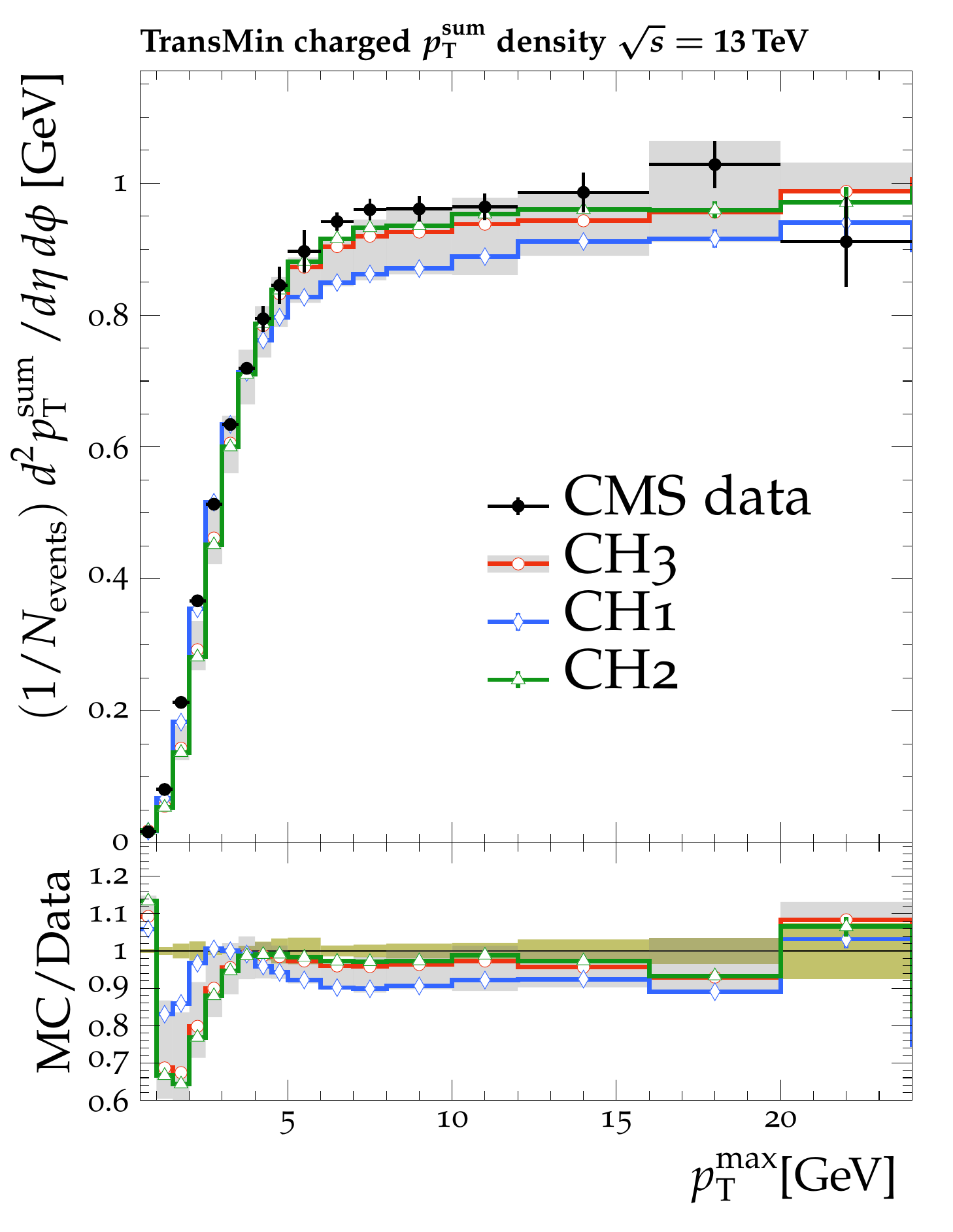}
  \includegraphics[width=0.49\textwidth]{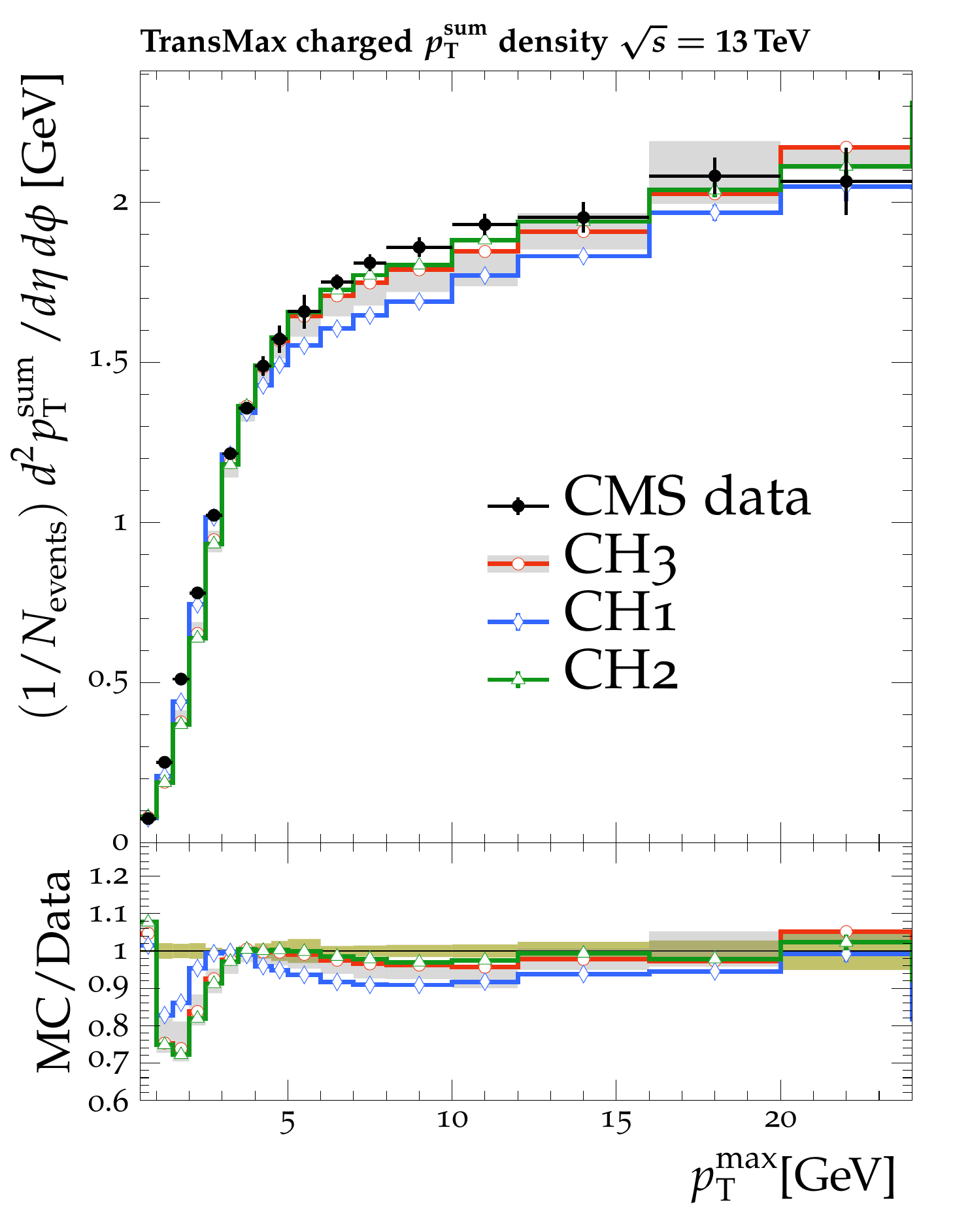} \\
  \includegraphics[width=0.49\textwidth]{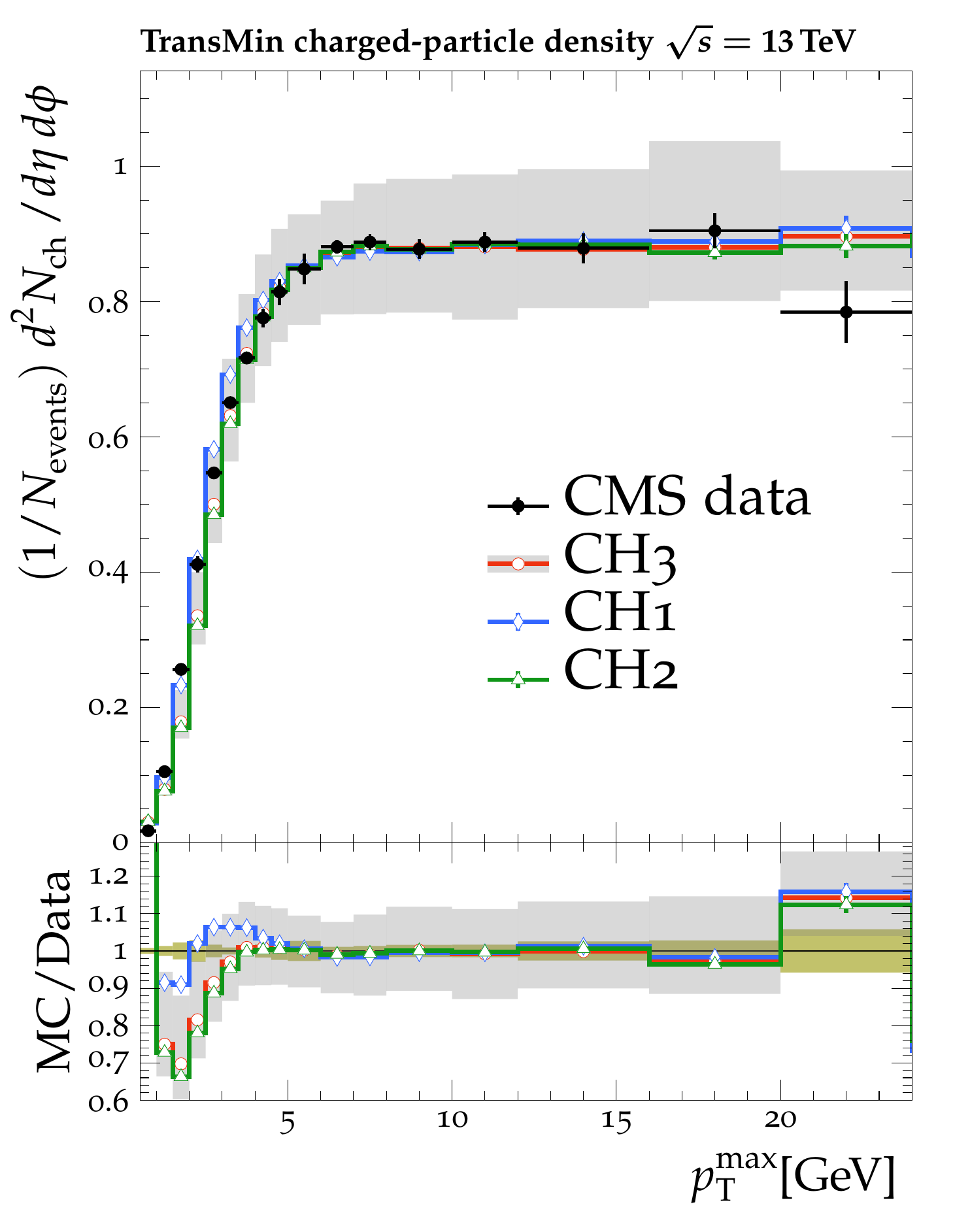}
  \includegraphics[width=0.49\textwidth]{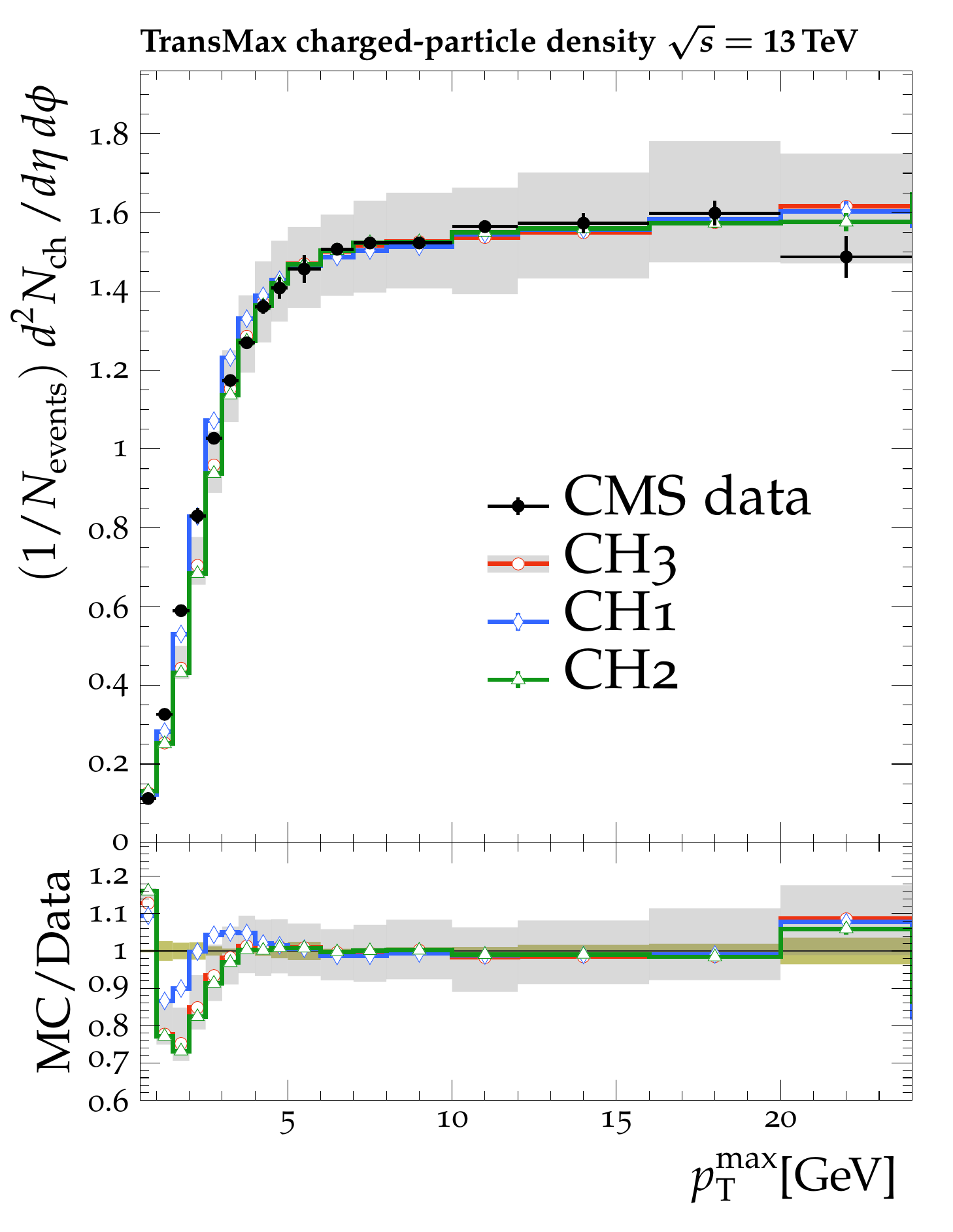}
  \caption{The \ptsum (upper) and \Nch (lower) density distributions in the transMin (left) and transMax (right) regions, as a function of the \pt of the leading track, \ptleadingtrack~\cite{CMS-PAS-FSQ-15-007}.  CMS MB data are compared with the predictions from \HerwigS, with the \CH tunes.  \captionColouredShadedBand  The grey-shaded band corresponds to the envelope of the ``up'' and ``down'' variations of the \CHth tune.
    }
  \label{fig:CMS_UE_13TeV_uncertainties}
\end{figure*}

\begin{figure}[tbh]
  \centering
  \includegraphics[width=0.49\textwidth]{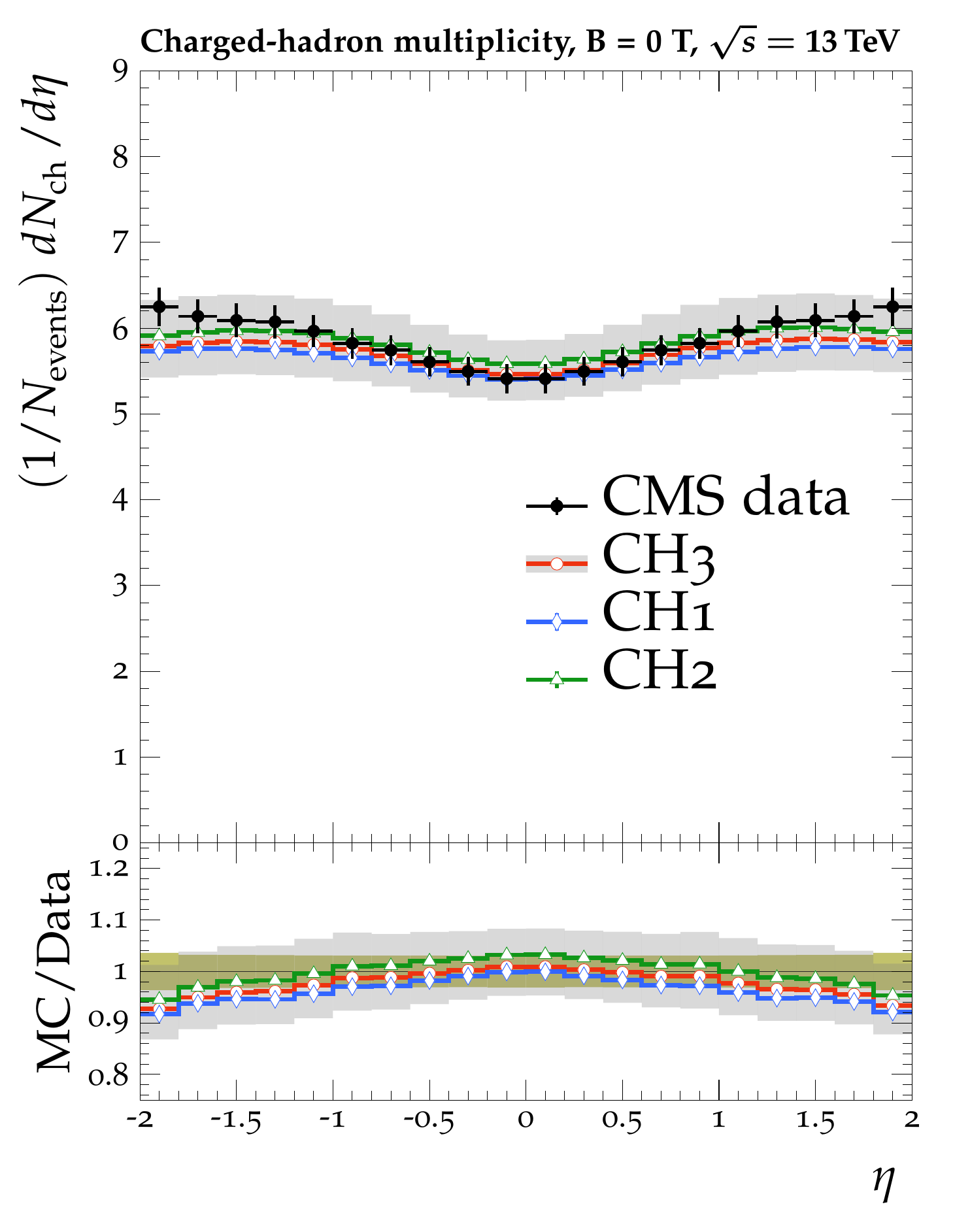} \\
  \caption{The normalized \dNdeta of charged hadrons as a function of $\eta$~\cite{CMSdNdeta}. CMS MB data are compared with the predictions from \HerwigS, with the \CH tunes.
  \captionColouredShadedBand  The grey-shaded band corresponds to the envelope of the ``up'' and ``down'' variations of the \CHth tune.  }
  \label{fig:CMS_MB_13TeV_uncertainties}
\end{figure}

Figures~\ref{fig:CMS_UE_13TeV_uncertainties} (normalized \ptsum and \Nch densities) and~\ref{fig:CMS_MB_13TeV_uncertainties} (normalized \dNdeta) show predictions from the \CH tunes.  The grey-shaded band corresponds to the envelope of the ``up'' and ``down'' variations, for the UE and MB observables used in the tuning procedure.  The differences between the \CHo and \CHt predictions and those from \CHth are within the uncertainty of \CHth, except for a small deviation at low \ptleadingtrack.

\clearpage

\section{Comparison with LEP data}
\label{sec:LepComparisons}

\begin{figure*}[tbp]
  \centering
  \includegraphics[width=0.49\textwidth]{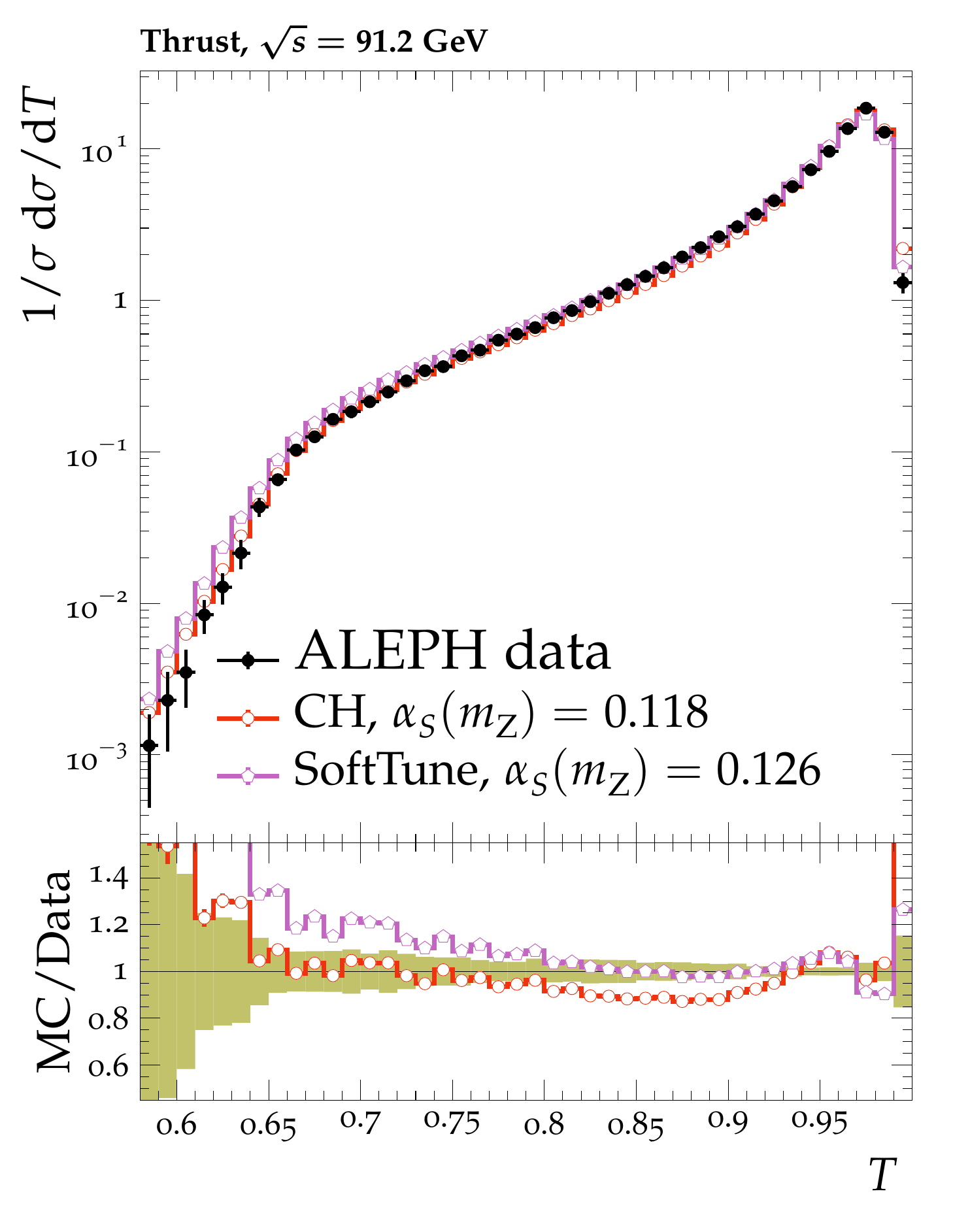}
  \includegraphics[width=0.49\textwidth]{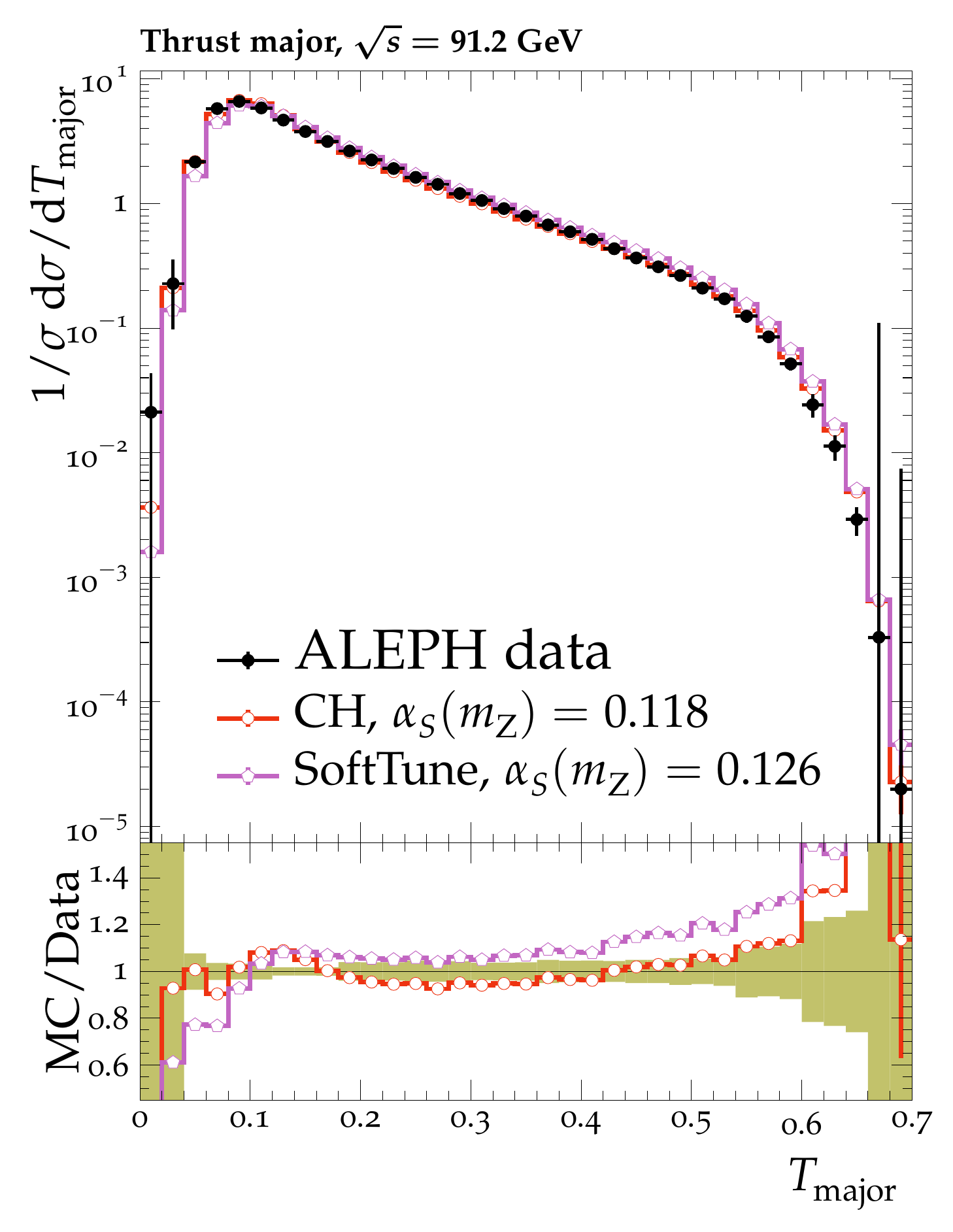} \\
  \includegraphics[width=0.49\textwidth]{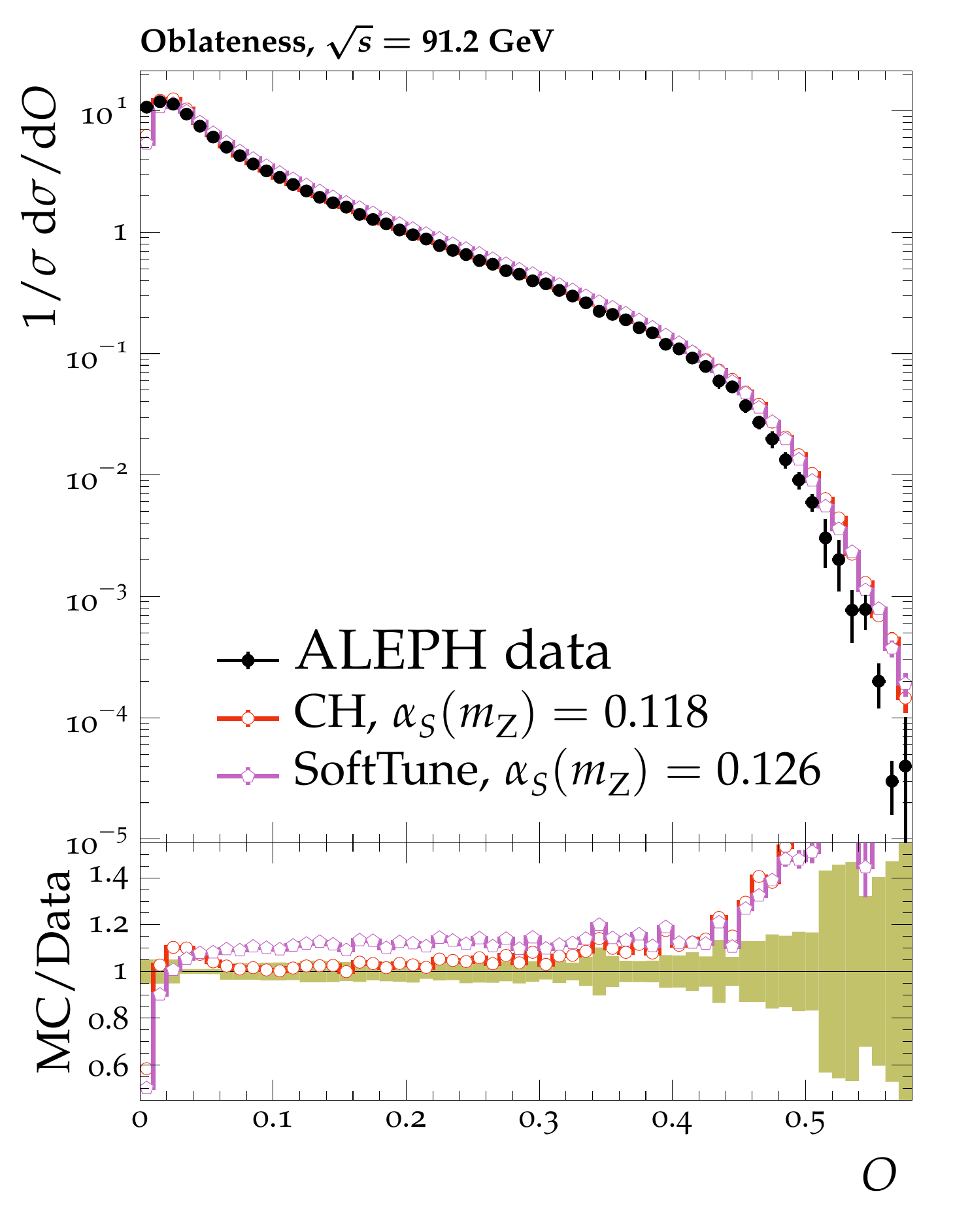}
  \includegraphics[width=0.49\textwidth]{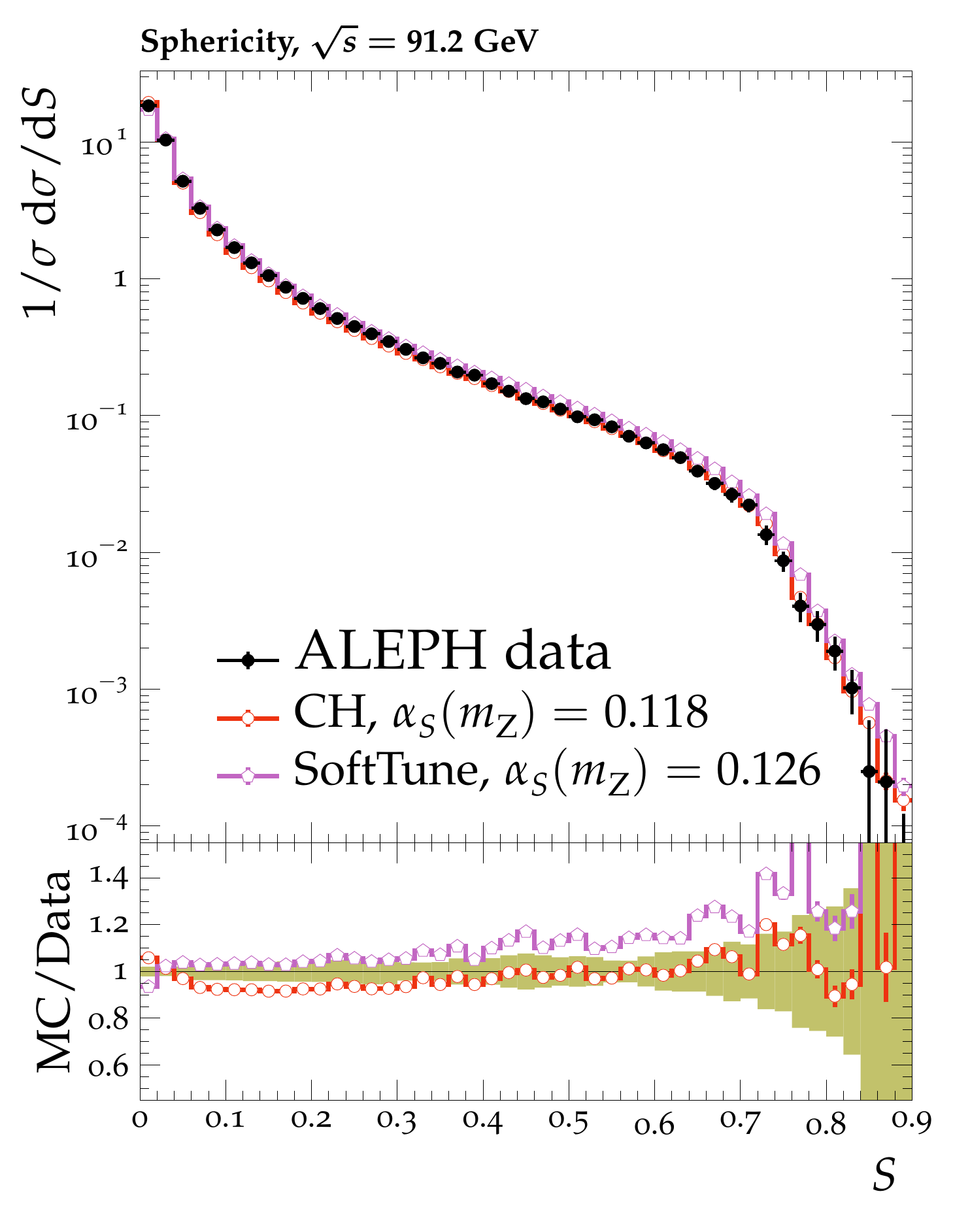}
  \caption{
  Normalized differential cross sections for \Pem\Pep~\cite{ALEPHEventShape} as a function of the variables \thrust (upper left), \thrustmajor (upper right), \oblateness (lower left), and \sphericity (lower right) for ALEPH data at $\sqrts=91.2\GeV$. ALEPH data are compared with the predictions from \HerwigS using the \SoftTune and \CH tunes.  The coloured band in the ratios of the different predictions from simulation to the data represents the total experimental uncertainty in the data.}
  \label{fig:ALEPH_eventShapes}
\end{figure*}

\HerwigS predictions are obtained in this section for event shape observables measured in LEP electron-positron collisions at $\sqrts=91.2\GeV$.  The predictions are obtained using NLO MEs implemented within \HerwigS.  Figure~\ref{fig:ALEPH_eventShapes} shows the thrust (\thrust), thrust major (\thrustmajor), oblateness (\oblateness), and sphericity (\sphericity) observables as measured by the ALEPH Collaboration~\cite{ALEPHEventShape}.  

{\tolerance=800 Because these observables are measured in collisions with a lepton-lepton initial state, the difference in choice of PDF and parameters of the MPI model in the three \CH tunes has no effect on the predictions.  Similarly, the only difference between the \CH tunes and \SoftTune is in the value of \alpSMZ.  The value of $\alpSMZ=0.118$ is used in the \CH tunes, and is consistent with the value used by the PDF set for the hard process and the PS when simulating proton-proton collisions.  A set of next-to-leading corrections to soft gluon emissions can be incorporated in the PS by using two-loop running of \alpS and including the Catani-Marchesini-Webber rescaling~\cite{CMW} of \alpSMZ from $\alpSMZ=0.118$ to $\alpSMZ=0.1262$, which corresponds to the value of \alpSMZ used in \SoftTune~\cite{Herwig7QGJets}. \par}

The \CH tunes underestimate the number of events with $0.80 < \thrust < 0.95$, whereas \SoftTune predicts too many isotropic events with lower values of $\thrust<0.8$ and with higher values of $\sphericity>0.4$.  The \CH tune provides a better overall description of the \thrustmajor observable compared with \SoftTune.  Both tunes predict too many planar events, as can be seen at larger values of \oblateness; however, the \CH tune provides a better description of the data at smaller values of \oblateness.

\section{Comparison with top quark pair production data}
\label{sec:TopComparisons}

Predictions using the \HerwigS tunes are compared in this section with observables measured in data containing top quark pairs.

The \POWHEG v2 generator is used to perform ME calculations in the hvq mode~\cite{Powheghvq} at NLO accuracy in QCD.  In the \POWHEG ME calculations, a value of $\alpSMZ=0.118$ with a two-loop evolution of \alpS is used, along with the \NNPDF NNLO PDF set, derived with a value of $\alpSMZ=0.118$.  The ME calculations are interfaced with \HerwigS for the simulation of the UE and PS.  The mass of the top quark is set to $\mt = 172.5\GeV$, and the value of the $\hdamp$ parameter, which controls the matching between the ME and PS, is set to $1.379~\mt$.  The value of $\hdamp$ in \POWHEG was derived from a fit to \ttbar data in the dilepton channel at $\sqrts=8\TeV$, where \POWHEG was interfaced with \PYTHIAEIGHT using the CP5 tune~\cite{TOP-16-021,GEN17001}.  

Samples are generated with the different \HerwigS tunes that use the same parton-level events for each tune.  For generating NLO matched samples such as these, an NLO (or NNLO) PDF set may be desirable for the simulation of the hard process.  In Ref.~\cite{ConsistentPDFMatching}, it is then advocated that the same PDF set and \alpSMZ value should be used in the PS.
However, one can still choose an LO PDF set for the simulation of the MPI and remnant handling in this case, such as the choices in the tunes \CHt and \CHth.  This configuration of PDF sets is not possible in \PYTHIA.

\begin{figure*}[tbp]
  \centering
  \includegraphics[width=0.49\textwidth]{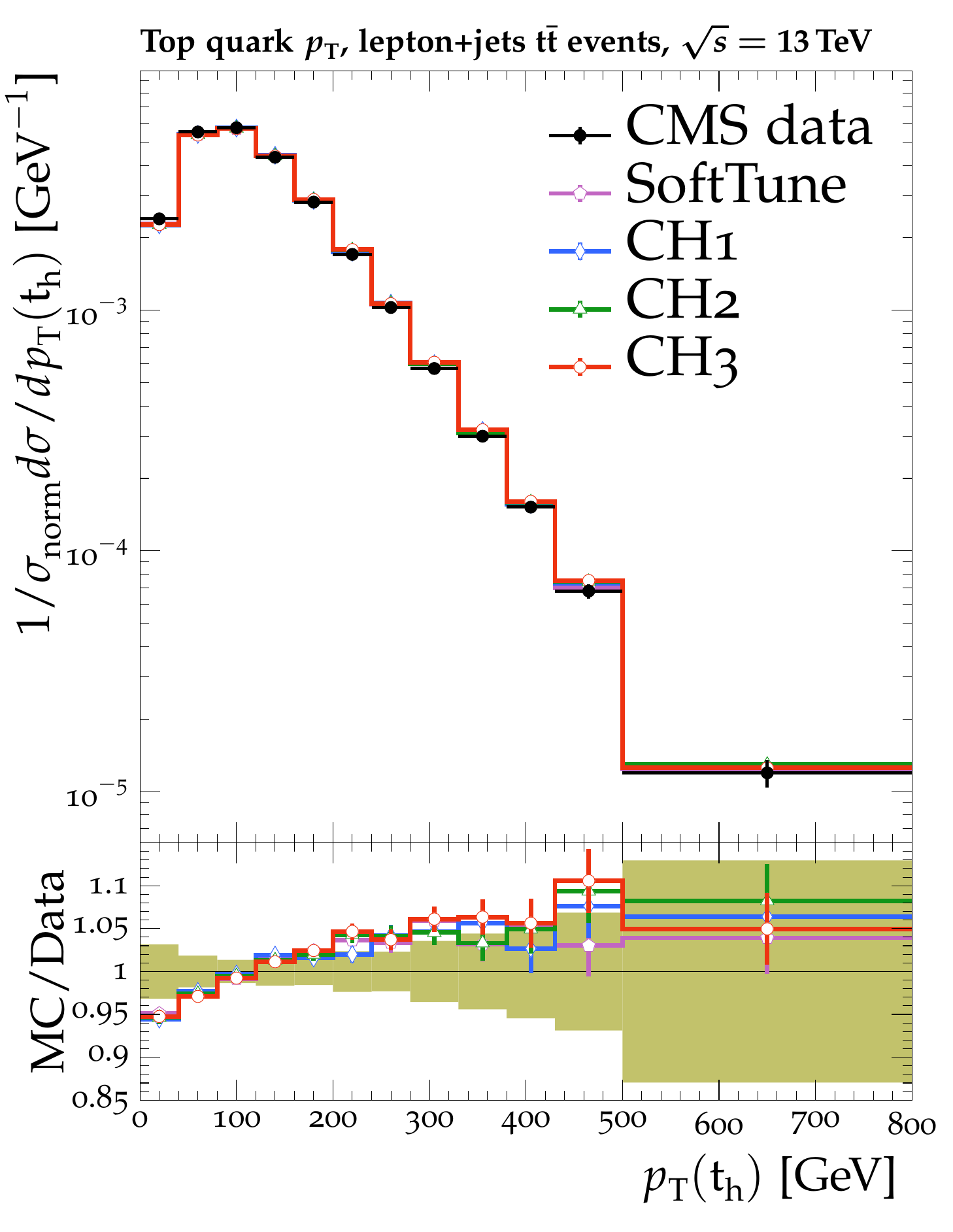}
  \includegraphics[width=0.49\textwidth]{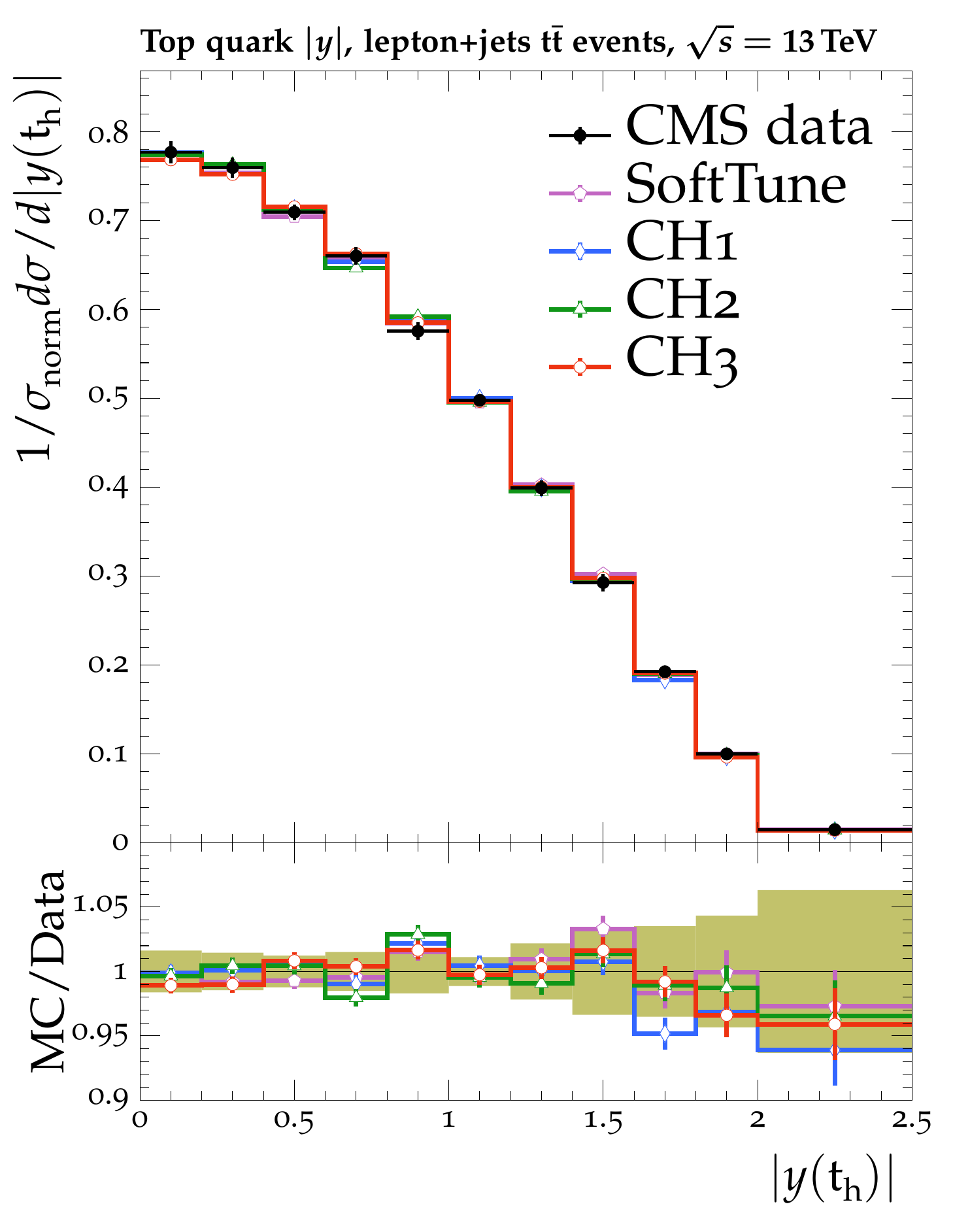} \\
  \includegraphics[width=0.49\textwidth]{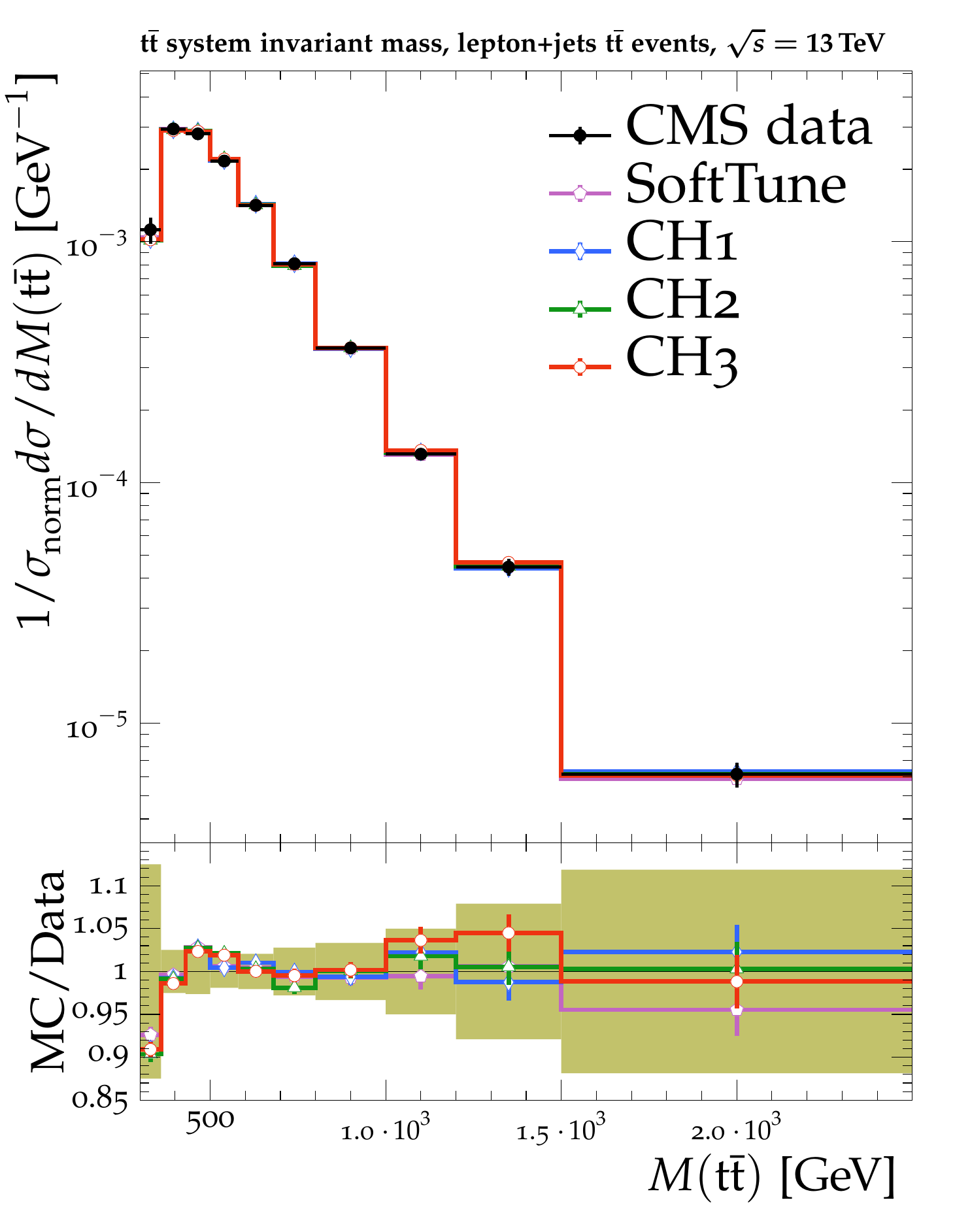}
  \includegraphics[width=0.49\textwidth]{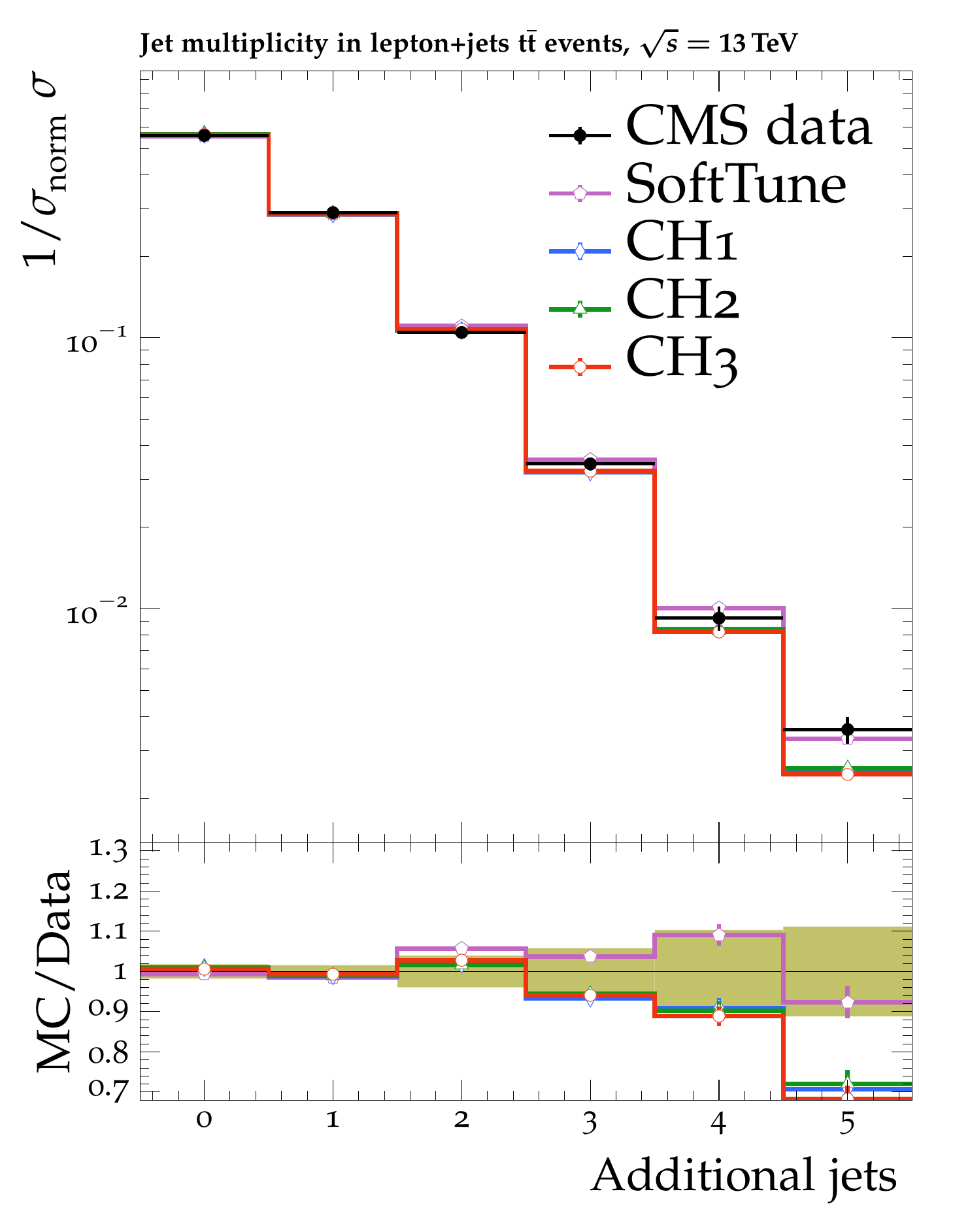}
  \caption{The differential cross sections are shown as functions of: the \pt (upper left) and rapidity (upper right) of the hadronically decaying top quark; the invariant mass of the \ttbar system (lower left); the additional jet multiplicity (lower right)~\cite{CMS-TOP-17-002}.  CMS \ttbar data are compared with the predictions from \POWHEGHERWIG, with the \SoftTune, \CHo, \CHt, and \CHth tunes.  \captionColouredShadedBand}
  \label{fig:CMS_ttbar_kinematics_13TeV}
\end{figure*}

\begin{figure*}[tbp]
  \centering
  \includegraphics[width=0.49\textwidth]{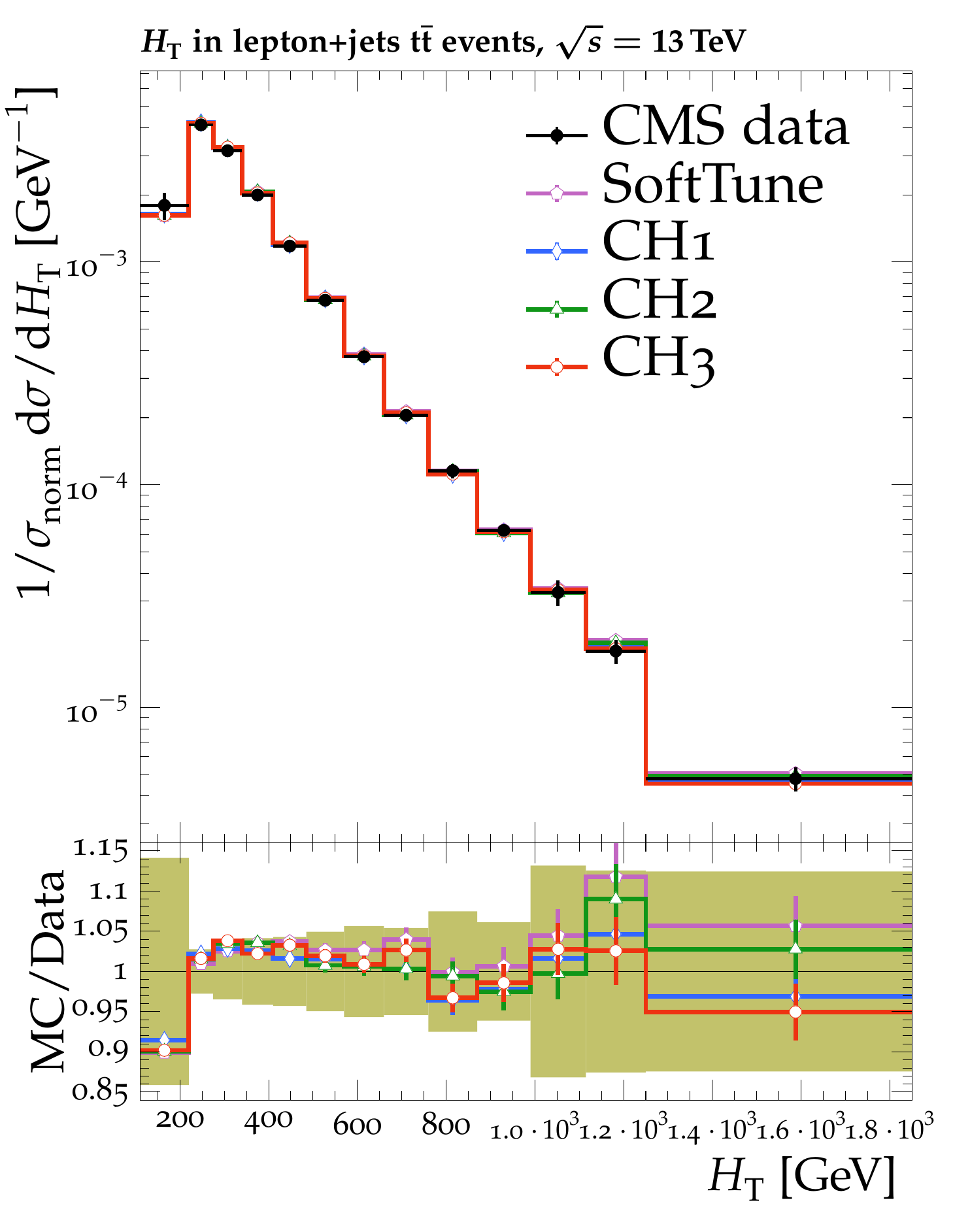}
  \includegraphics[width=0.49\textwidth]{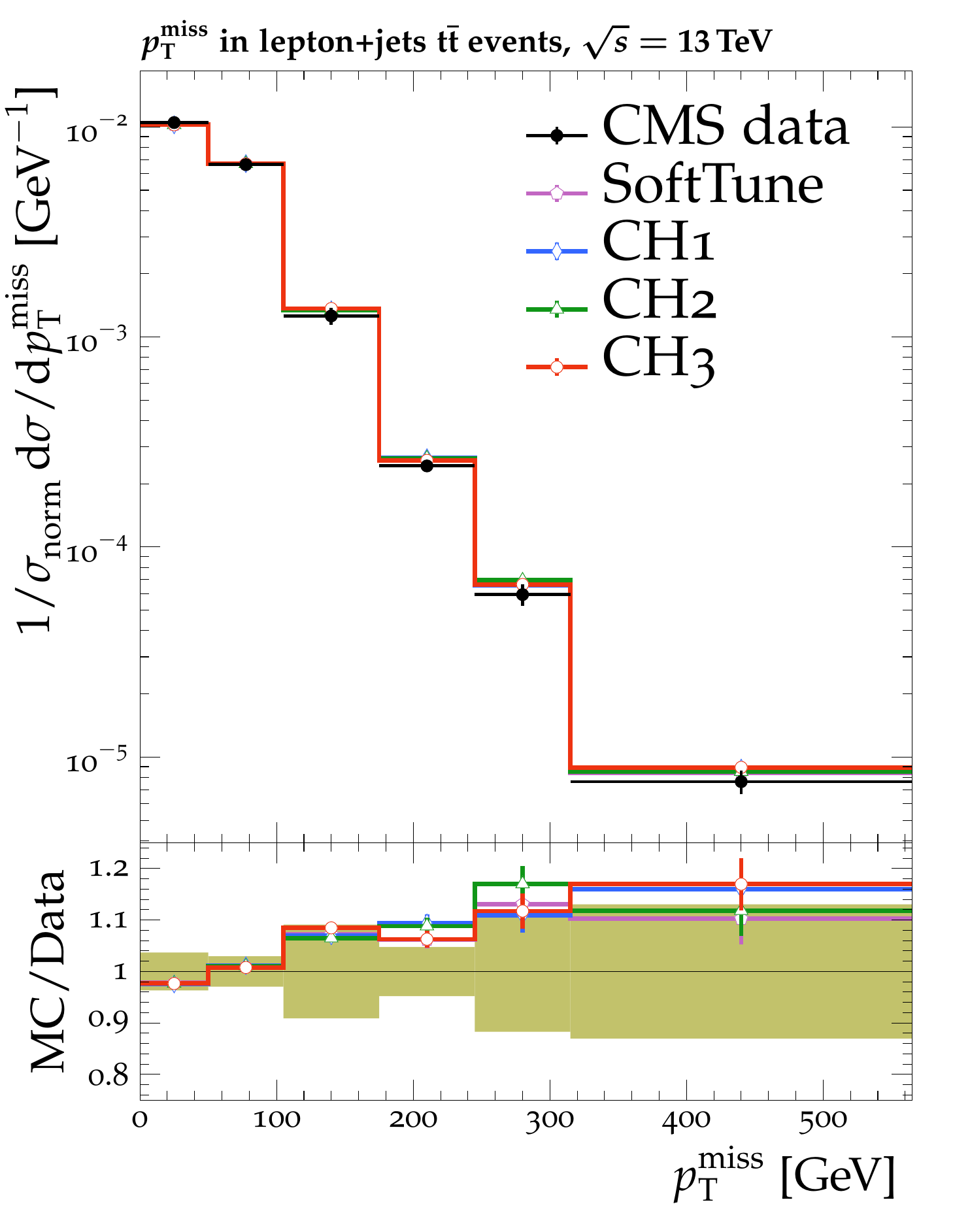}
  \caption{The differential cross sections are shown as functions of \HT (left) and \ptmiss (right)~\cite{CMS-TOP-16-014}.  CMS \ttbar data are compared with the predictions from \POWHEGHERWIG, with the \SoftTune, \CHo, \CHt, and \CHth tunes.  \captionColouredShadedBand}
  \label{fig:CMS_ttbar_kinematics_HT_MET_13TeV}
\end{figure*}

First, kinematic properties of the \ttbar system are compared with $\sqrts=13\TeV$ CMS data in the single-lepton channel~\cite{CMS-TOP-17-002}. Figure~\ref{fig:CMS_ttbar_kinematics_13TeV} presents normalized differential cross sections as functions of the \pt and rapidity $y$ of the particle-level hadronically decaying top quark.  The invariant mass of the reconstructed \ttbar system and the number of additional jets with $\pt > 30\GeV$ in the event are also shown, where the jets are reconstructed using the \antikt algorithm~\cite{antikt1,antikt2} with a distance parameter of 0.4.  Normalized cross sections as a function of global event variables, namely \HT,  the scalar \pt sum of all jets, and \ptmiss, the magnitude of the missing transverse momentum vector~\cite{CMS-TOP-16-014} are shown in Fig.~\ref{fig:CMS_ttbar_kinematics_HT_MET_13TeV}.

The predictions from the different simulations are mostly compatible with each other, indicating a small effect of the tune on these observables. The only notable difference is seen in the additional jet multiplicity, originating from the smaller \alpSMZ value used in the simulations with \HerwigS\,\CH tunes. The simulated events with the \CH tunes describe the CMS data well up to 4 additional jets, but slightly underestimate the multiplicity for a higher number of jets.  The differences between the predictions with the \CH tunes and the tune \SoftTune are comparable with the typical size of the theoretical uncertainties in the ME calculation, as studied in Ref.~\cite{TOP-16-021}.

\begin{figure*}[tbp]
  \centering
  \includegraphics[width=0.49\textwidth]{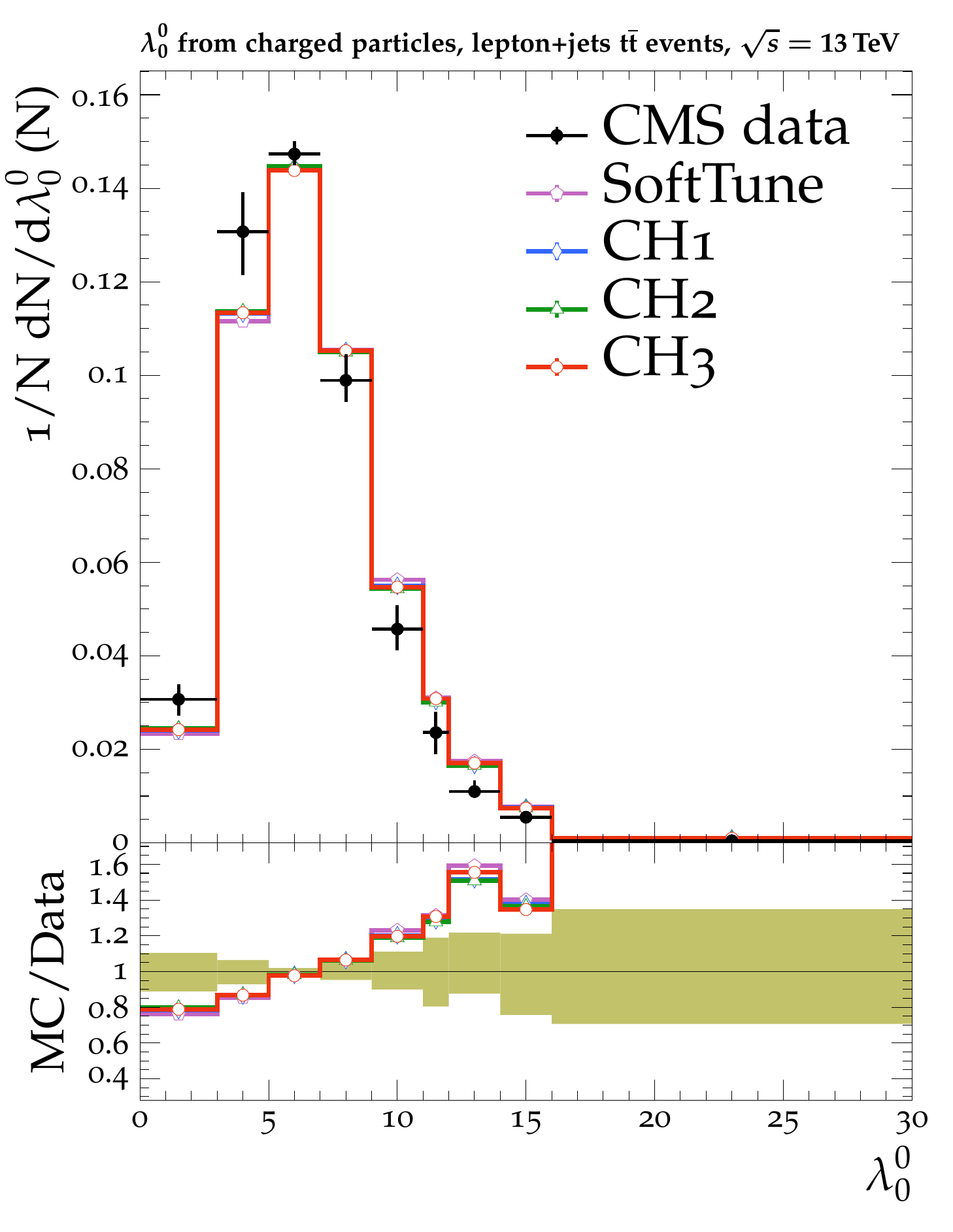}
  \includegraphics[width=0.49\textwidth]{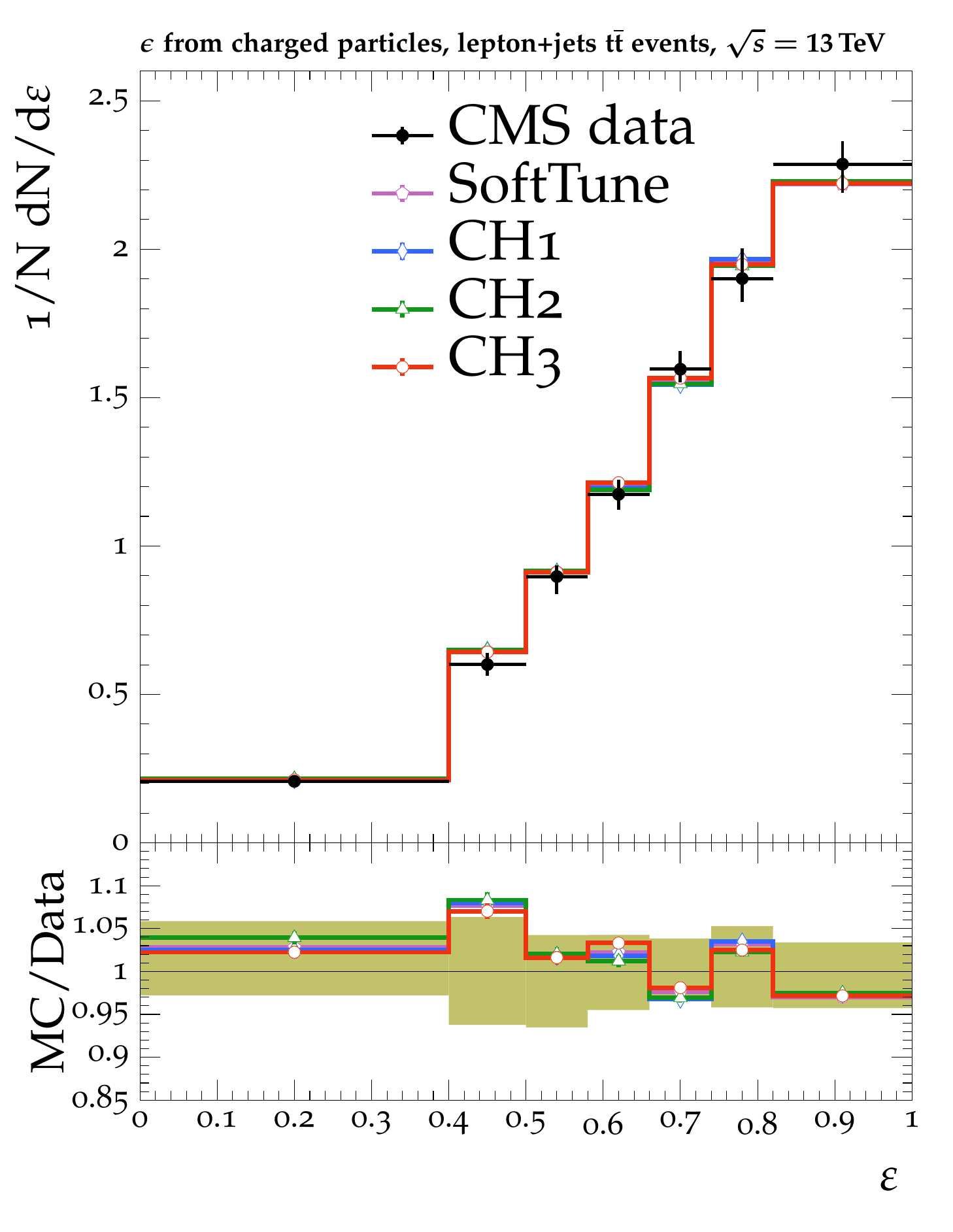} \\
  \includegraphics[width=0.49\textwidth]{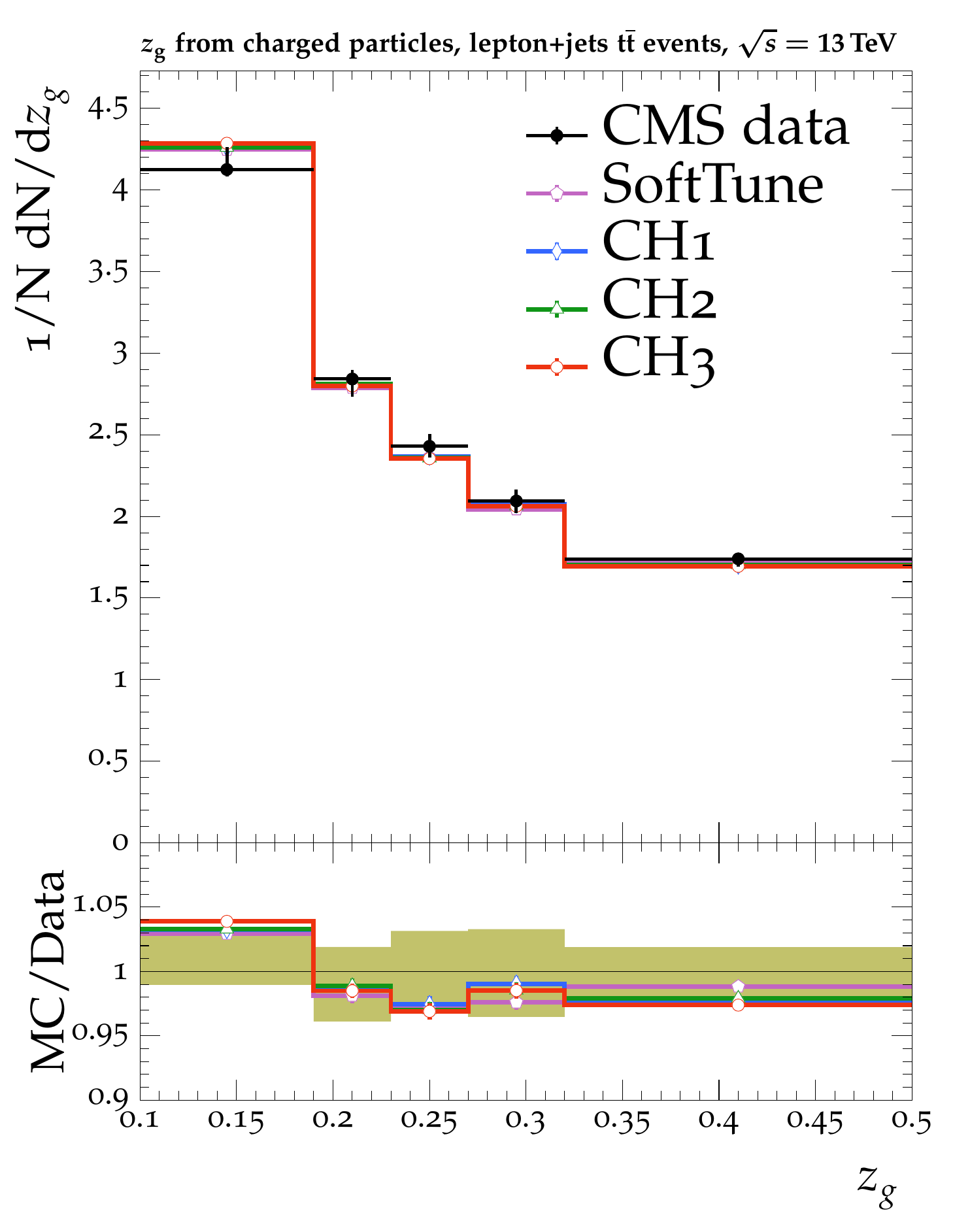}
  \includegraphics[width=0.49\textwidth]{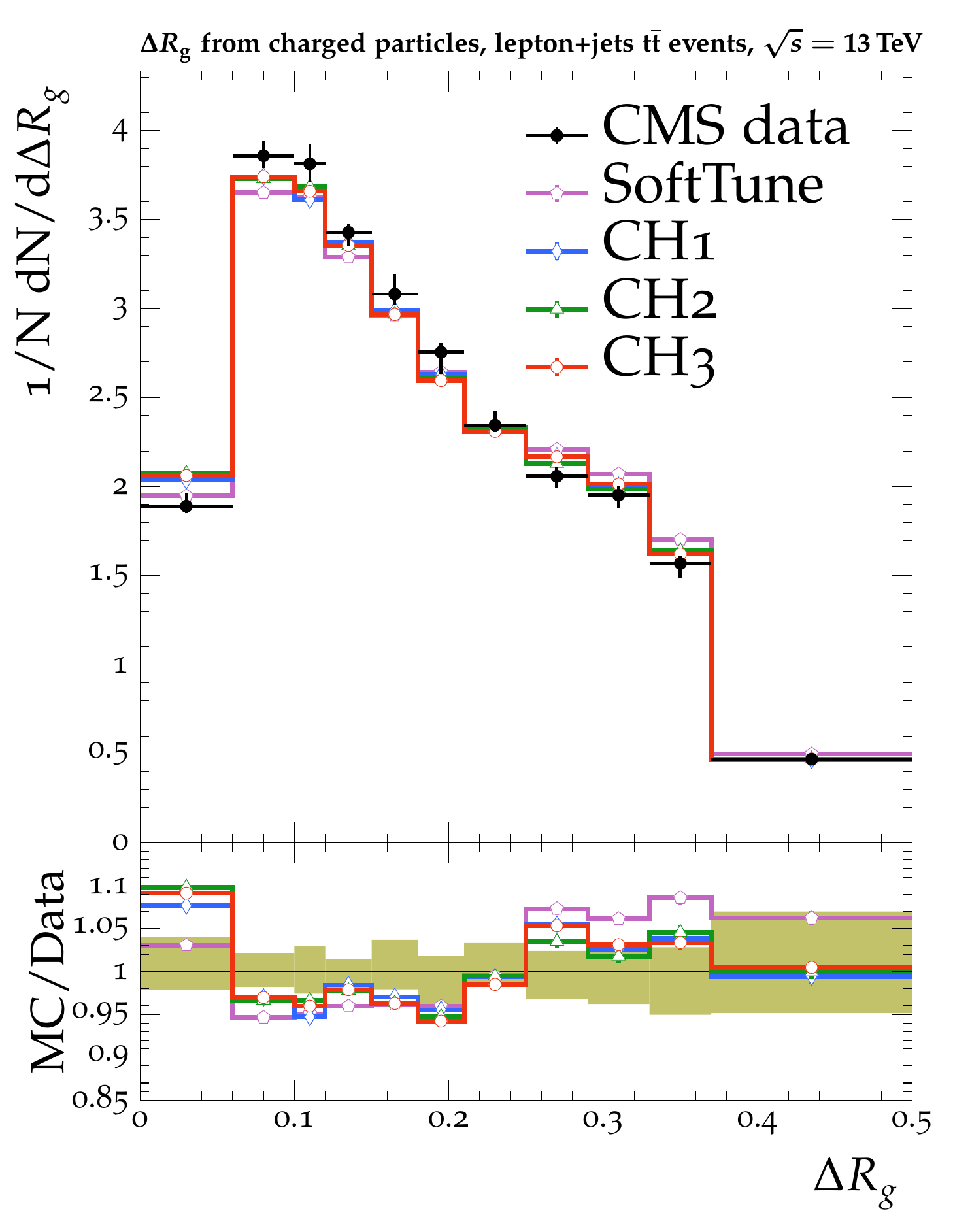}
  \caption{The normalized jet substructure observables in single-lepton events: the charged-particle multiplicity (upper left); the eccentricity (upper right); the groomed momentum fraction (lower left); and the angle between the groomed subjects (lower right)~\cite{CMS-TOP-17-013}. CMS \ttbar data are compared with the predictions from \POWHEGHERWIG, with the \SoftTune, \CHo, \CHt, and \CHth tunes.  \captionColouredShadedBand}
  \label{fig:CMS_ttbar_jetsubstructure_13TeV}
\end{figure*}

Next, jet substructure observables are compared to $\sqrts=13\TeV$ CMS data in the single-lepton channel~\cite{CMS-TOP-17-013}. Normalized number of jets as a function of four variables with relatively low correlations amongst themselves are shown in Fig.~\ref{fig:CMS_ttbar_jetsubstructure_13TeV}. The variables presented are the charged-particle multiplicity (\lambdazerozero), the eccentricity (\eccentricity) calculated from the charged jet constituents, the groomed momentum fraction (\zg), and the angle between the groomed subjets (\deltaRg). 

The choice of tune has little effect on most of the jet substructure observables.  All choices of \HerwigS tune overestimate \lambdazerozero, which was also observed in Ref.~\cite{CMS-TOP-17-013}.
The predictions for \eccentricity and \zg distributions agree closely with the data in all cases. The \deltaRg spectrum at very low values is somewhat less well described by the simulation employing the \CH tunes, whereas for high values the description is better for the \CH tune samples than with \SoftTune. Since the \deltaRg observable is strongly dependent on the amount of final-state radiation~\cite{CMS-TOP-17-013}, the difference comes mostly from the choice of \alpSMZ, with the choice of \alpSMZ in the \CH tunes preferred to that in \SoftTune.

\section{Comparisons with inclusive jet events}
\label{sec:JetMetComparisons}

The predictions of \HerwigS with the various tunes for inclusive jet production are investigated in this section.  In particular, the substructure of the jets is considered.  Events are generated with the LO QCD two-to-two MEs implemented in \HerwigS.  Although a comparison of the substructure of jets in \ttbar events was already presented in Section~\ref{sec:TopComparisons}, 
the comparison based on inclusive jet events is complementary because it probes a wider range of jet \pt.

\begin{figure*}[tbp]
  \centering
  \includegraphics[width=0.49\textwidth]{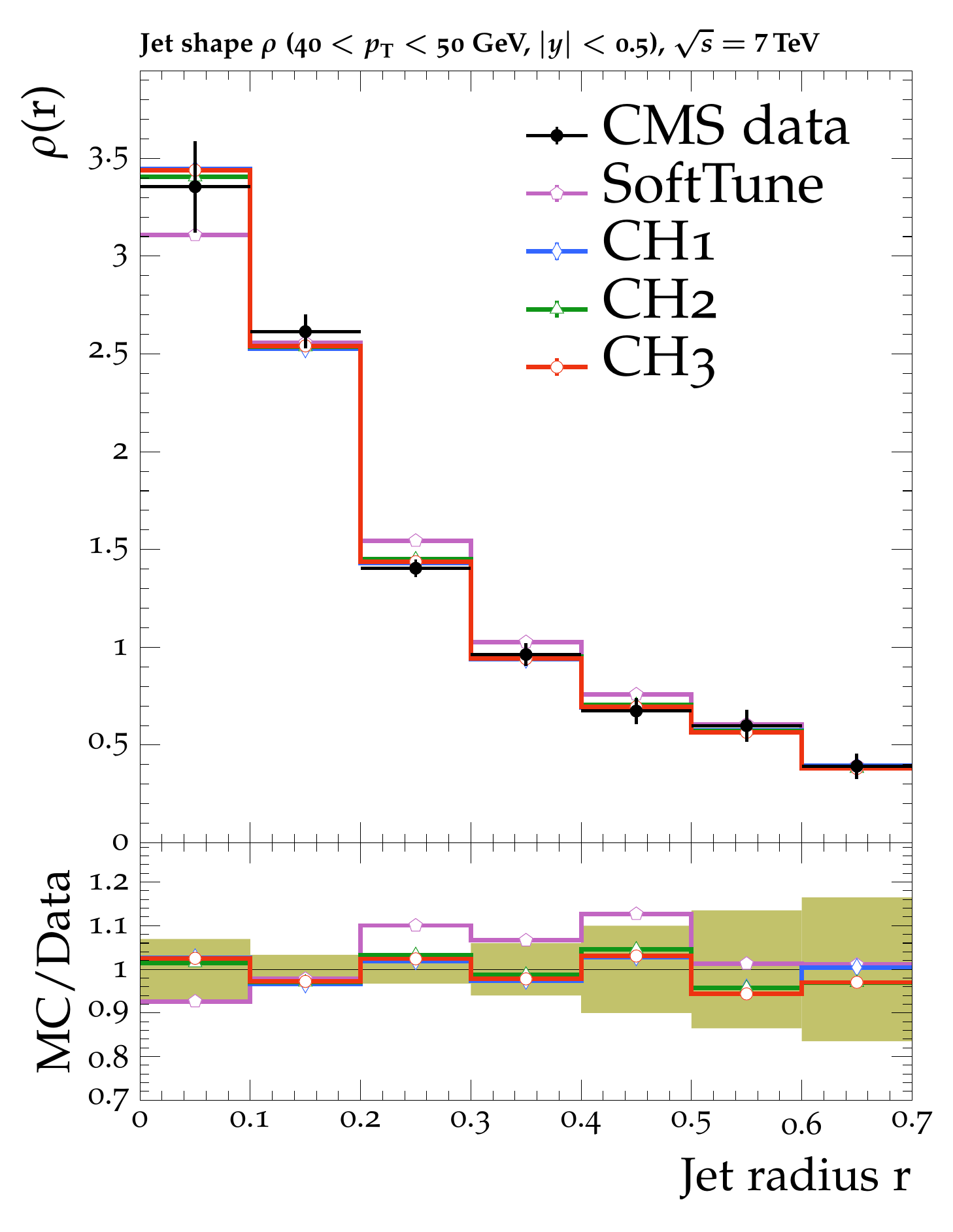}
  \includegraphics[width=0.49\textwidth]{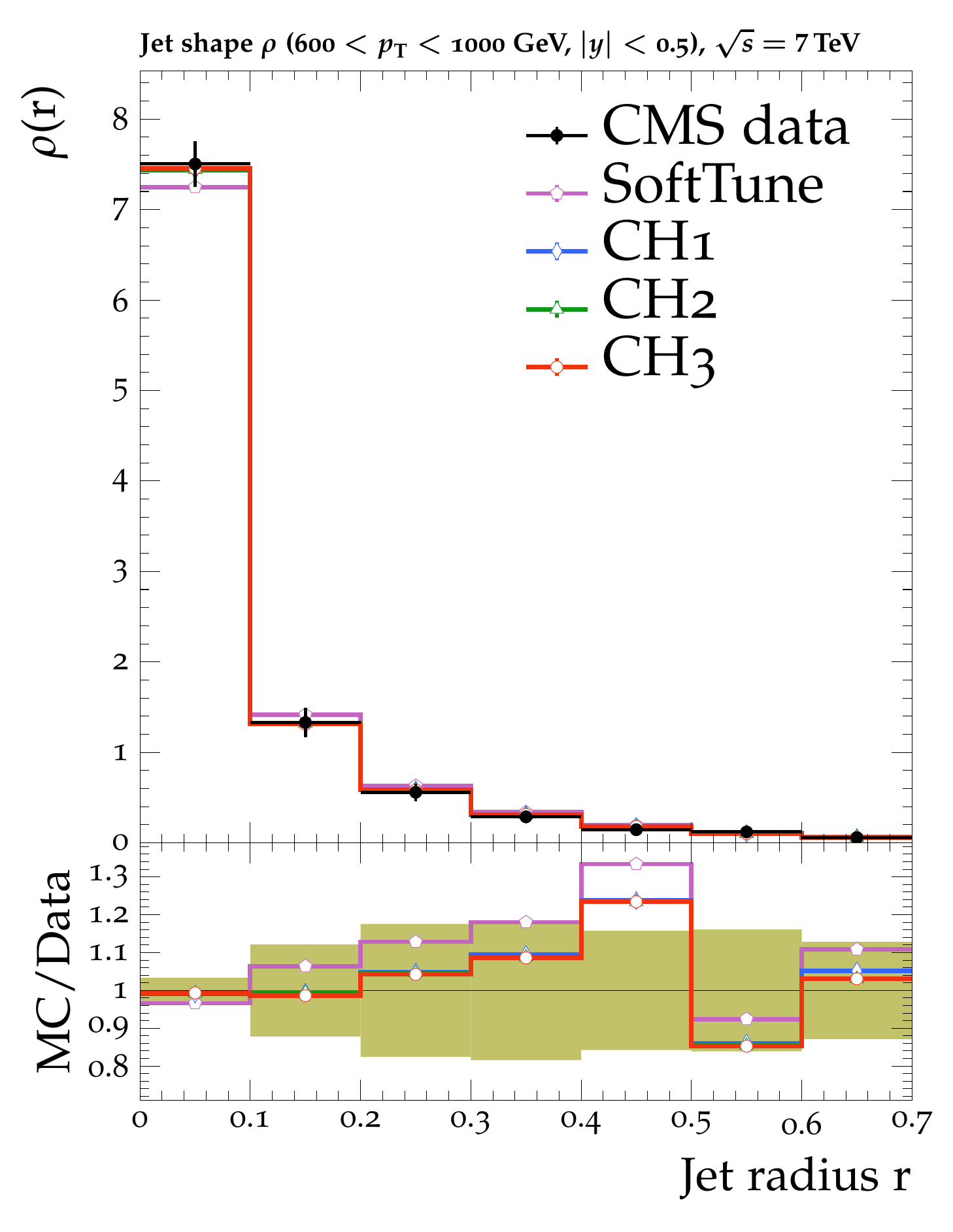} \\
  \includegraphics[width=0.49\textwidth]{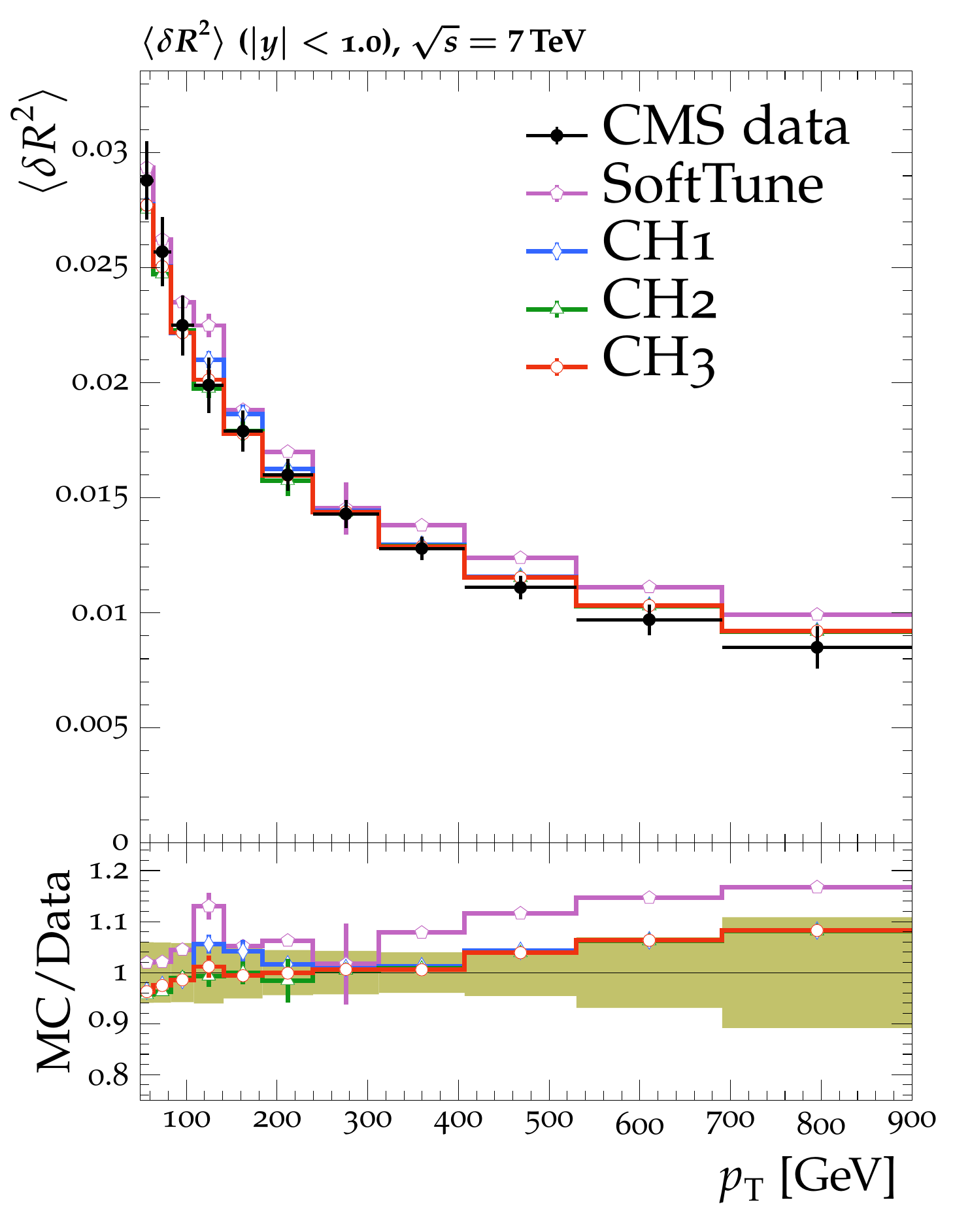}
  \caption{The differential jet shape \rhor (upper left and right) and the second moment of the jet transverse width \jetmoment in inclusive jet events~\cite{CMSJetShape}. CMS inclusive jet data are compared with the predictions from \HerwigS, with the \SoftTune and \CH tunes.  \captionColouredShadedBand }
  \label{fig:CMSJetShape}
\end{figure*}

Figure~\ref{fig:CMSJetShape} shows the differential jet shape, \rhor, as measured by the CMS experiment at $\sqrts=7\TeV$~\cite{CMSJetShape} for two bins of ranges of jet \pt (\ptjet): $40 < \ptjet < 50 \GeV$ and $600 < \ptjet < 1000 \GeV$.  The observable \rhor is defined as the average fraction of the \pt of the jet constituents contained inside an annulus with inner radius $\mathrm{r}-0.1$ and outer radius $\mathrm{r}+0.1$.  The second moment of the jet transverse width, \jetmoment, is also shown.
The jets are clustered with the \antikt algorithm with a distance parameter of 0.7 for the jet shape observables, and 0.5 for the \jetmoment observable.
The predictions from the three \CH tunes are very similar for all distributions, and agree with the data.  
On the other hand, the prediction from \SoftTune differs from the \CH tunes, and also does not agree well with the \jetmoment distribution in data.

Additional comparisons of the predictions for various tunes of \HerwigS tunes with the substructure of jets collected by the ATLAS experiment are shown in Appendix~\ref{app:ATLASJetSub}.

\section{Comparison with \texorpdfstring{\PZ}{Z} and \texorpdfstring{\PW}{W} boson production data}
\label{sec:VJetsComparisons}

{\tolerance=800 In this section, the performance of the \HerwigS tunes is compared with $\sqrts=13\TeV$ data on \PZ and \PW boson production.  Predictions for \PZ and \PW boson production are obtained with \MGaMC v2.6.7~\cite{MG5aMCatNLO} for ME calculations at NLO, which are interfaced with \HerwigS using the the \FXFX merging scheme~\cite{FXFX}, with the merging scale set to 30\GeV.  Up to two additional partons in the final state are included in the NLO ME calculations.  The PDF in the ME calculations is \NNPDF NNLO, and the value of \alpSMZ in the ME calculations is set to $\alpSMZ=0.118$ in all the predictions considered here. \par}

\begin{figure*}[tbp]
  \centering
  \includegraphics[width=0.45\textwidth]{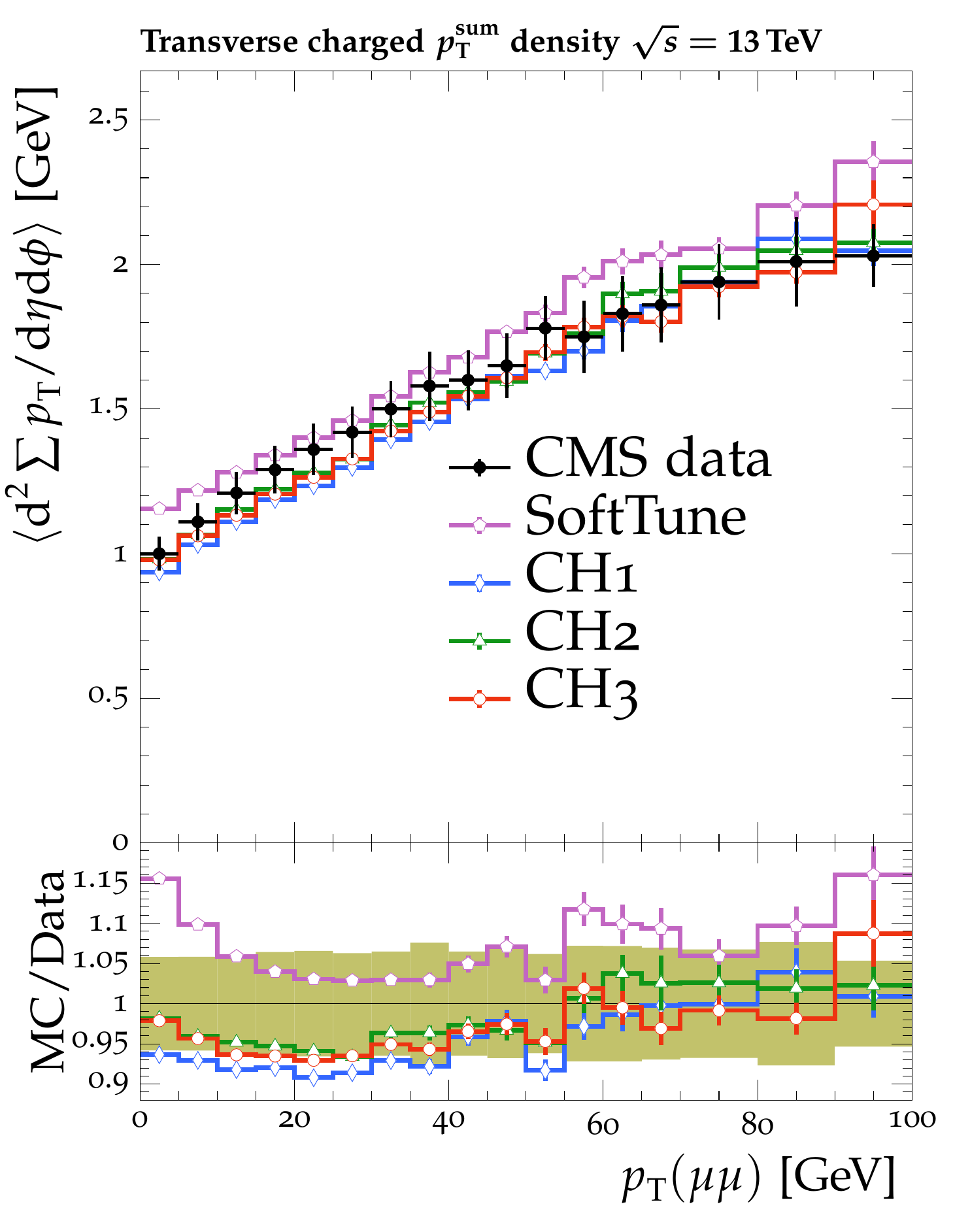}
  \includegraphics[width=0.45\textwidth]{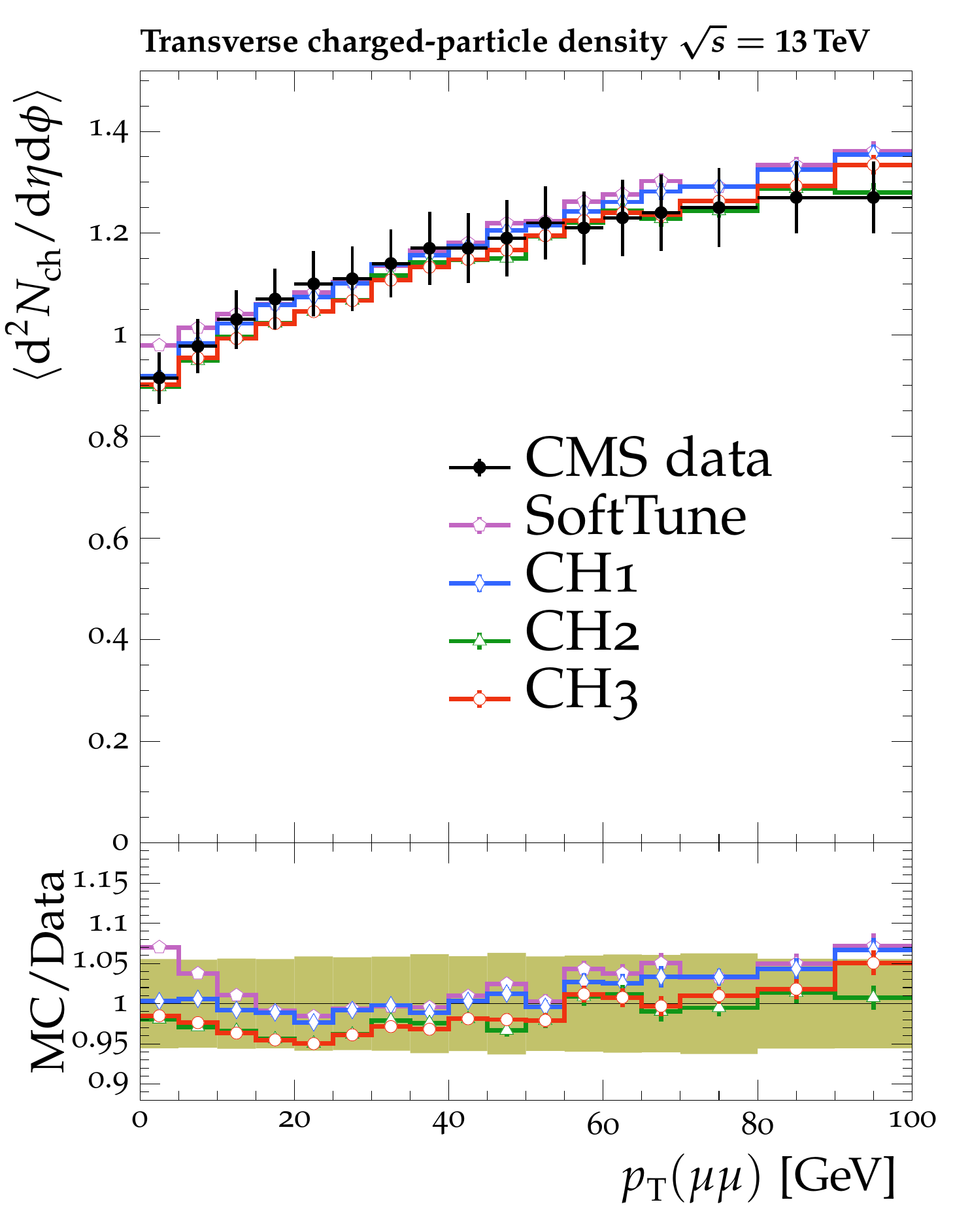} \\
  \caption{The \ptsum (left) and \Nch (right) density distributions in the transverse region, as a function of the \pt of the two muons, \ptmumu~\cite{CMSZUE}.  The transverse region is defined with respect to \ptmumu, where the two muons are required to have an invariant mass close the the mass of the \PZ boson.  CMS \PZ boson data are compared with the predictions from \MGaMCHerwigS, with the \SoftTune and \CH tunes.  \captionColouredShadedBand}
  \label{fig:CMS_ZJet_FXFX_UE}
\end{figure*}

\begin{figure*}[tbp]
  \centering
  \includegraphics[width=0.45\textwidth]{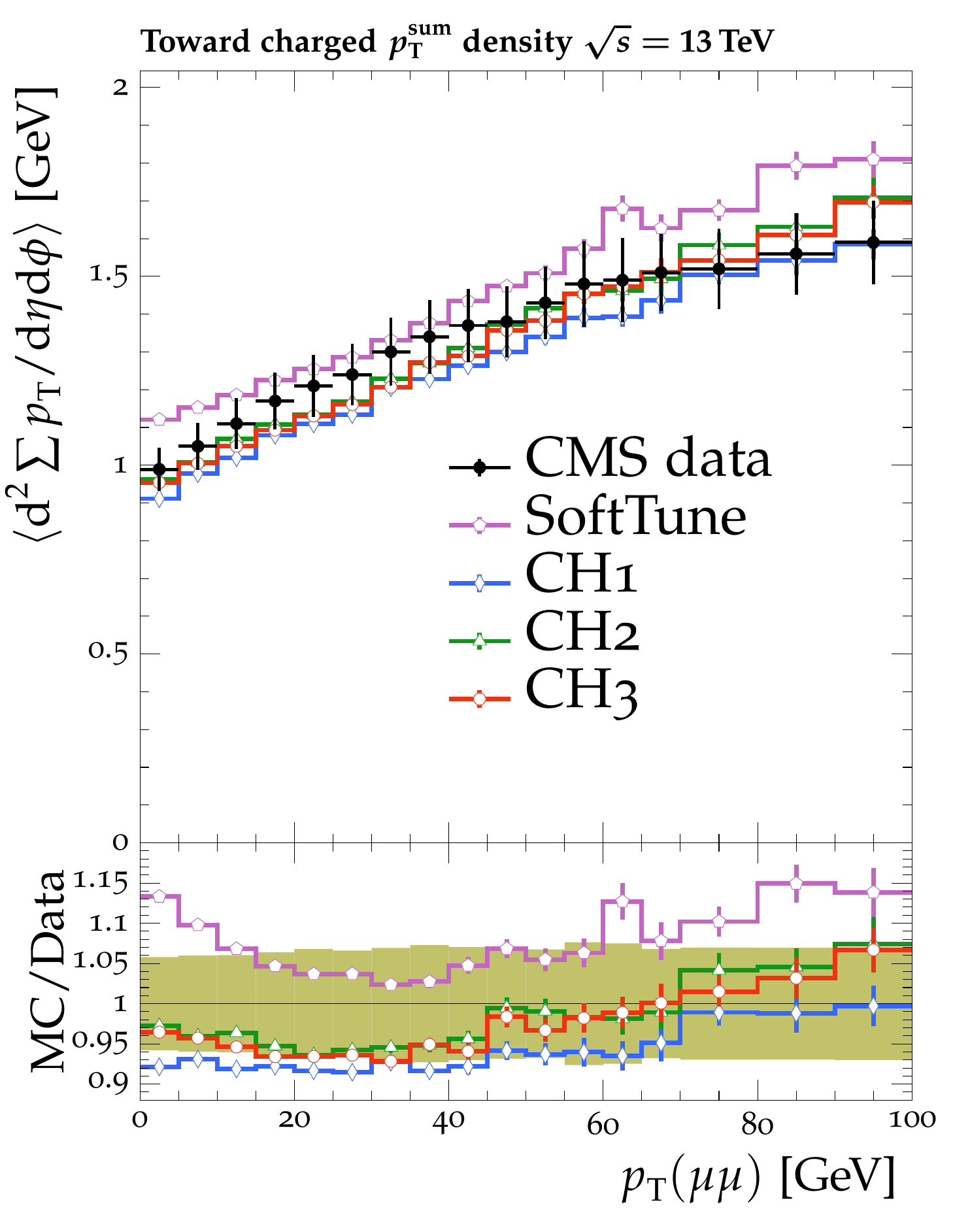}
  \includegraphics[width=0.45\textwidth]{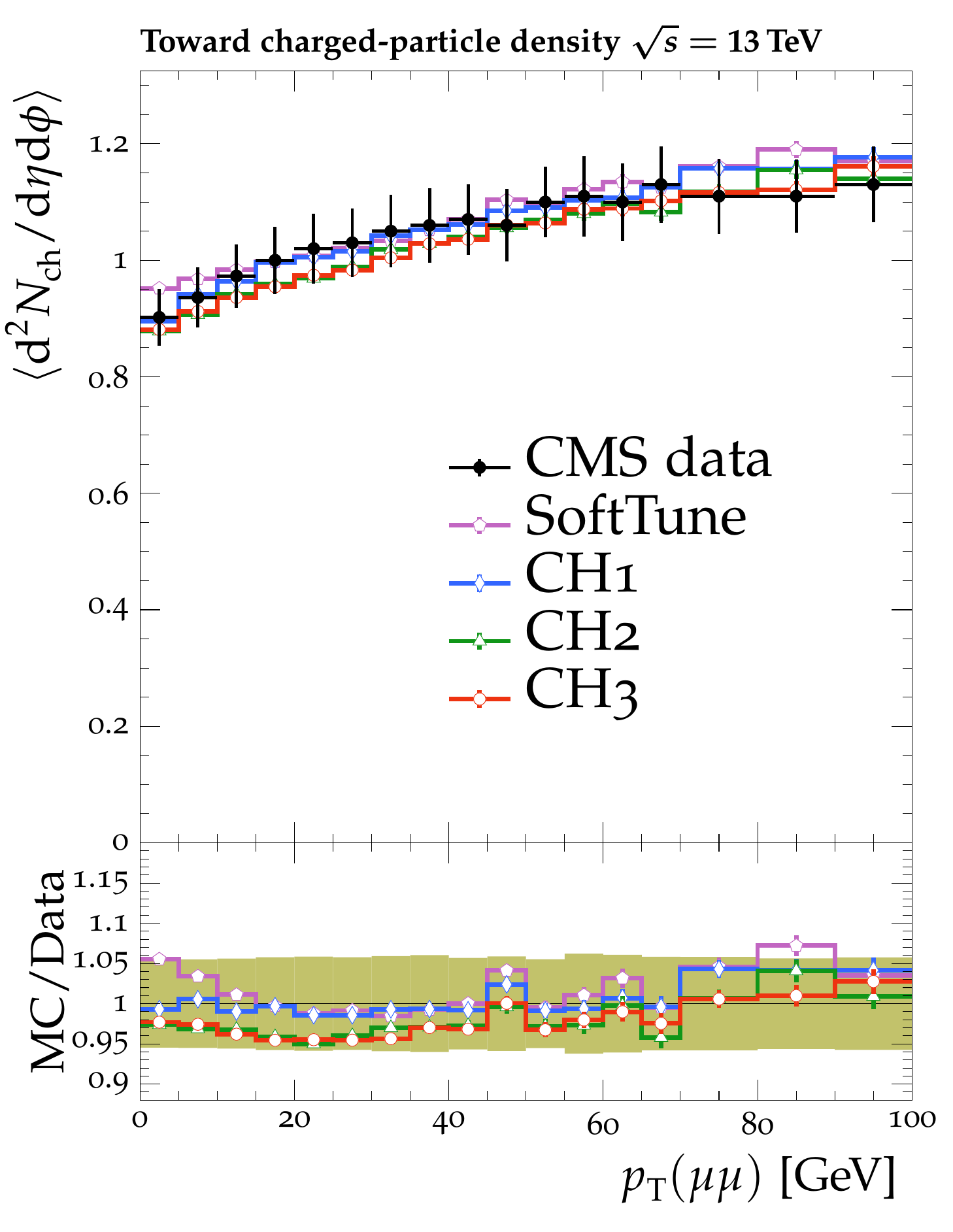} \\
  \includegraphics[width=0.45\textwidth]{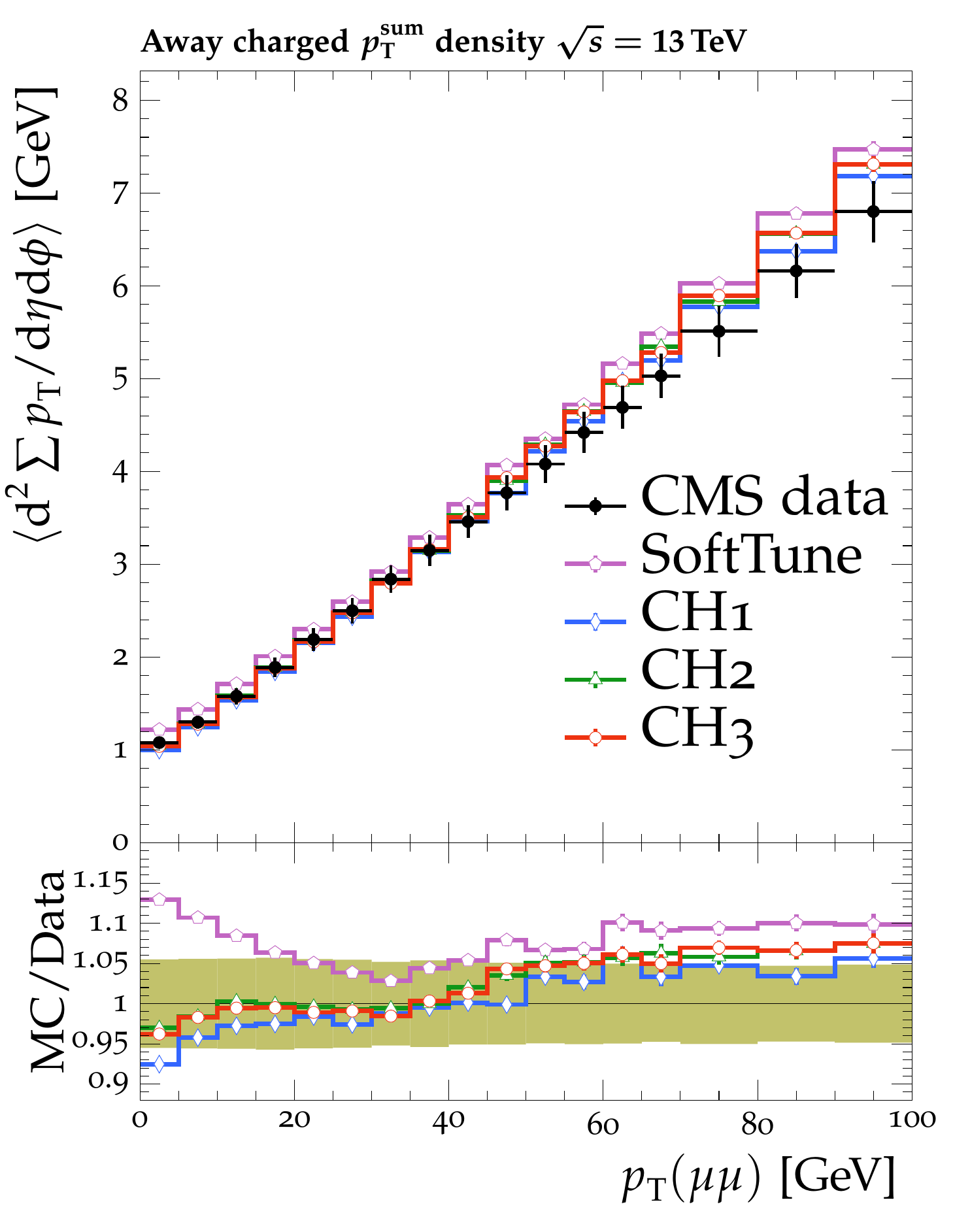}
  \includegraphics[width=0.45\textwidth]{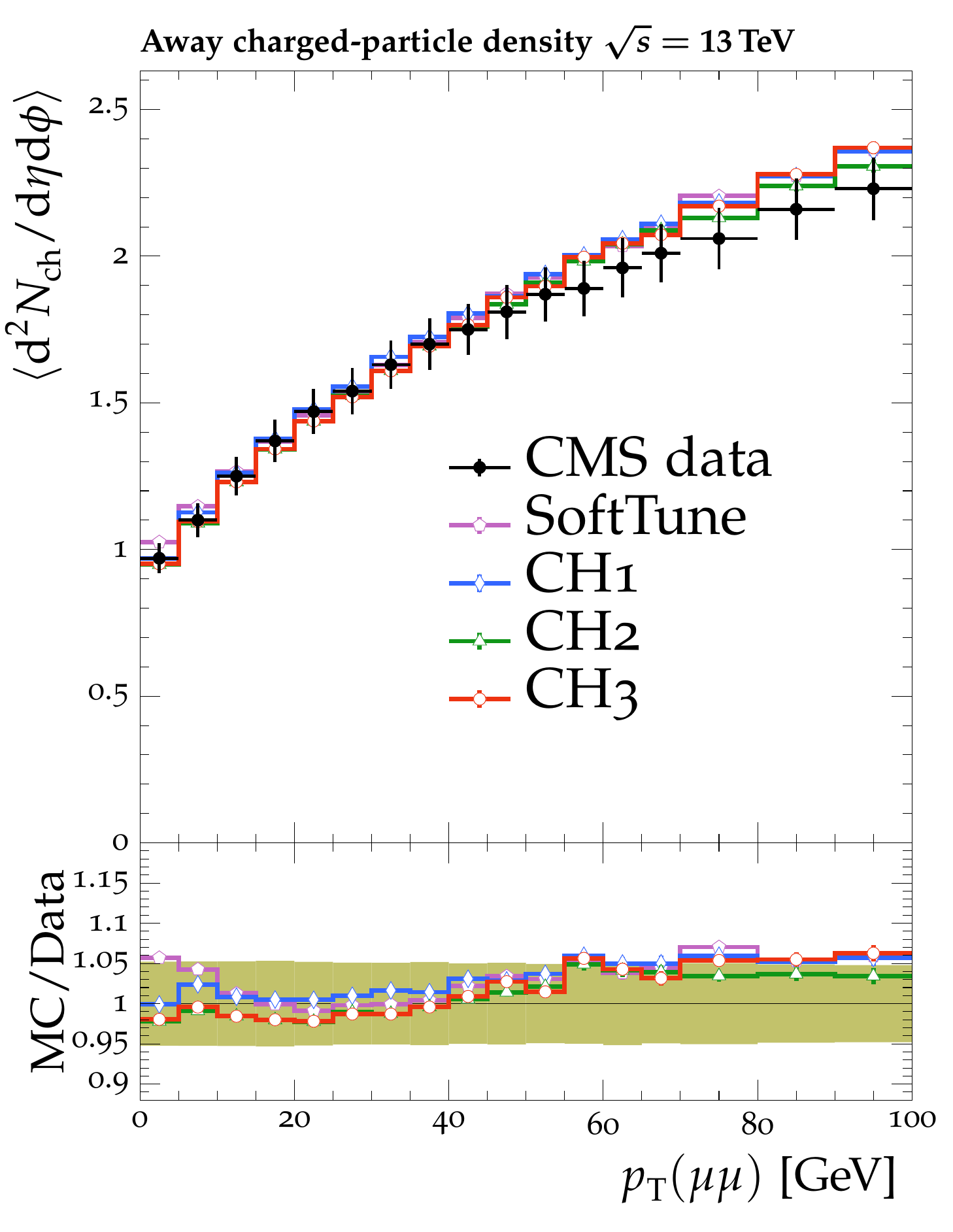} \\
  \caption{The \ptsum (left) and \Nch (right) density distributions in the toward (upper), and away (lower) regions, as a function of the \pt of the two muons, \ptmumu~\cite{CMSZUE}.  The toward and away regions are defined with respect to \ptmumu, where the two muons are required to have an invariant mass close the the mass of the \PZ boson.  CMS \PZ boson data are compared with the predictions from \MGaMCHerwigS, with the \SoftTune and \CH tunes.  \captionColouredShadedBand}
  \label{fig:CMS_ZJet_FXFX_UE_Toward_Away}
\end{figure*}

First, the \ptsum and \Nch distributions characterizing the UE in \PZ boson production~\cite{CMSZUE} are compared to simulation in Figs.~\ref{fig:CMS_ZJet_FXFX_UE} and~\ref{fig:CMS_ZJet_FXFX_UE_Toward_Away}.  Events are required to have two muons with an invariant mass between 81 and 101\GeV to select events within the \PZ boson mass peak.  The \ptsum and \Nch distributions are measured in the transverse region as shown in Fig.~\ref{fig:CMS_ZJet_FXFX_UE}, and in the toward and away regions as shown in Fig.~\ref{fig:CMS_ZJet_FXFX_UE_Toward_Away}, in analogy to the corresponding distributions measured in MB data introduced in Section~\ref{sec:TuneProcedure}.  The regions are defined with respect to the \pt of the \PZ boson, calculated from the \pt of the two muons.  The \CH tunes describe the data well, and are typically similar to each other.  However, the configuration with \SoftTune fails to give a simultaneous description of the \ptsum and \Nch distributions in any region at low \ptmumu.

\begin{figure*}[tbp]
  \centering
  \includegraphics[width=0.49\textwidth]{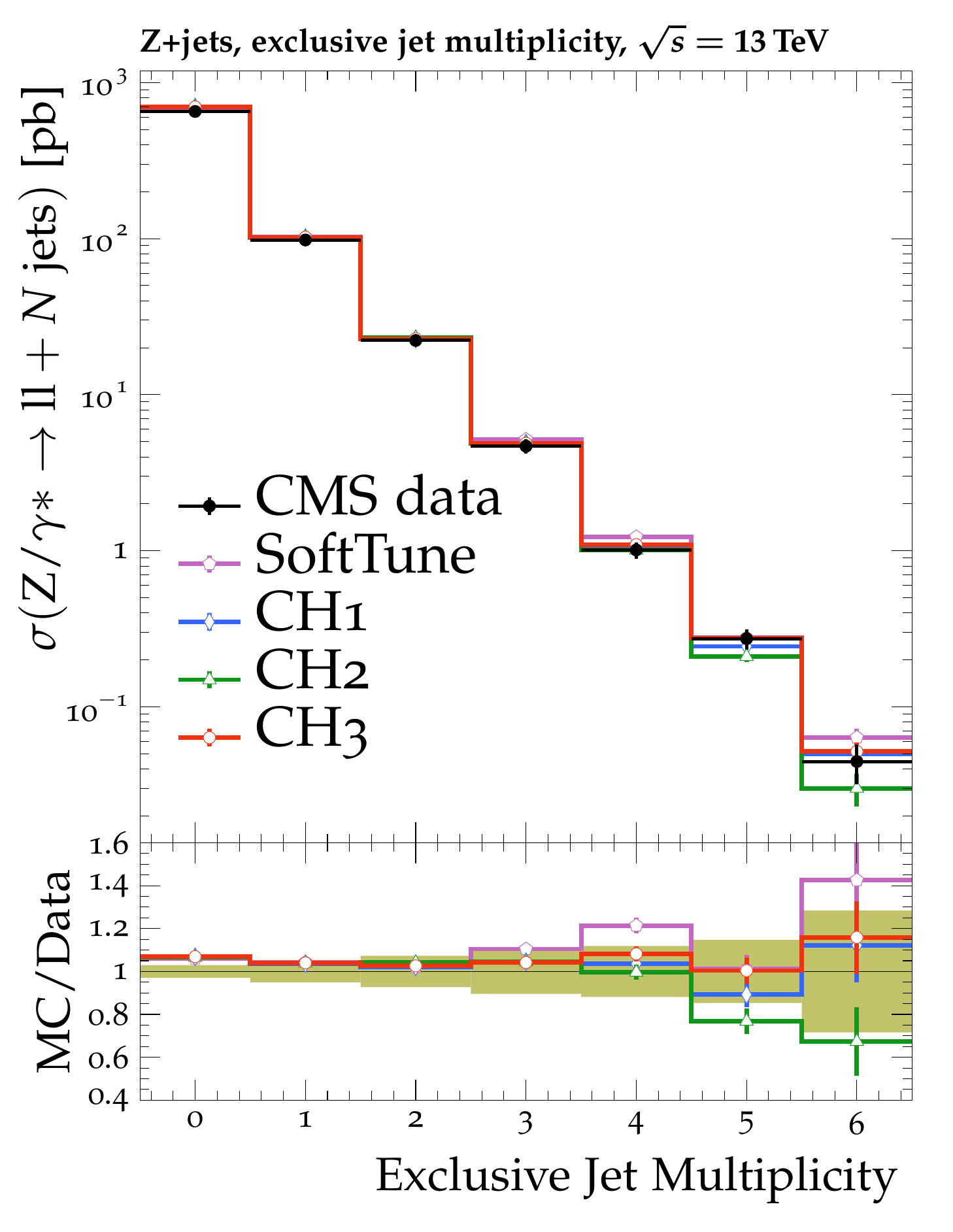}
  \includegraphics[width=0.49\textwidth]{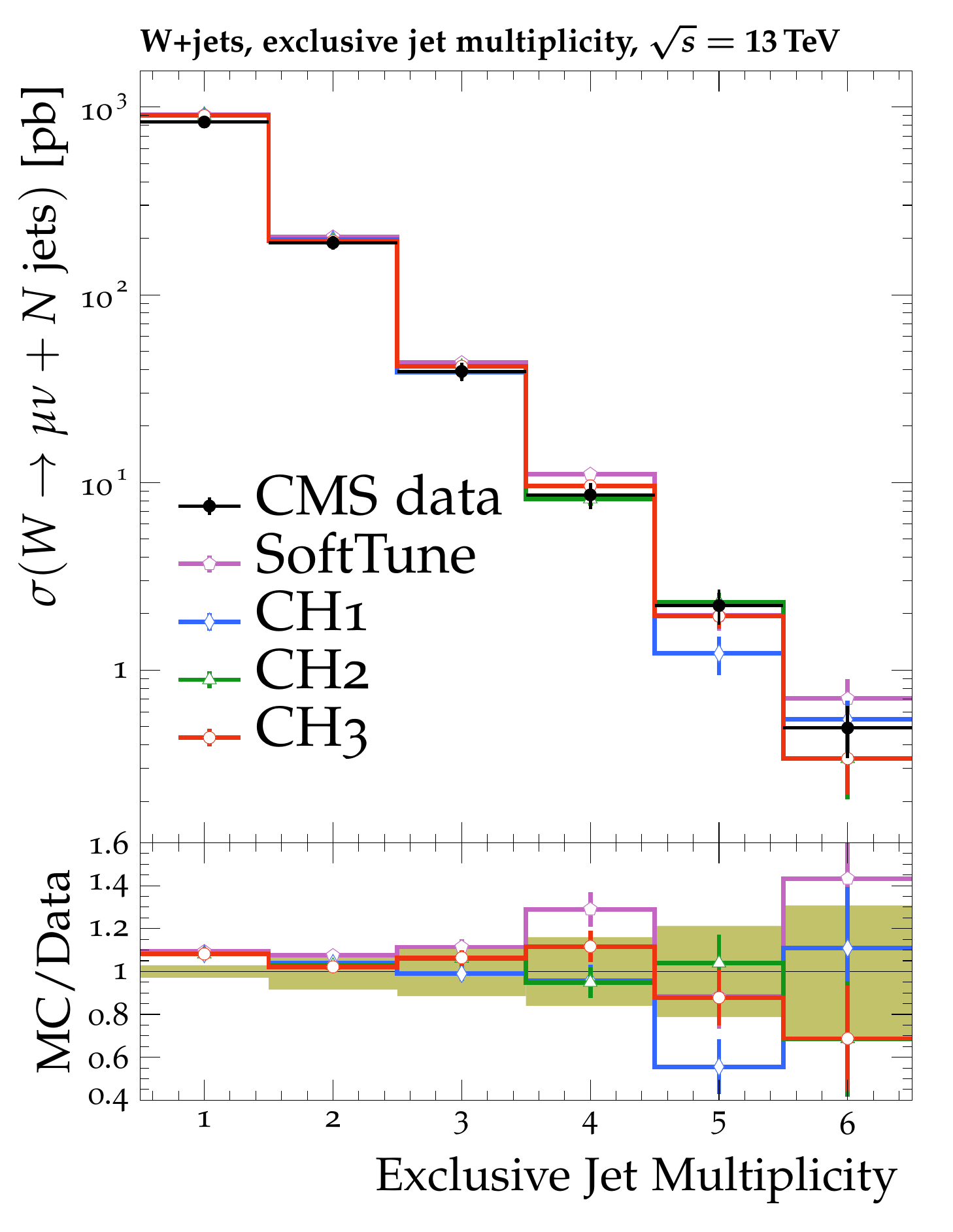} \\
  \caption{The exclusive jet multiplicity in \PZ (left) and \PW (right) boson events, measured by CMS at $\sqrts=13\TeV$~\cite{ZJet13TeV,WJet13TeV}.  CMS \PZ boson and \PW boson data are compared with the predictions from \MGaMCHerwigS, with the \SoftTune and \CH tunes.  \captionColouredShadedBand  }
  \label{fig:CMS_ZJet_FXFX_NJets}
\end{figure*}

Next, the exclusive jet multiplicity distributions in \PZ and \PW boson events are shown in Fig.~\ref{fig:CMS_ZJet_FXFX_NJets}~\cite{ZJet13TeV,WJet13TeV}.  Events in the \PZ boson sample contain at least two electrons or muons with $\pt>20\GeV$ and $\abs{\eta}<2.4$, and the invariant mass of the two highest \pt electrons or muons must have an invariant mass within 20\GeV of the \PZ boson mass.  In the \PW boson measurement, only final states with a muon of $\pt>25\GeV$ and $\abs{\eta}<2.4$ are considered.  The transverse mass of the \PW boson candidate, defined as \transverseMass, where $\cos(\Delta\phi_{\mu,\ptvecmiss})$ is the difference in azimuthal angle between the direction of the muon momentum and \ptmiss, must satisfy $\mT>50\GeV$.  In both \PZ and \PW events jets are reconstructed using the \antikt algorithm with a distance parameter of 0.4, and are required to satisfy $\pt>30\GeV$ and $\abs{y}<2.4$.  Jets must also be separated from any lepton by $\deltaRDefn>0.4$, where $\phi$ is in radians.  The jet multiplicity is well described by all tunes in both \PZ and \PW boson events at both low multiplicities, where the ME calculations dominate, and high multiplicities, where the PS is important.

\begin{figure*}[tbp]
  \centering
  \includegraphics[width=0.49\textwidth]{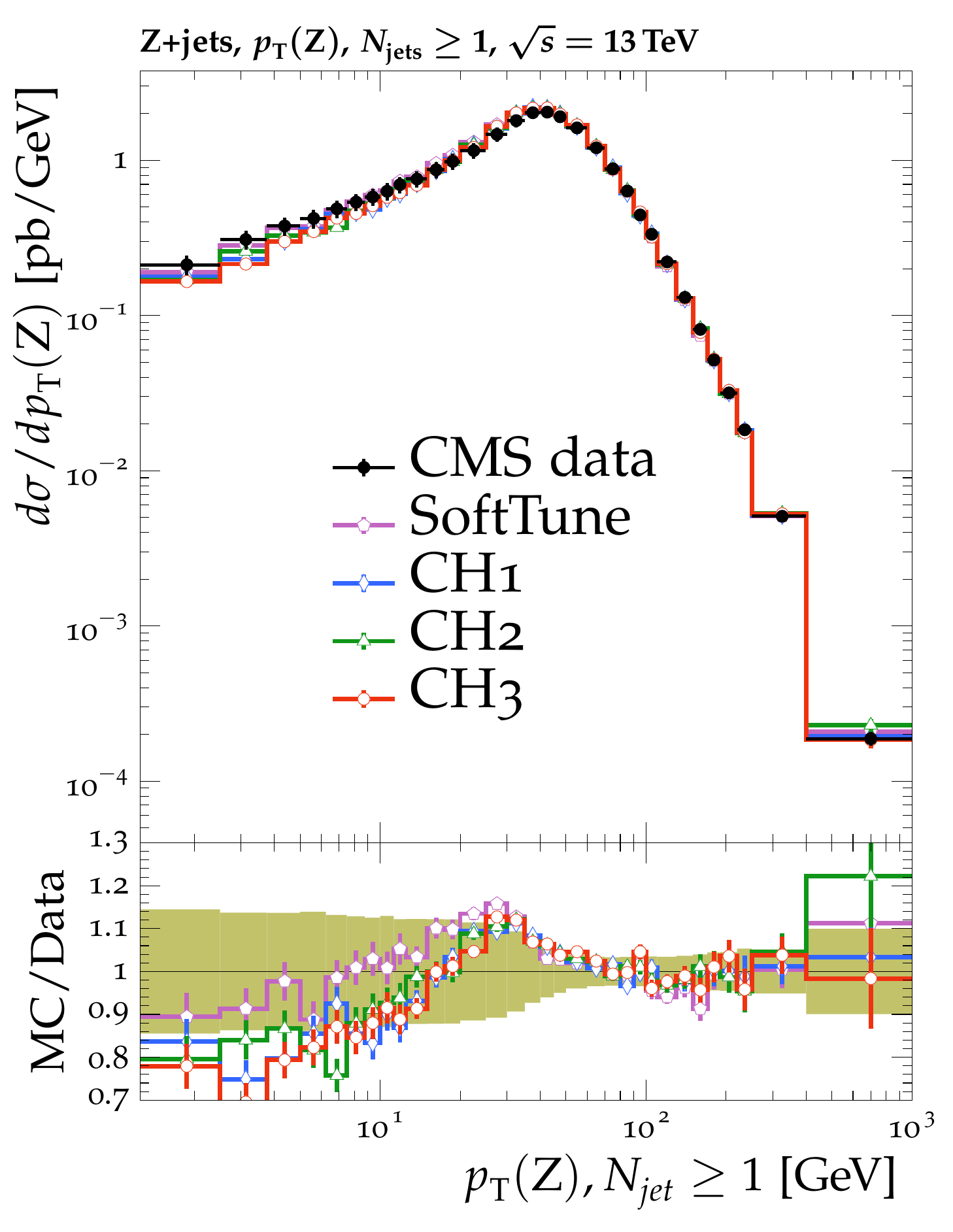}
  \includegraphics[width=0.49\textwidth]{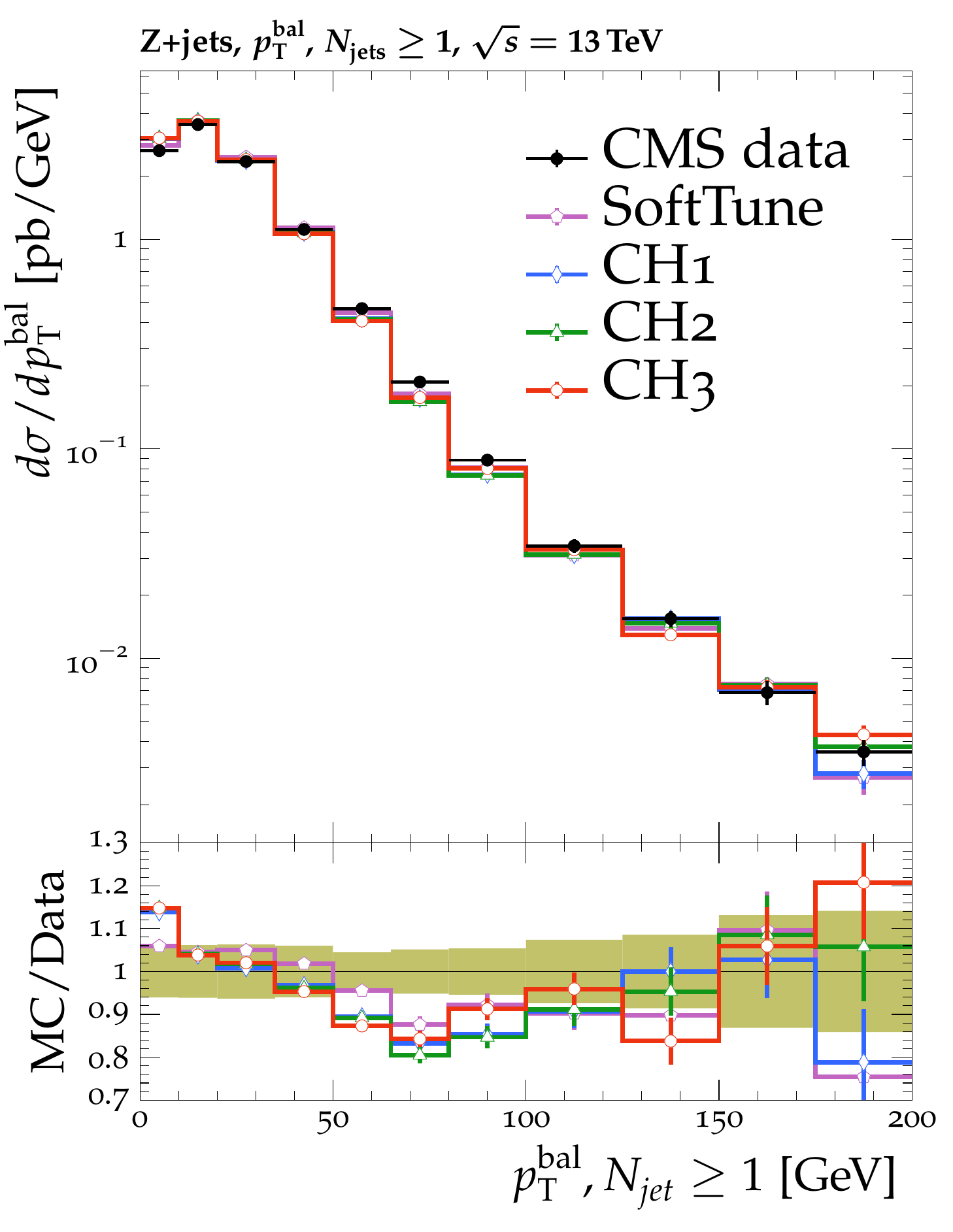} \\
  \includegraphics[width=0.49\textwidth]{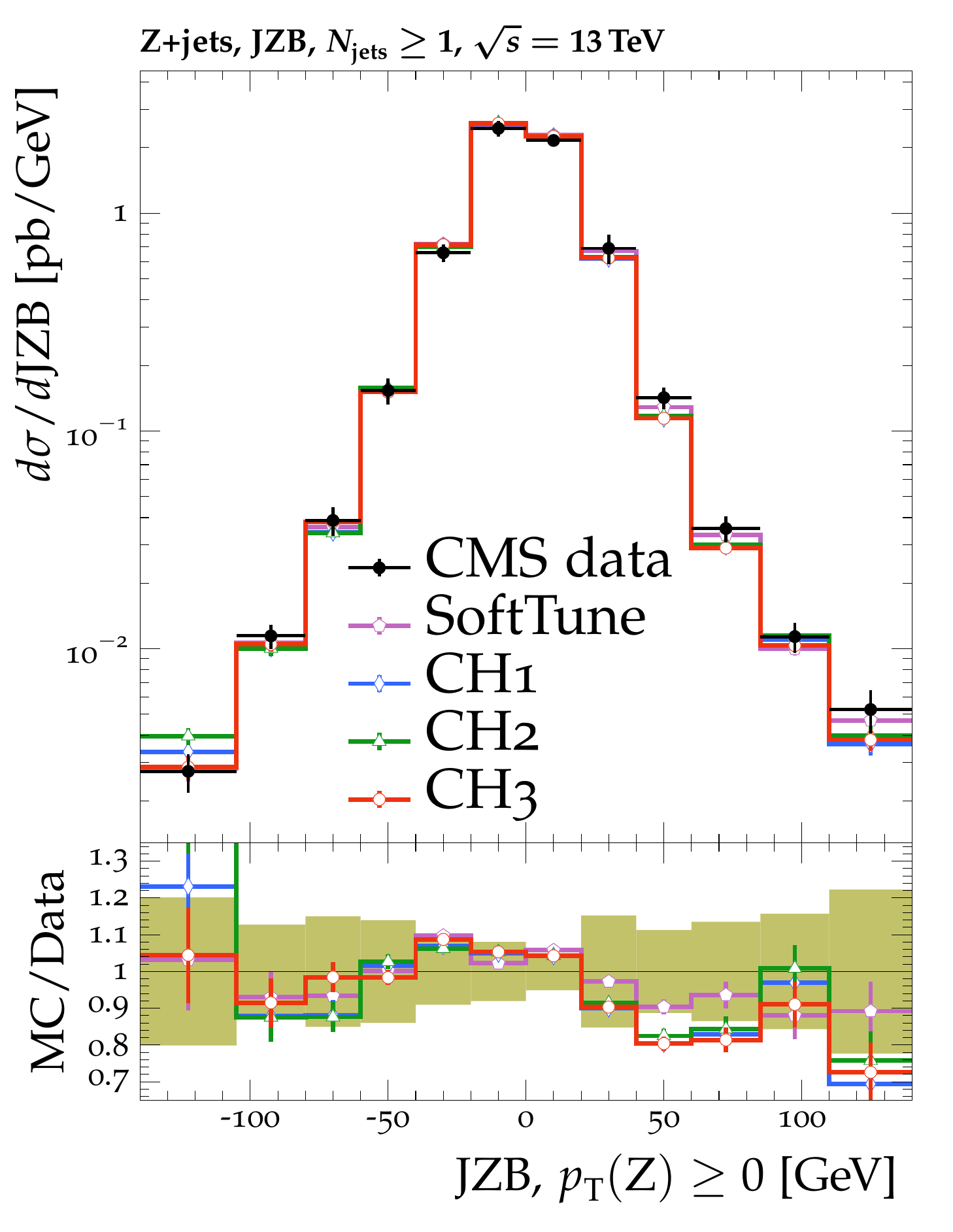} \\
  \caption{Differential cross sections as a function of \ptZ (upper left), \ptbal (upper right), and \JZB (lower)~\cite{ZJet13TeV}.  CMS \PZ boson data are compared with the predictions from \MGaMCHerwigS, with the \SoftTune and \CH tunes.  \captionColouredShadedBand}
  \label{fig:CMS_ZJet_FXFX_ptZ}
\end{figure*}

Finally, in Fig.~\ref{fig:CMS_ZJet_FXFX_ptZ}, the \ptZ and \ptbal distributions are shown, both for final states containing at least one additional jet.  The \ptbal variable is defined as $\ptbal = \abs{\ptvecZ+\sum_{\mathrm{jets}}\ptvec(\mathrm{j})}$.  The so-called jet-\PZ balance (\JZB) variable, defined as $\JZB = \abs{\sum_{\mathrm{jets}}\ptvec(\mathrm{j})}-\abs{\ptvecZ}$, is also shown in Fig.~\ref{fig:CMS_ZJet_FXFX_ptZ}.  All distributions are measured for events with at least one additional jet.  The \ptZ predictions for all tunes are similar for $\ptZ>30\GeV$, where the predictions are driven by the ME calculations.  At lower \ptZ, where events contain additional hadronic activity that is not clustered into jets, the predictions with the \CH tunes are similar to each other, and differ slightly from the prediction with \SoftTune, which provides a closer description of the data at very low $\ptZ<10\GeV$. The \ptbal and \JZB distributions are also sensitive to additional hadronic activity not clustered into jets.  For \ptbal, all tunes are compatible with each other, except at $\ptbal<10\GeV$, where the prediction with \SoftTune differs from the predictions with the \CH tunes.  The \JZB distributions are well described by all the predictions.

\section{Summary}
\label{sec:Summary}
{\tolerance=800 Three new tunes for the multiple-parton interaction (MPI) model of the \HerwigS (version 7.1.4) generator have been derived from minimum-bias (MB) data collected by the CMS experiment.  All of the \CH (``CMS \HERWIG{}'') tunes, \CHo, \CHt, and \CHth, are based on the next-to-next-to-leading-order (NNLO) \NNPDF PDF set for the simulation of the parton shower (PS) in \HerwigS; the value of the strong coupling at a scale equal to the \PZ boson mass is $\alpSMZ=0.118$ with a two-loop evolution of \alpS.  The configuration of the tunes differs in the PDF used for the simulation of MPI and beam remnants.  The tune \CHo uses the same NNLO PDF set for these aspects of the \HerwigS simulation, whereas \CHt and \CHth use leading-order (LO) versions of the PDF set.  The tune \CHt is based on an LO PDF set that was derived assuming $\alpSMZ=0.118$, and \CHth on an LO PDF set assuming $\alpSMZ=0.130$. \par}

The parameters of the MPI model were optimized for each tune with the \Professor framework to describe the underlying event (UE) in MB data collected by CMS.
The predictions using the tune \CHt or \CHth provide a better description of the data than those using \CHo or \SoftTune.  Furthermore, the differences in the predictions of \CHt and \CHth are observed to be small.  The configuration of PDF sets in the tune \CHth, where the LO PDF used for the simulation of MPI, was derived with a value of $\alpSMZ$ typically associated with LO PDF sets, is the preferred choice over \CHt.  Two alternative tunes representing the uncertainties in the fitted parameters of \CHth are also derived, based on the tuning procedure provided by \Professor.

Predictions using the three \CH tunes are compared with a range of data beyond MB events: event shape data from LEP; proton-proton data enriched in top quark pairs, \PZ bosons and \PW bosons; and inclusive jet data.  This validated the performance of \HerwigS using these tunes against a wide range of data sets sensitive to various aspects of the modelling by \HerwigS, and in particular the modelling of the UE.
The event shape observables measured at LEP, which are sensitive to the modelling of final-state radiation, are well described by \HerwigS with the new tunes.
Predictions using the new tunes are also shown to describe the UE in events containing \PZ bosons, demonstrating the universality of the UE modelling in \HerwigS.  The kinematics of top quark events, and the modelling of jets in \ttbar, \PZ boson, \PW boson, and inclusive jet data are also well described by predictions using the new tunes.  In general, predictions with the new \CH tunes derived in this paper provide a better description of measured observables than those using \SoftTune, the default tune available in \HerwigS.
\ifthenelse{\boolean{cms@external}}{\clearpage}{}
\begin{acknowledgments}
  We congratulate our colleagues in the CERN accelerator departments for the excellent performance of the LHC and thank the technical and administrative staffs at CERN and at other CMS institutes for their contributions to the success of the CMS effort. In addition, we gratefully acknowledge the computing centres and personnel of the Worldwide LHC Computing Grid for delivering so effectively the computing infrastructure essential to our analyses. Finally, we acknowledge the enduring support for the construction and operation of the LHC and the CMS detector provided by the following funding agencies: BMBWF and FWF (Austria); FNRS and FWO (Belgium); CNPq, CAPES, FAPERJ, FAPERGS, and FAPESP (Brazil); MES (Bulgaria); CERN; CAS, MoST, and NSFC (China); COLCIENCIAS (Colombia); MSES and CSF (Croatia); RIF (Cyprus); SENESCYT (Ecuador); MoER, ERC PUT and ERDF (Estonia); Academy of Finland, MEC, and HIP (Finland); CEA and CNRS/IN2P3 (France); BMBF, DFG, and HGF (Germany); GSRT (Greece); NKFIA (Hungary); DAE and DST (India); IPM (Iran); SFI (Ireland); INFN (Italy); MSIP and NRF (Republic of Korea); MES (Latvia); LAS (Lithuania); MOE and UM (Malaysia); BUAP, CINVESTAV, CONACYT, LNS, SEP, and UASLP-FAI (Mexico); MOS (Montenegro); MBIE (New Zealand); PAEC (Pakistan); MSHE and NSC (Poland); FCT (Portugal); JINR (Dubna); MON, RosAtom, RAS, RFBR, and NRC KI (Russia); MESTD (Serbia); SEIDI, CPAN, PCTI, and FEDER (Spain); MOSTR (Sri Lanka); Swiss Funding Agencies (Switzerland); MST (Taipei); ThEPCenter, IPST, STAR, and NSTDA (Thailand); TUBITAK and TAEK (Turkey); NASU (Ukraine); STFC (United Kingdom); DOE and NSF (USA).
  
  \hyphenation{Rachada-pisek} Individuals have received support from the Marie-Curie programme and the European Research Council and Horizon 2020 Grant, contract Nos.\ 675440, 724704, 752730, and 765710 (European Union); the Leventis Foundation; the A.P.\ Sloan Foundation; the Alexander von Humboldt Foundation; the Belgian Federal Science Policy Office; the Fonds pour la Formation \`a la Recherche dans l'Industrie et dans l'Agriculture (FRIA-Belgium); the Agentschap voor Innovatie door Wetenschap en Technologie (IWT-Belgium); the F.R.S.-FNRS and FWO (Belgium) under the ``Excellence of Science -- EOS" -- be.h project n.\ 30820817; the Beijing Municipal Science \& Technology Commission, No. Z191100007219010; the Ministry of Education, Youth and Sports (MEYS) of the Czech Republic; the Deutsche Forschungsgemeinschaft (DFG) under Germany's Excellence Strategy -- EXC 2121 ``Quantum Universe" -- 390833306; the Lend\"ulet (``Momentum") Programme and the J\'anos Bolyai Research Scholarship of the Hungarian Academy of Sciences, the New National Excellence Program \'UNKP, the NKFIA research grants 123842, 123959, 124845, 124850, 125105, 128713, 128786, and 129058 (Hungary); the Council of Science and Industrial Research, India; the HOMING PLUS programme of the Foundation for Polish Science, cofinanced from European Union, Regional Development Fund, the Mobility Plus programme of the Ministry of Science and Higher Education, the National Science Center (Poland), contracts Harmonia 2014/14/M/ST2/00428, Opus 2014/13/B/ST2/02543, 2014/15/B/ST2/03998, and 2015/19/B/ST2/02861, Sonata-bis 2012/07/E/ST2/01406; the National Priorities Research Program by Qatar National Research Fund; the Ministry of Science and Higher Education, project no. 02.a03.21.0005 (Russia); the Tomsk Polytechnic University Competitiveness Enhancement Program; the Programa Estatal de Fomento de la Investigaci{\'o}n Cient{\'i}fica y T{\'e}cnica de Excelencia Mar\'{\i}a de Maeztu, grant MDM-2015-0509 and the Programa Severo Ochoa del Principado de Asturias; the Thalis and Aristeia programmes cofinanced by EU-ESF and the Greek NSRF; the Rachadapisek Sompot Fund for Postdoctoral Fellowship, Chulalongkorn University and the Chulalongkorn Academic into Its 2nd Century Project Advancement Project (Thailand); the Kavli Foundation; the Nvidia Corporation; the SuperMicro Corporation; the Welch Foundation, contract C-1845; and the Weston Havens Foundation (USA).
\end{acknowledgments}
\bibliography{auto_generated}
\appendix
\numberwithin{figure}{section}
\section{Comparison with ATLAS MB data}
\label{app:ATLASMB}

\begin{figure*}[htbp]
  \centering
  \includegraphics[width=0.49\textwidth]{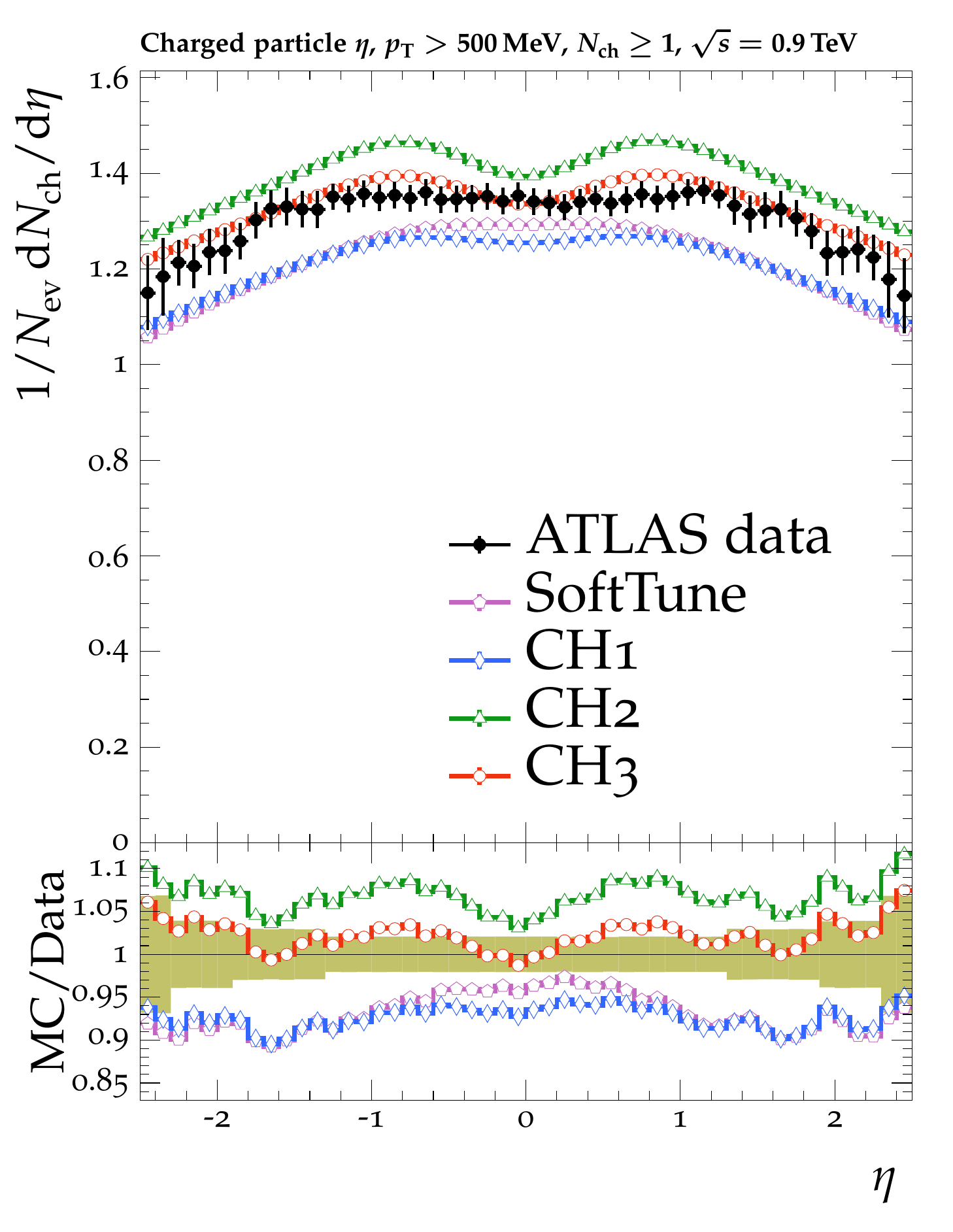}
  \includegraphics[width=0.49\textwidth]{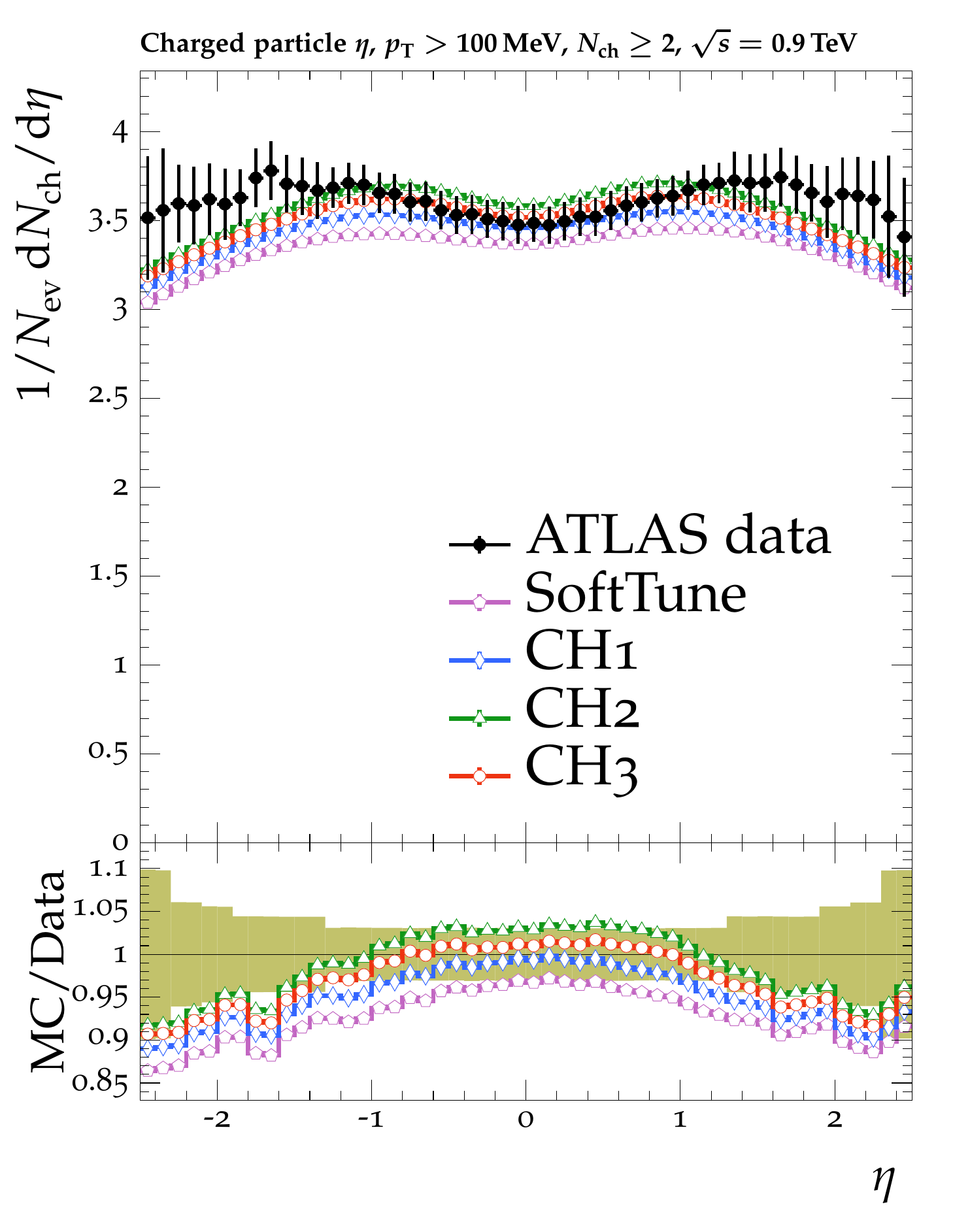} \\
  \includegraphics[width=0.49\textwidth]{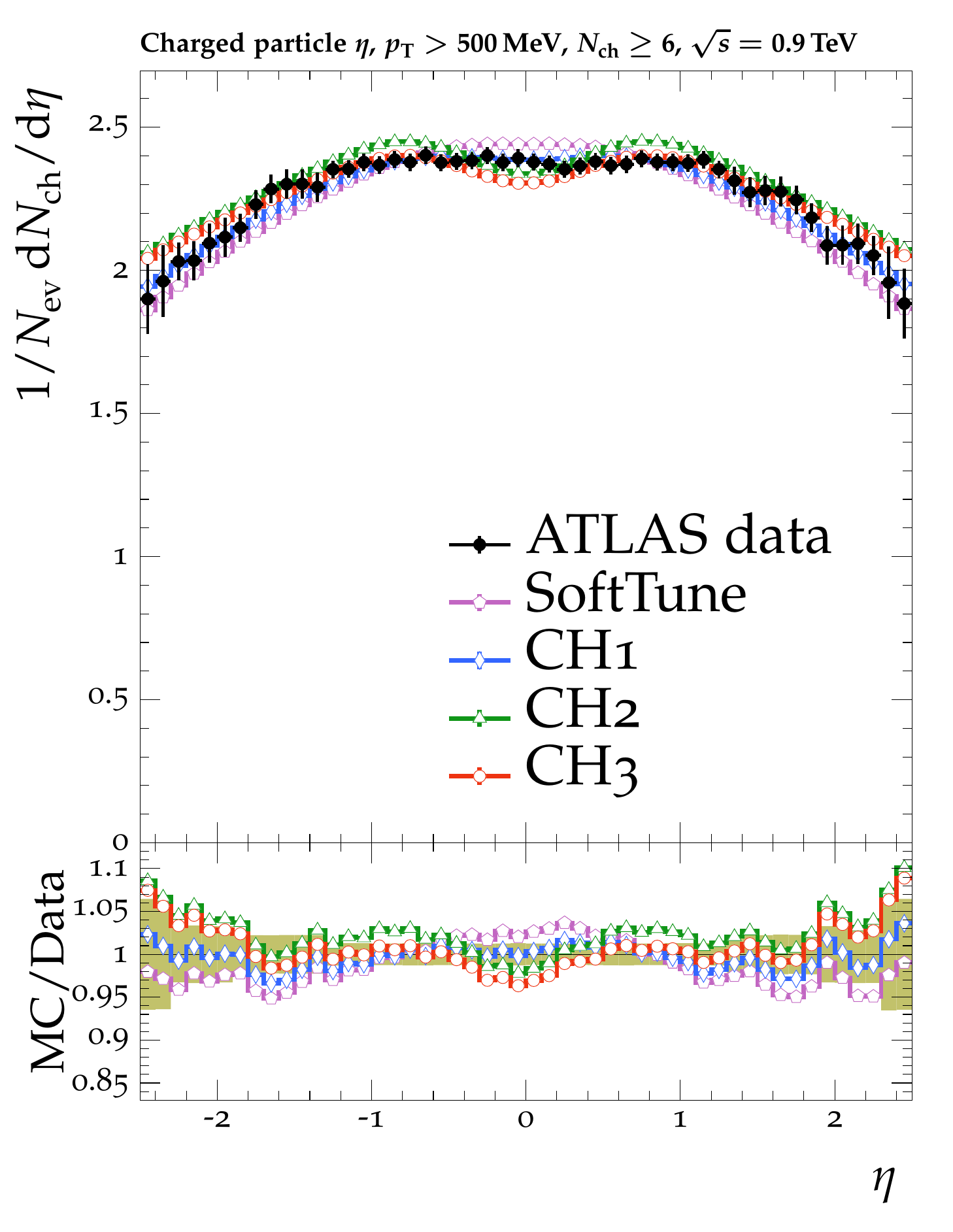}
  \includegraphics[width=0.49\textwidth]{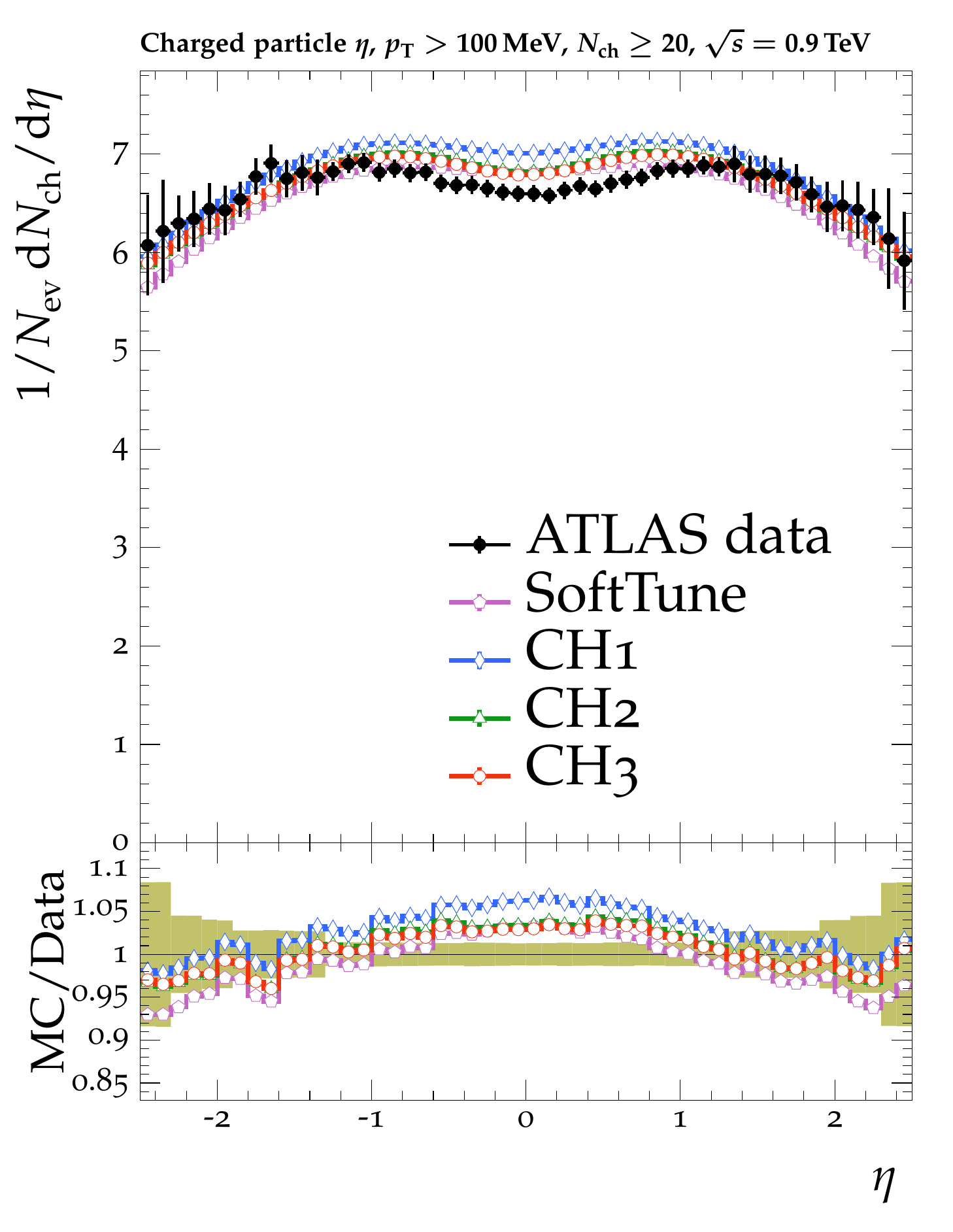}
  \caption{Normalized plots~\cite{ATLASdNdEta0p9And7TeV} for the pseudorapidity of charged particles for $\Nch\ge1$ (upper left), and $\Nch\ge6$ (lower left), for charged particles with $\pt>500\MeV$.  The figure on the upper right shows a similar distribution for $\Nch\ge2$, and the lower right for $\Nch\ge20$, where the charged particles have $\pt>100\MeV$.  ATLAS MB data are compared with the predictions from \HerwigS, with the \SoftTune and \CH tunes.  \captionColouredShadedBand}
  \label{fig:ATLAS_dNdeta_0p9TeV}
\end{figure*}

\begin{figure*}[htbp]
  \centering
  \includegraphics[width=0.49\textwidth]{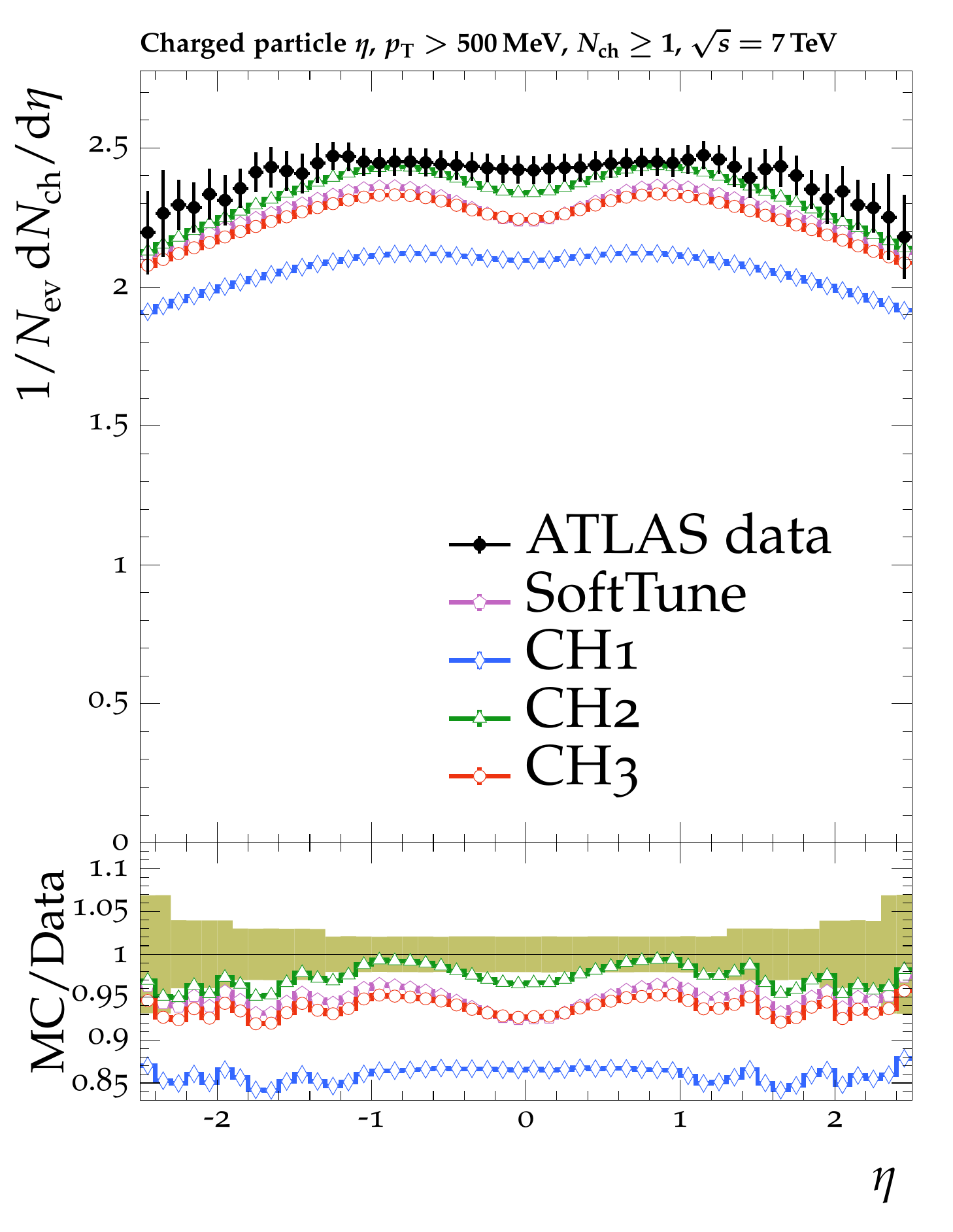}
  \includegraphics[width=0.49\textwidth]{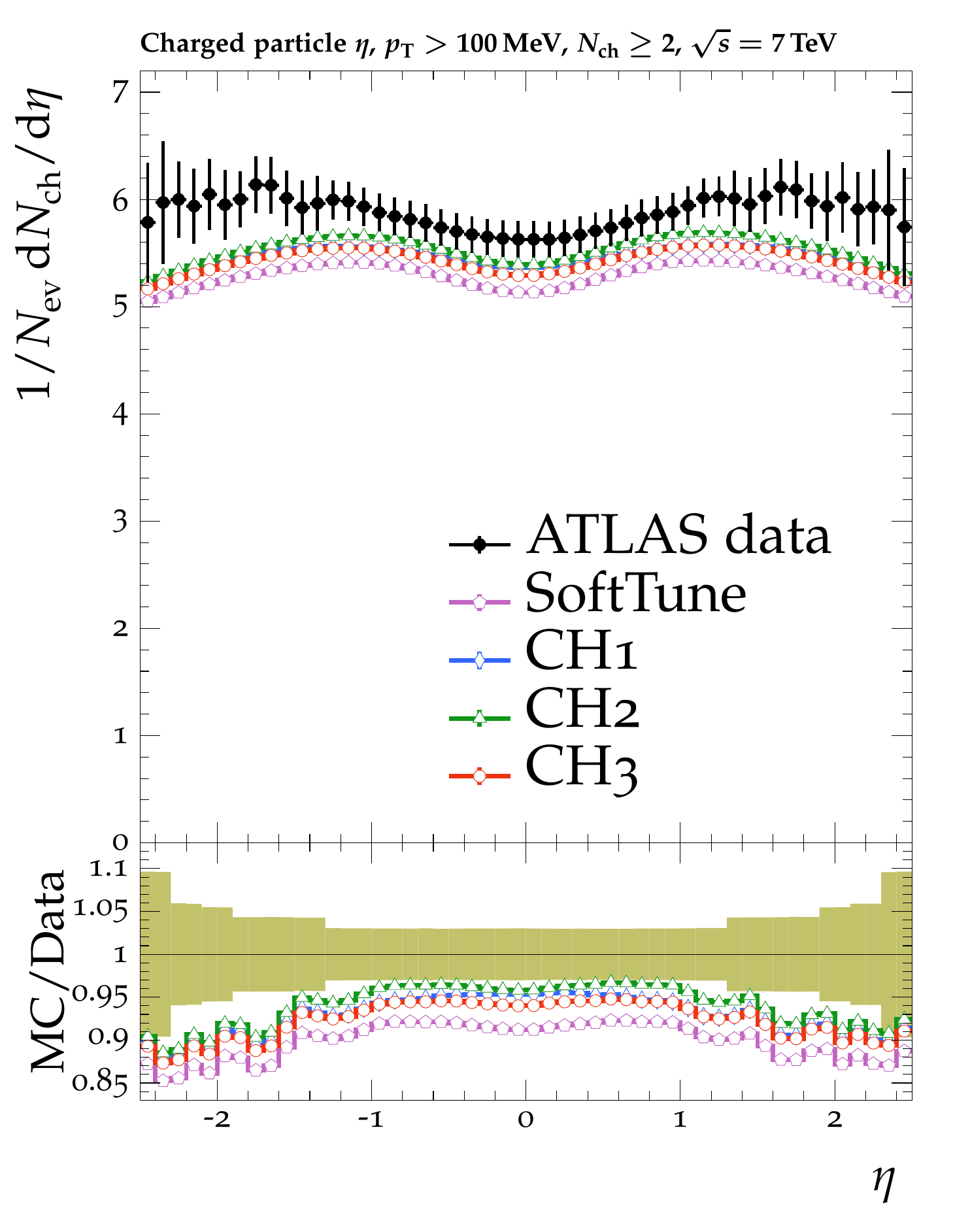} \\
  \includegraphics[width=0.49\textwidth]{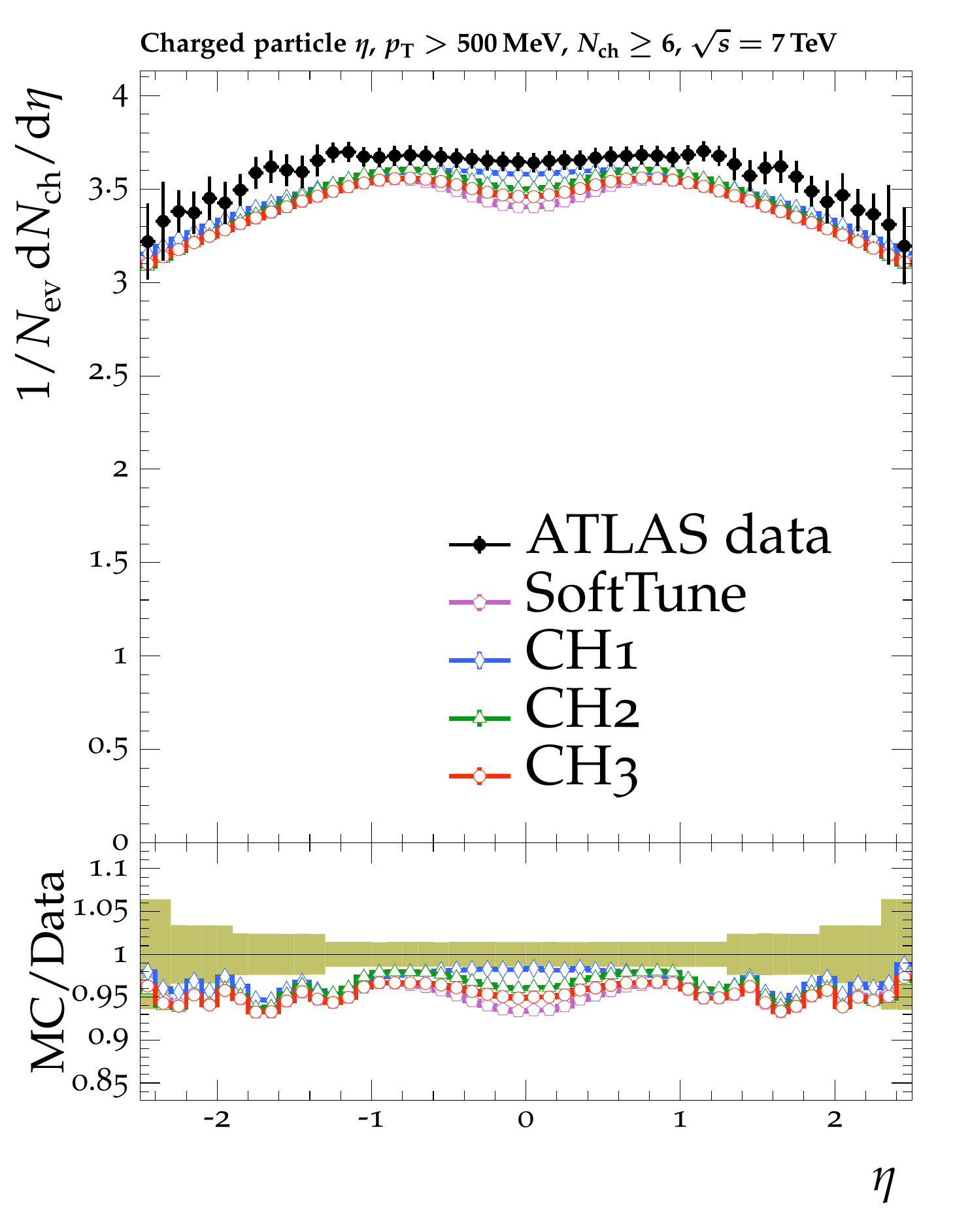}
  \includegraphics[width=0.49\textwidth]{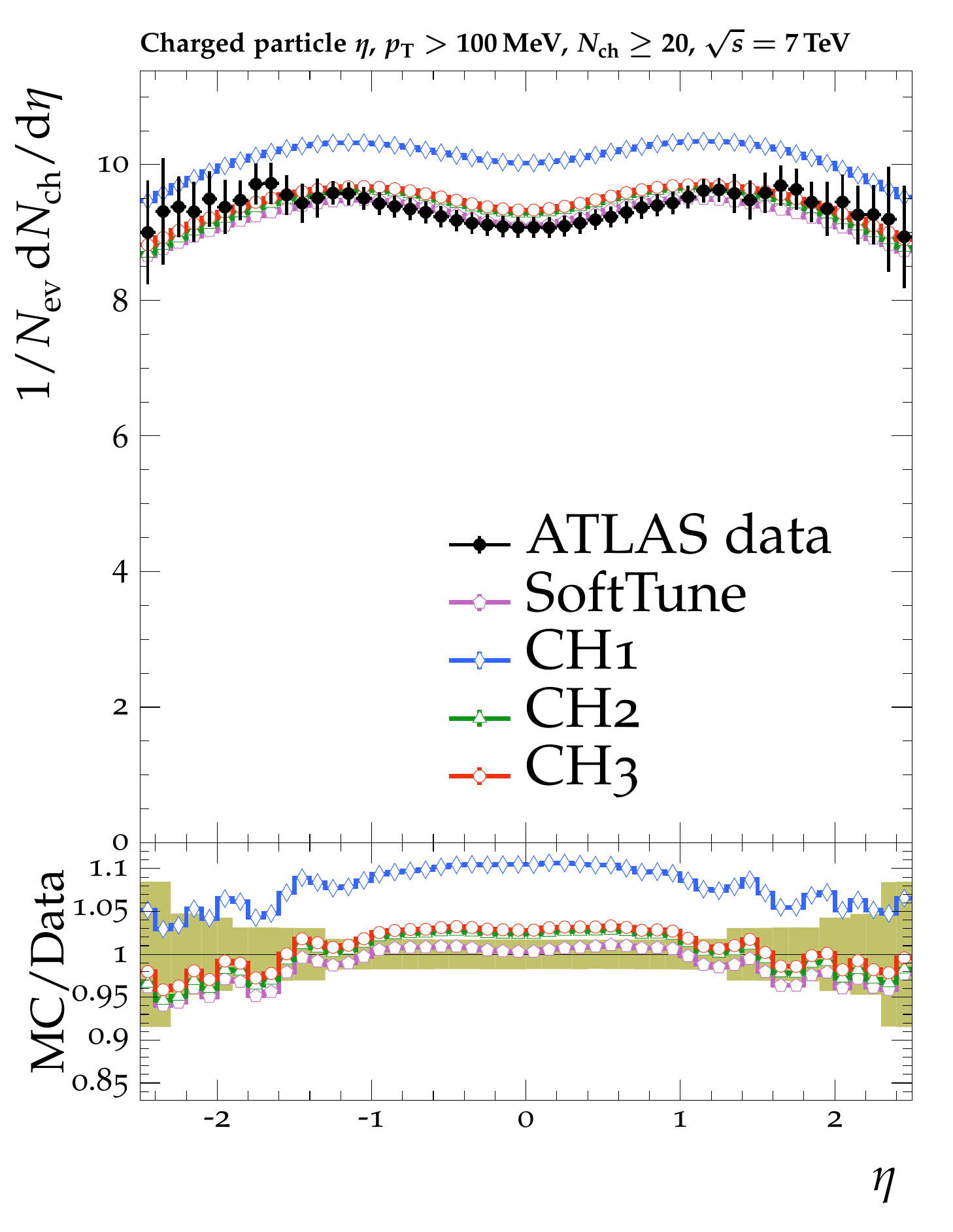}
  \caption{Normalized plots~\cite{ATLASdNdEta0p9And7TeV} for the pseudorapidity of charged particles for $\Nch\ge1$ (upper left), and $\Nch\ge6$ (lower left), for charged particles with $\pt>500\MeV$.  The figure on the upper right shows a similar distribution for $\Nch\ge2$, and the lower right for $\Nch\ge20$, where the charged particles have $\pt>100\MeV$.  ATLAS MB data are compared with the predictions from \HerwigS, with the \SoftTune and \CH tunes.  \captionColouredShadedBand}
  \label{fig:ATLAS_dNdeta_7TeV}
\end{figure*}

\begin{figure*}[htbp]
  \centering
  \includegraphics[width=0.49\textwidth]{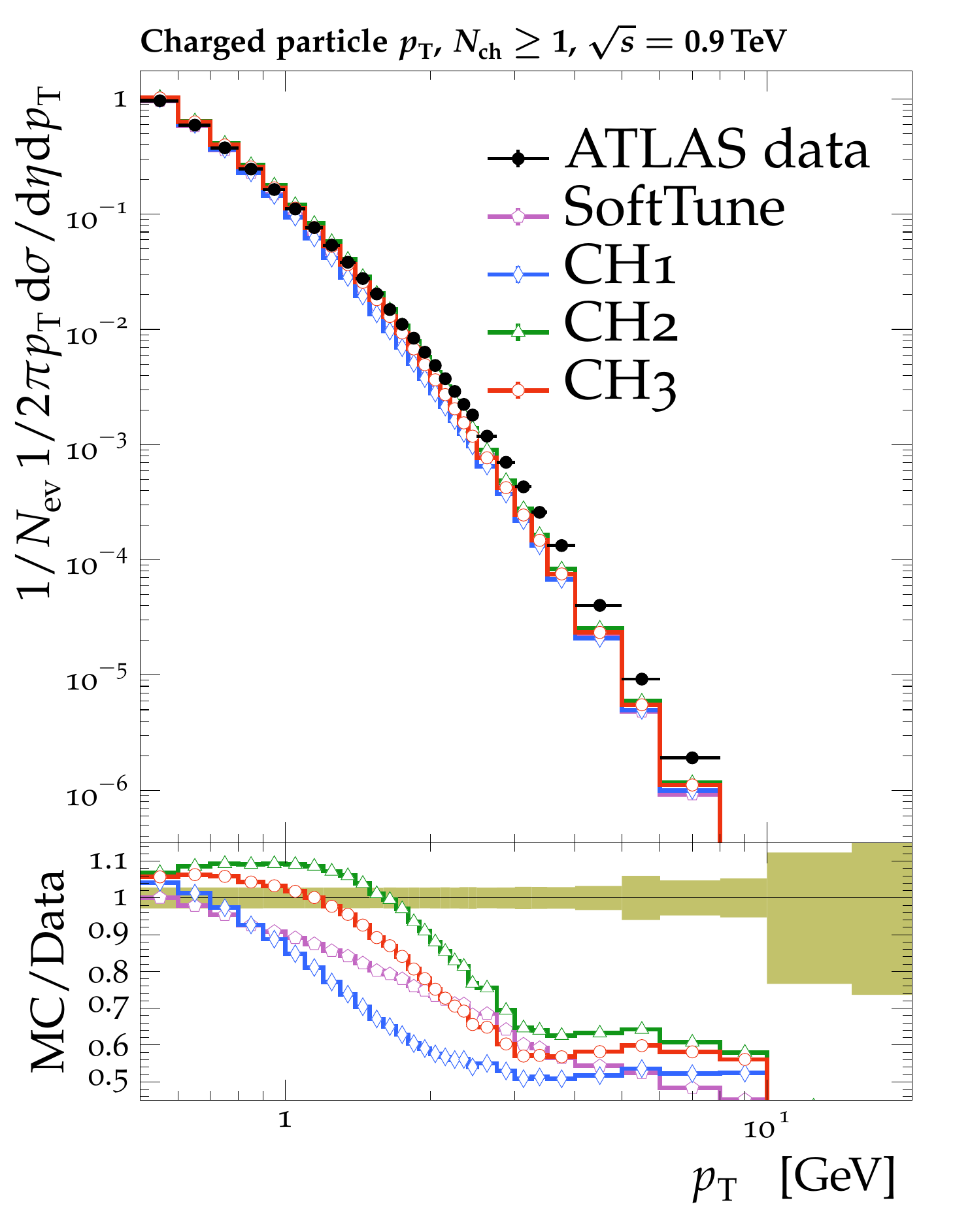}
  \includegraphics[width=0.49\textwidth]{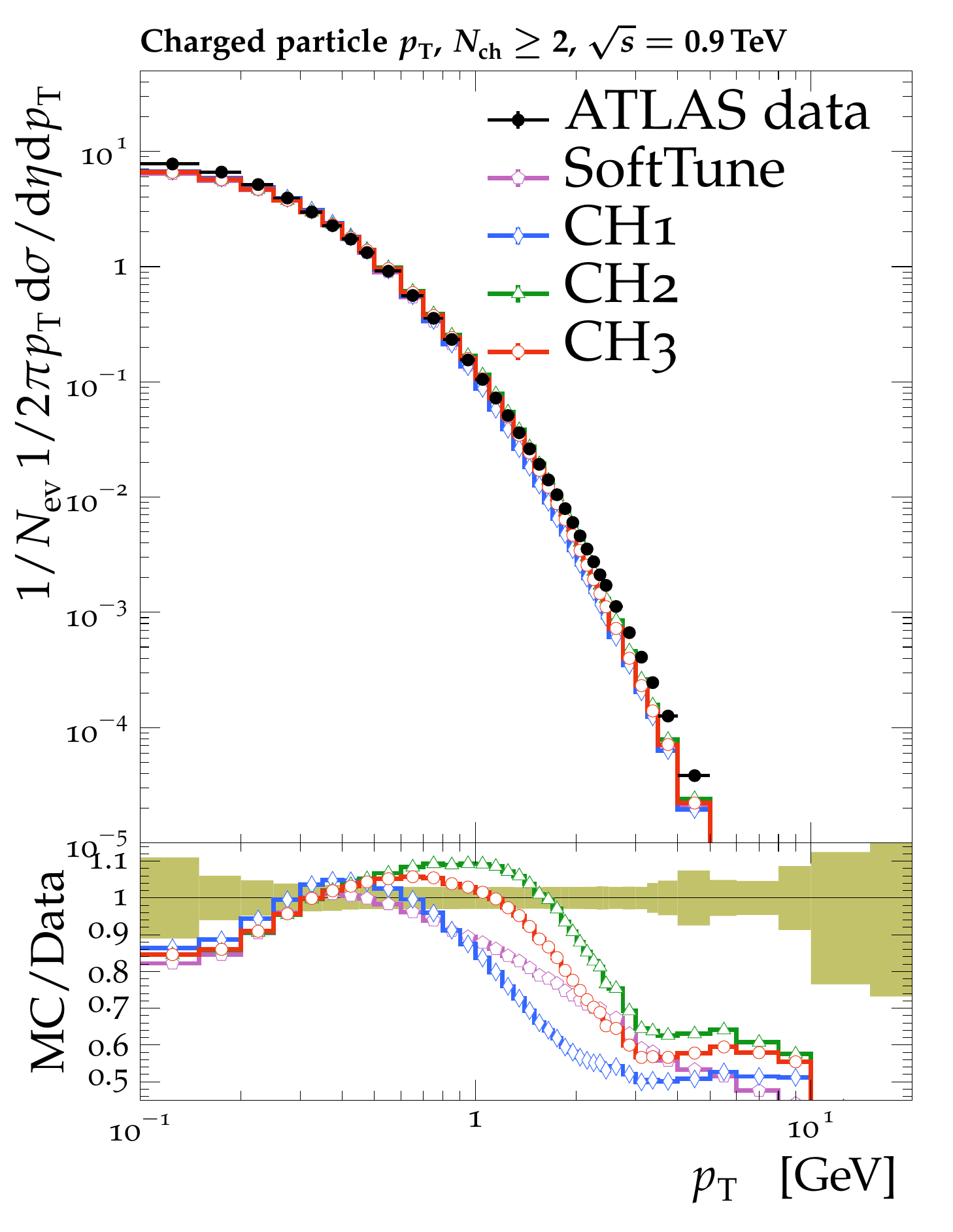} \\
  \includegraphics[width=0.49\textwidth]{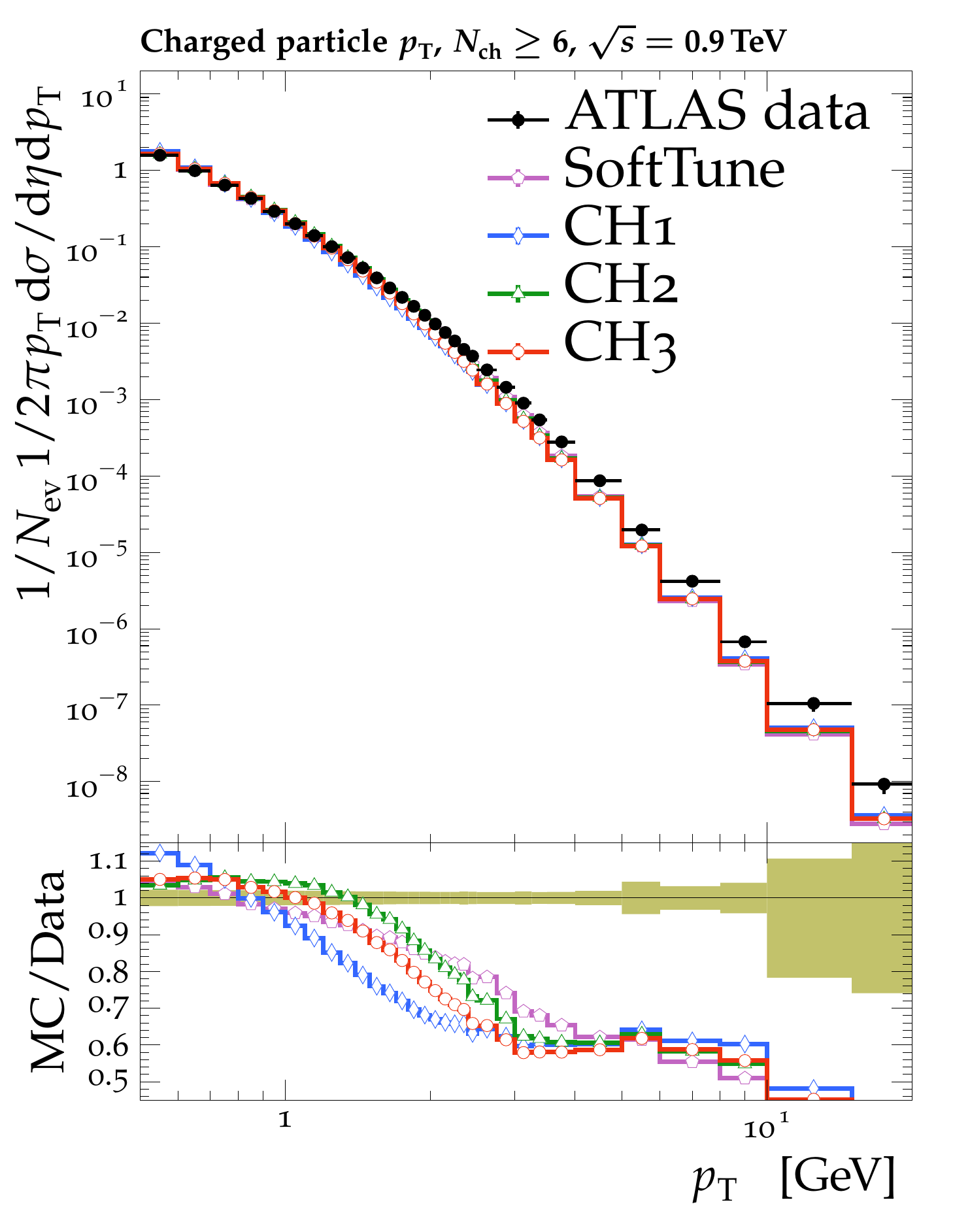}
  \includegraphics[width=0.49\textwidth]{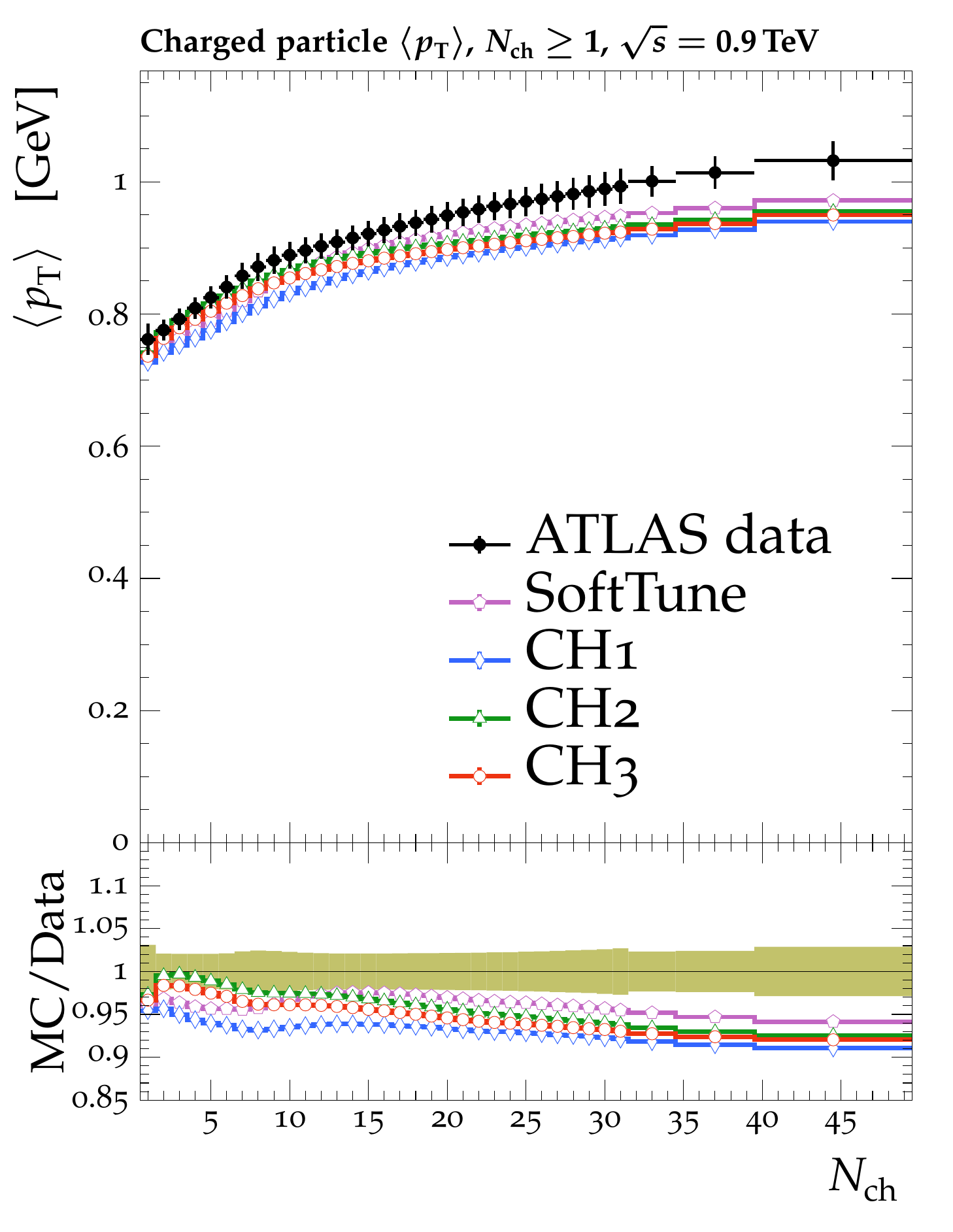}
  \caption{Normalized plots~\cite{ATLASdNdEta0p9And7TeV} for the charged-particle \pt for $\Nch\ge1$ (upper left), $\Nch\ge2$ (upper right), and $\Nch\ge6$ (lower left).  The mean charged-particle \pt as a function of the charged-particle multiplicity is also shown (lower right).  ATLAS MB data are compared with the predictions from \HerwigS, with the \SoftTune and \CH tunes.  \captionColouredShadedBand}
  \label{fig:ATLAS_dNdpt_0p9TeV}
\end{figure*}

\begin{figure*}[htbp]
  \centering
  \includegraphics[width=0.49\textwidth]{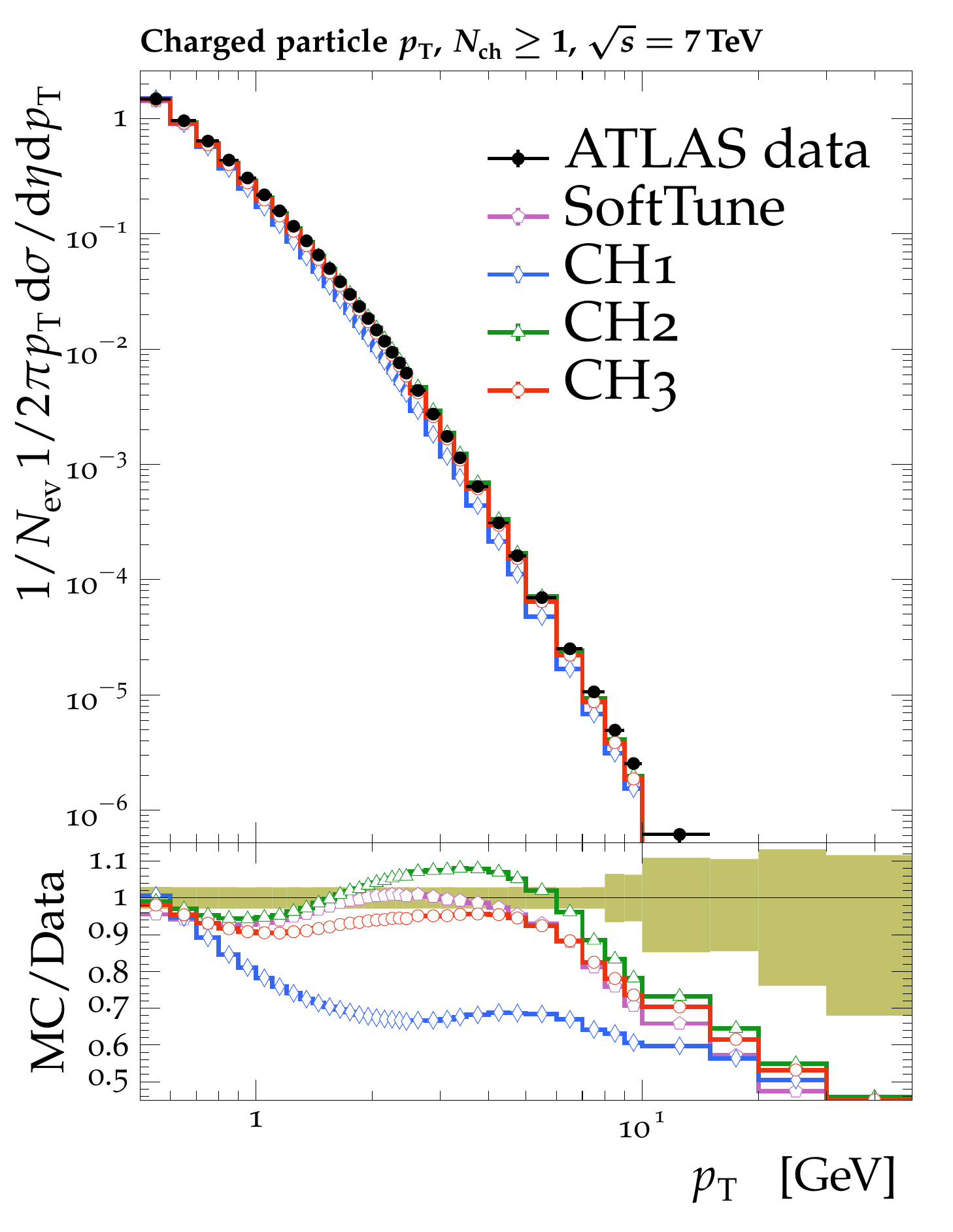}
  \includegraphics[width=0.49\textwidth]{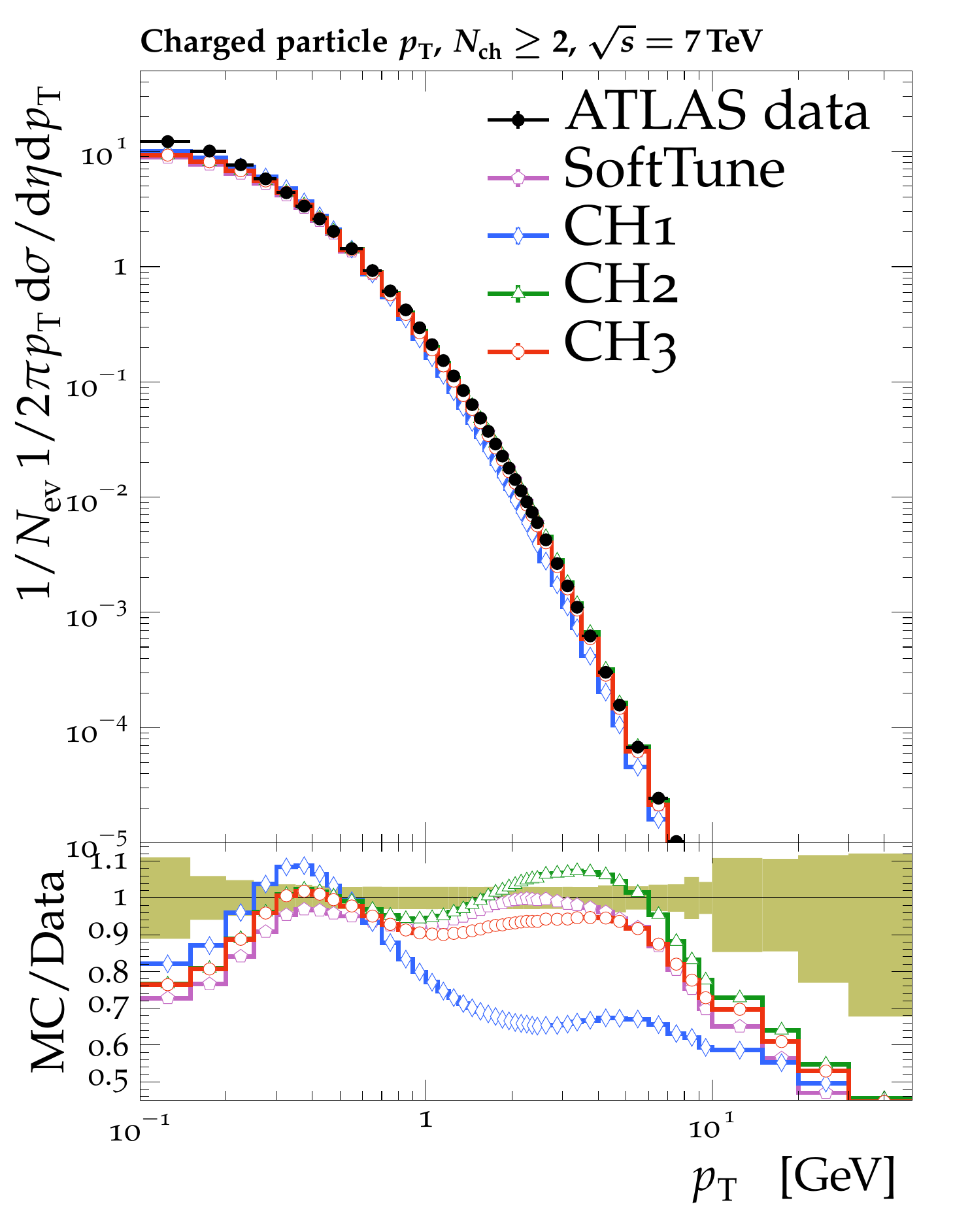} \\
  \includegraphics[width=0.49\textwidth]{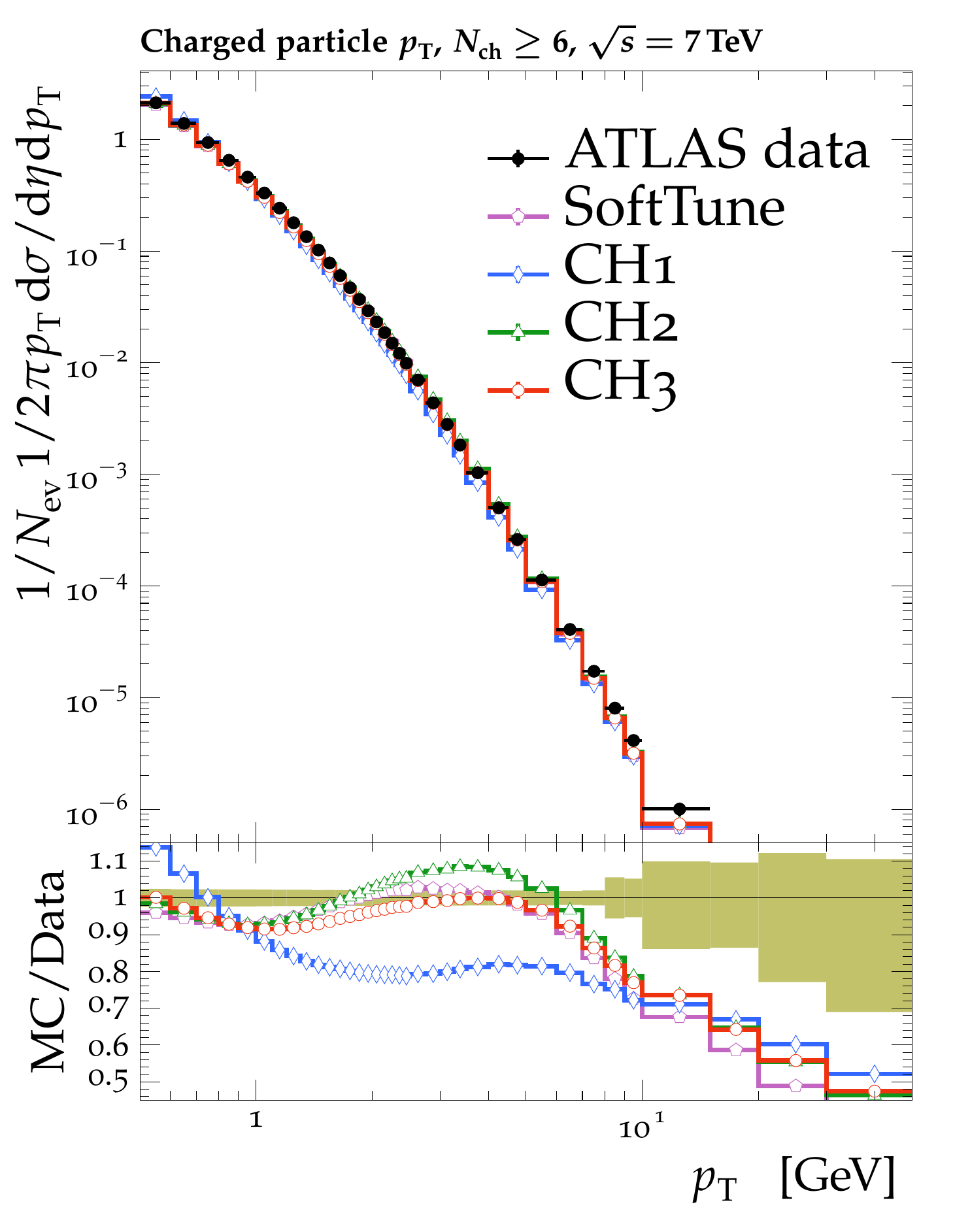}
  \includegraphics[width=0.49\textwidth]{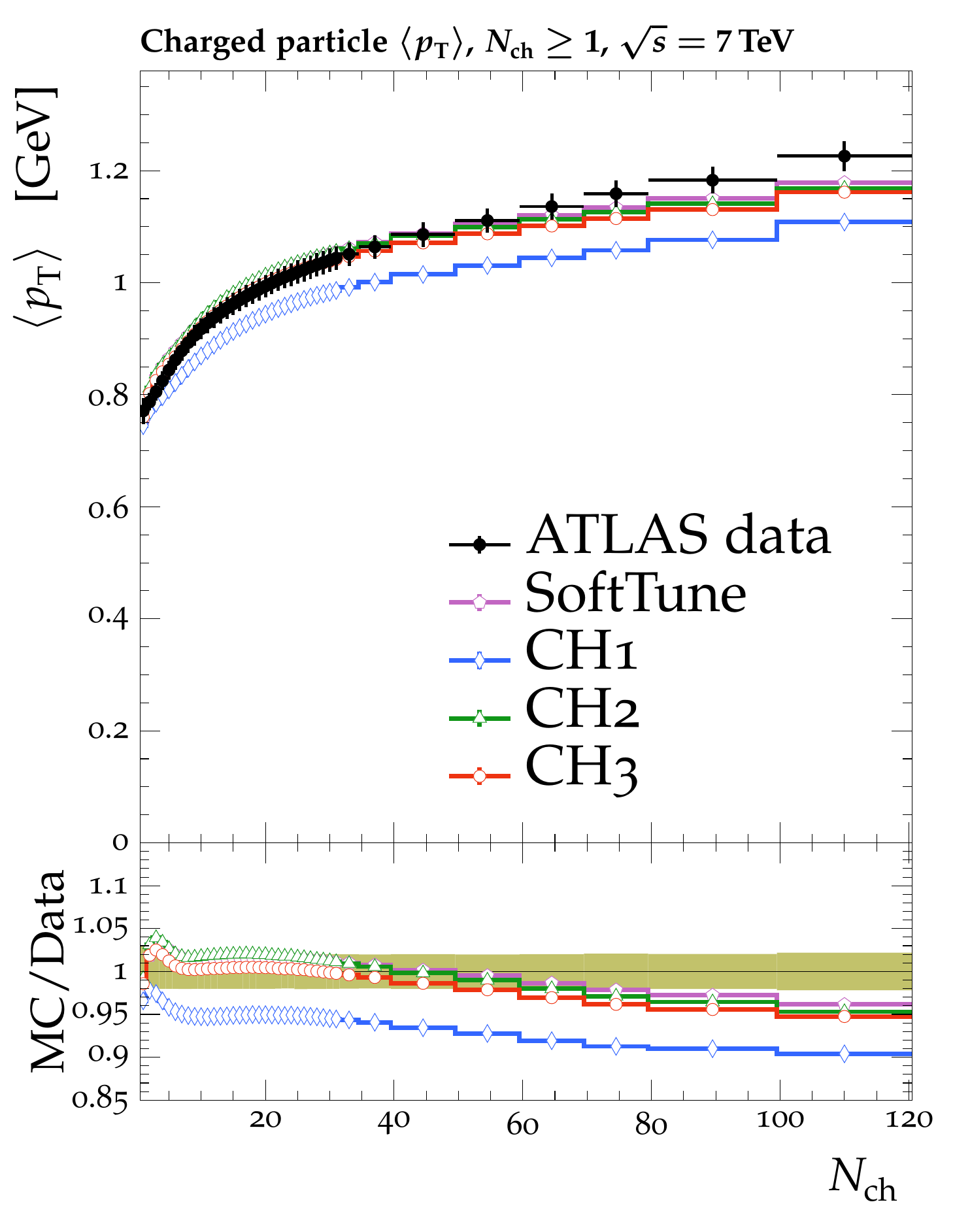}
  \caption{Normalized plots~\cite{ATLASdNdEta0p9And7TeV} for the charged-particle \pt for $\Nch\ge1$ (upper left), $\Nch\ge2$ (upper right), and $\Nch\ge6$ (lower left).  The mean charged-particle \pt as a function of the charged-particle multiplicity is also shown (lower right).  ATLAS MB data are compared with the predictions from \HerwigS, with the \SoftTune and \CH tunes.  \captionColouredShadedBand}
  \label{fig:ATLAS_dNdpt_7TeV}
\end{figure*}

\begin{figure*}[htbp]
  \centering
  \includegraphics[width=0.45\textwidth]{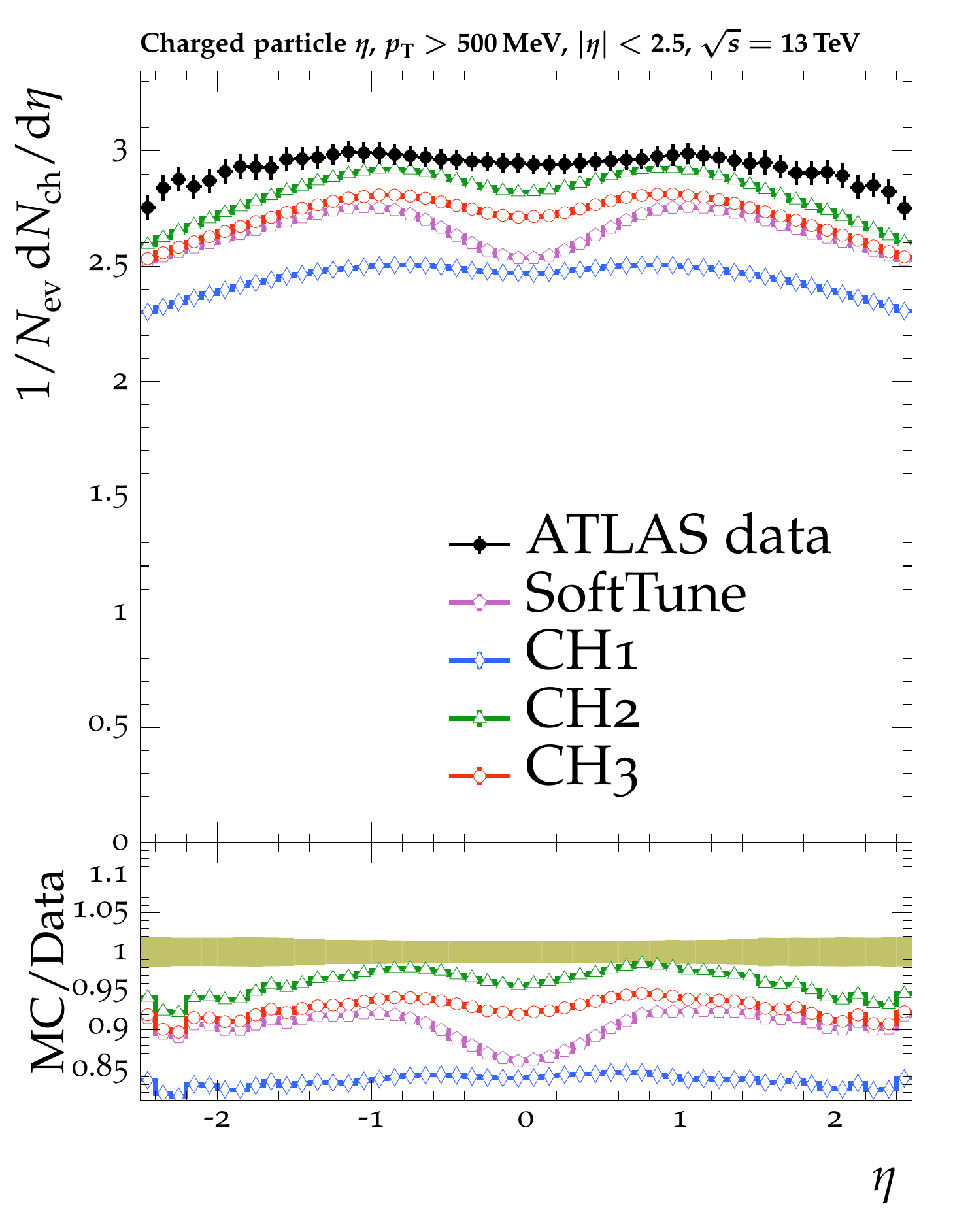}
  \includegraphics[width=0.45\textwidth]{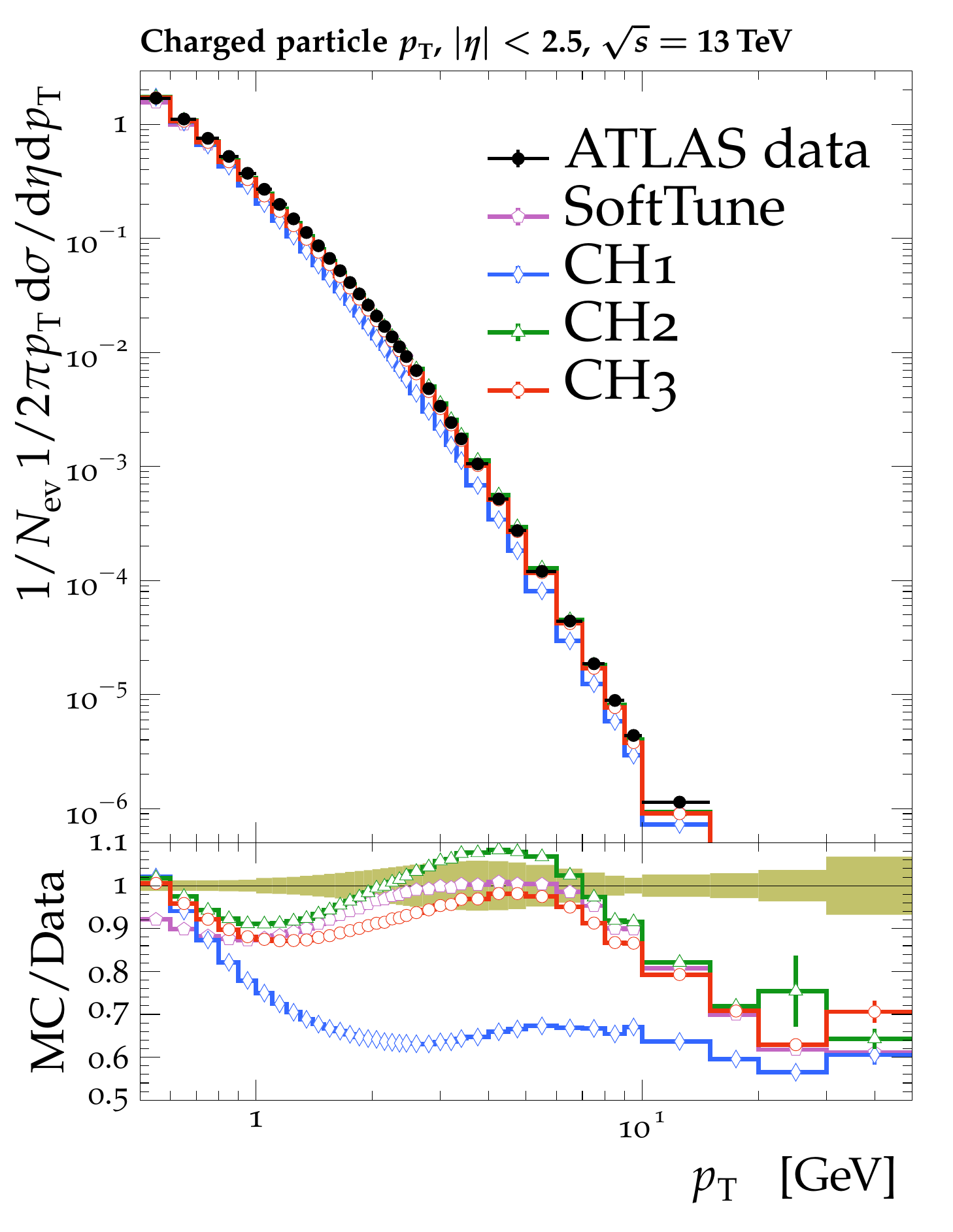}\\
  \includegraphics[width=0.45\textwidth]{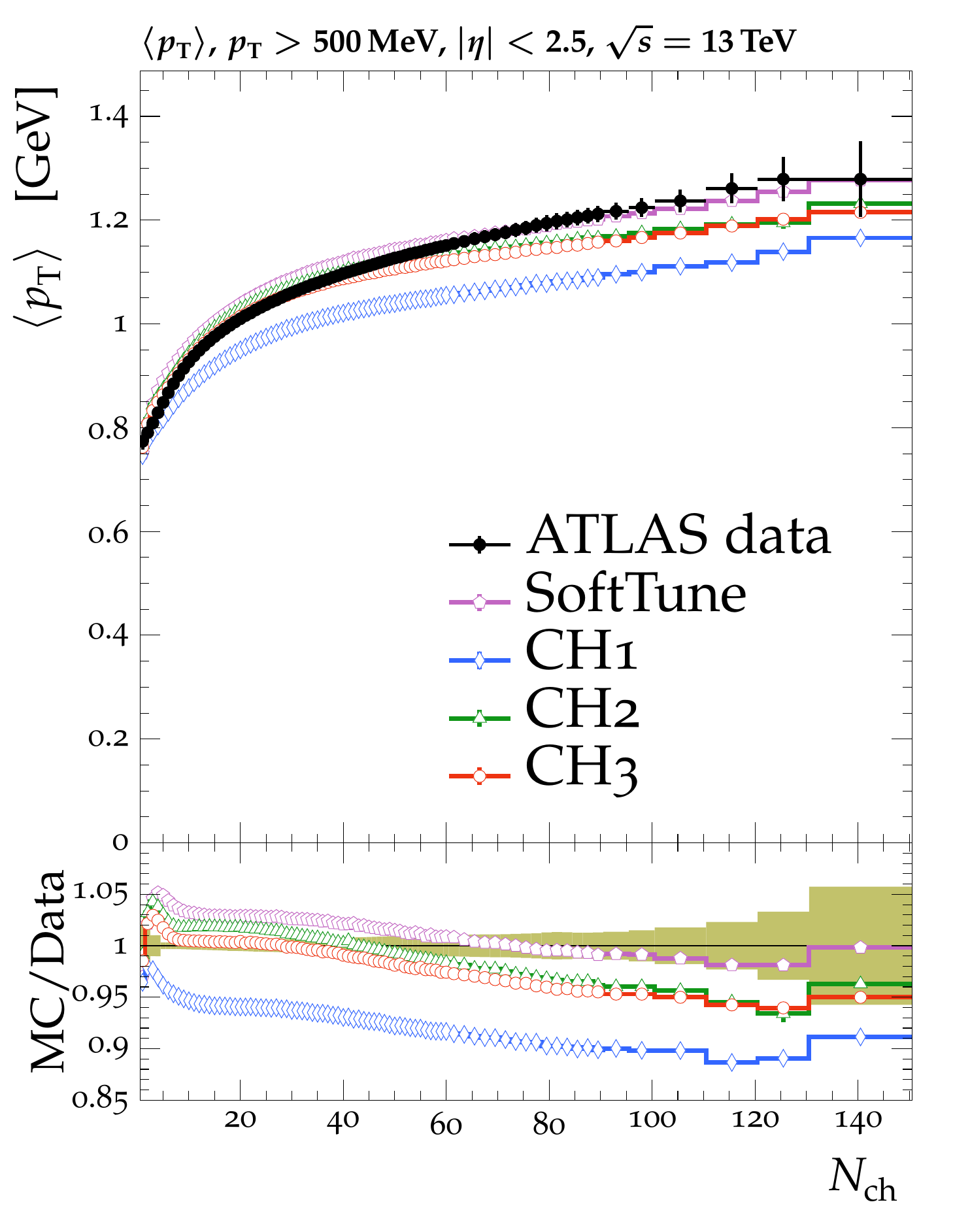}
  \caption{Normalized plots~\cite{ATLASdNdEta13TeV} for the pseudorapidity of charged particles (upper left), charged-particle \pt distribution (upper left), and the mean charged-particle \pt distribution as a function of the charged-particle multiplicity (lower), all for $\abs{\eta}<2.5$.  ATLAS MB data are compared with the predictions from \HerwigS, with the \SoftTune and \CH tunes.  \captionColouredShadedBand}
  \label{fig:ATLAS_13TeV_eta2p5}
\end{figure*}

\begin{figure*}[htbp]
  \centering
  \includegraphics[width=0.45\textwidth]{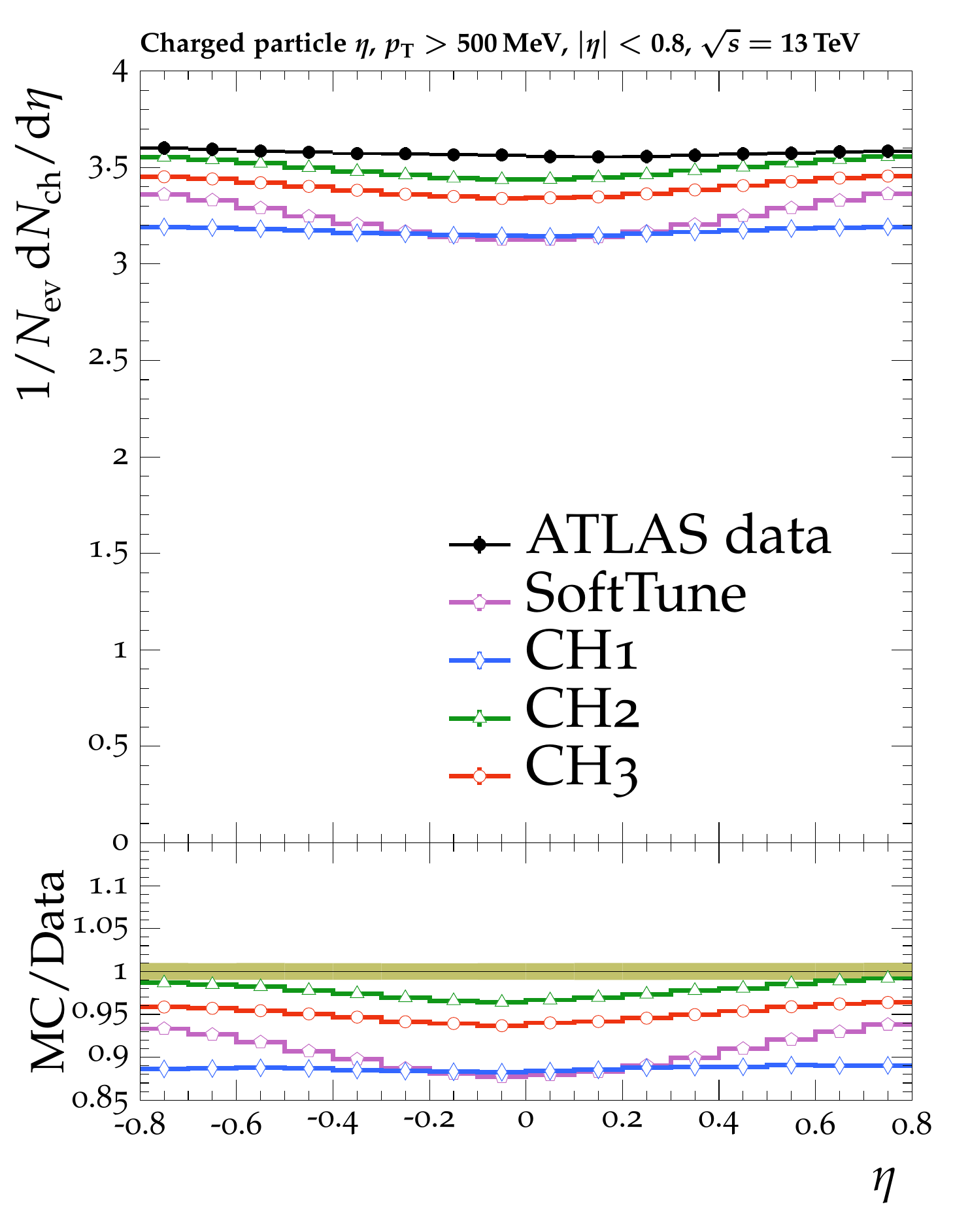}
  \includegraphics[width=0.45\textwidth]{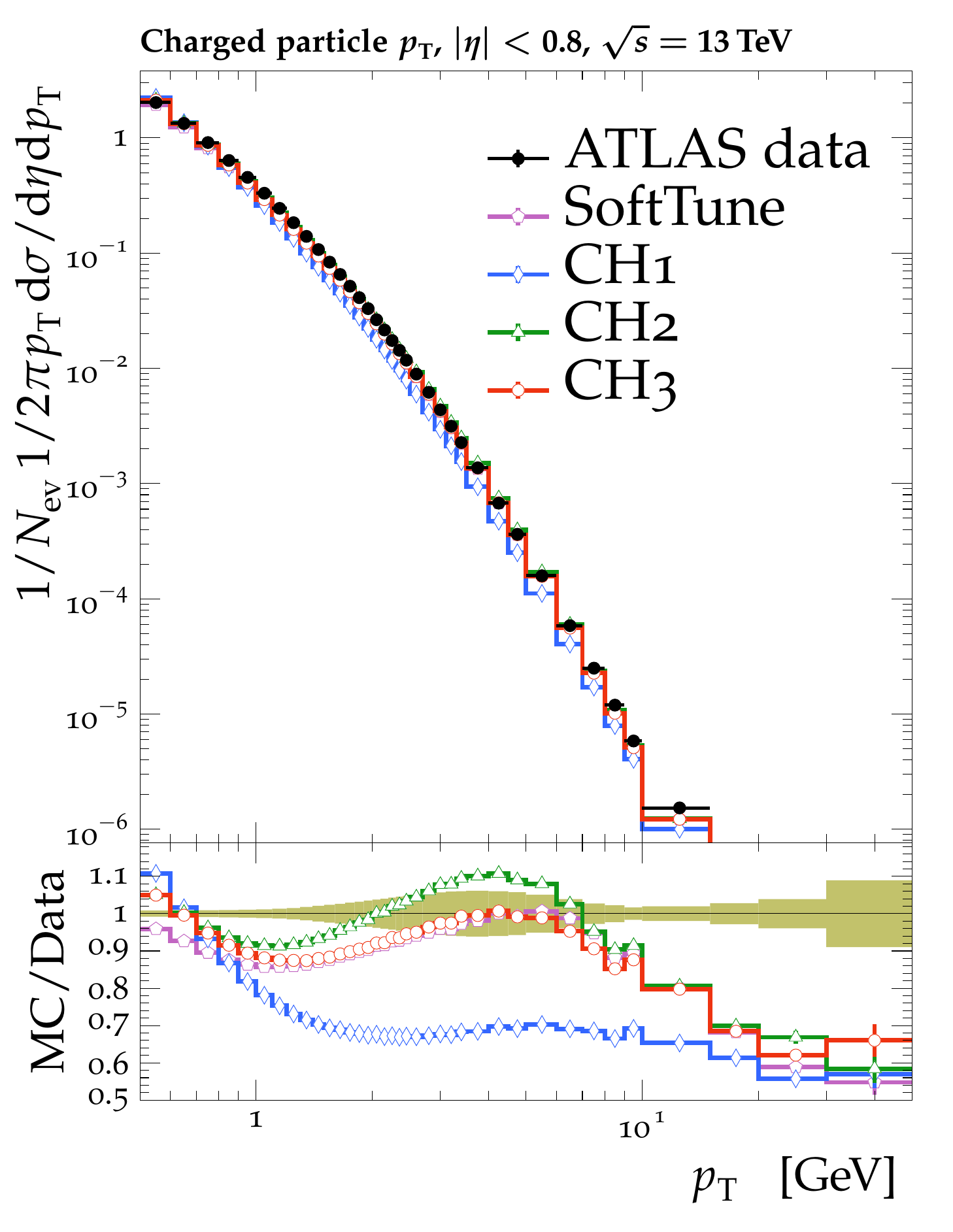} \\
  \includegraphics[width=0.45\textwidth]{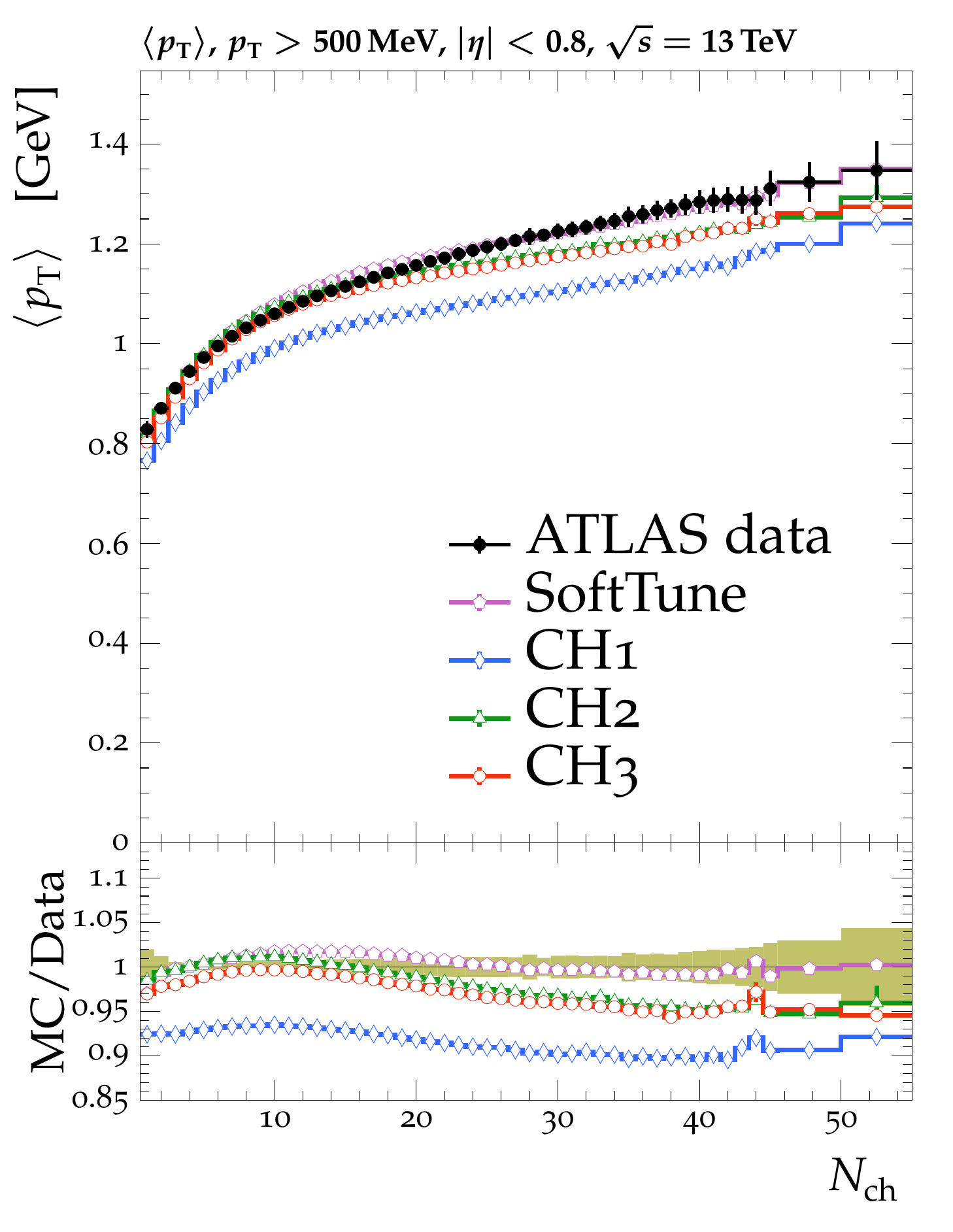}
  \caption{Normalized plots~\cite{ATLASdNdEta13TeV} for the pseudorapidity of charged particles (upper left), charged-particle \pt distribution (upper left), and the mean charged-particle \pt distribution as a function of the charged-particle multiplicity (lower), all for $\abs{\eta}<0.8$.  ATLAS MB data are compared with the predictions from \HerwigS, with the \SoftTune and \CH tunes.  \captionColouredShadedBand}
  \label{fig:ATLAS_13TeV_eta0p8}
\end{figure*}

Figures~\ref{fig:ATLAS_dNdeta_0p9TeV} to~\ref{fig:ATLAS_13TeV_eta0p8} show comparisons of the tune predictions with MB data collected by the ATLAS experiment at $\sqrts=0.9,~7,$ and $13\TeV$, which were used in deriving the parameters of \SoftTune.  Figures~\ref{fig:ATLAS_dNdeta_0p9TeV} and~\ref{fig:ATLAS_dNdeta_7TeV} show the pseudorapidity distributions of charged particles at $\sqrts=0.9$ and $7\TeV$ respectively, for various minimum \Nch.  Figures~\ref{fig:ATLAS_dNdpt_0p9TeV} and~\ref{fig:ATLAS_dNdpt_7TeV} show the charged-particle \pt distributions at $\sqrts=0.9$ and $7\TeV$ respectively, for various minimum \Nch.  The distributions of mean charged-particle \pt as a function of the charged-particle multiplicity are also shown in Figs.~\ref{fig:ATLAS_dNdpt_0p9TeV} and~\ref{fig:ATLAS_dNdpt_7TeV}.  Figures~\ref{fig:ATLAS_13TeV_eta2p5} and~\ref{fig:ATLAS_13TeV_eta0p8} show the pseudorapidity and charged-particle \pt distributions at $\sqrts=13\TeV$, for $\abs{\eta}<2.5$ and $\abs{\eta}<0.8$ respectively.  The corresponding distributions of the mean charged-particle \pt as a function of the charged-particle multiplicity are also shown in Figs.~\ref{fig:ATLAS_13TeV_eta2p5} and~\ref{fig:ATLAS_13TeV_eta0p8}.

\clearpage
\section{Comparison with ATLAS inclusive jet events}
\label{app:ATLASJetSub}

\begin{figure*}[htbp]
  \centering
  \includegraphics[width=0.49\textwidth]{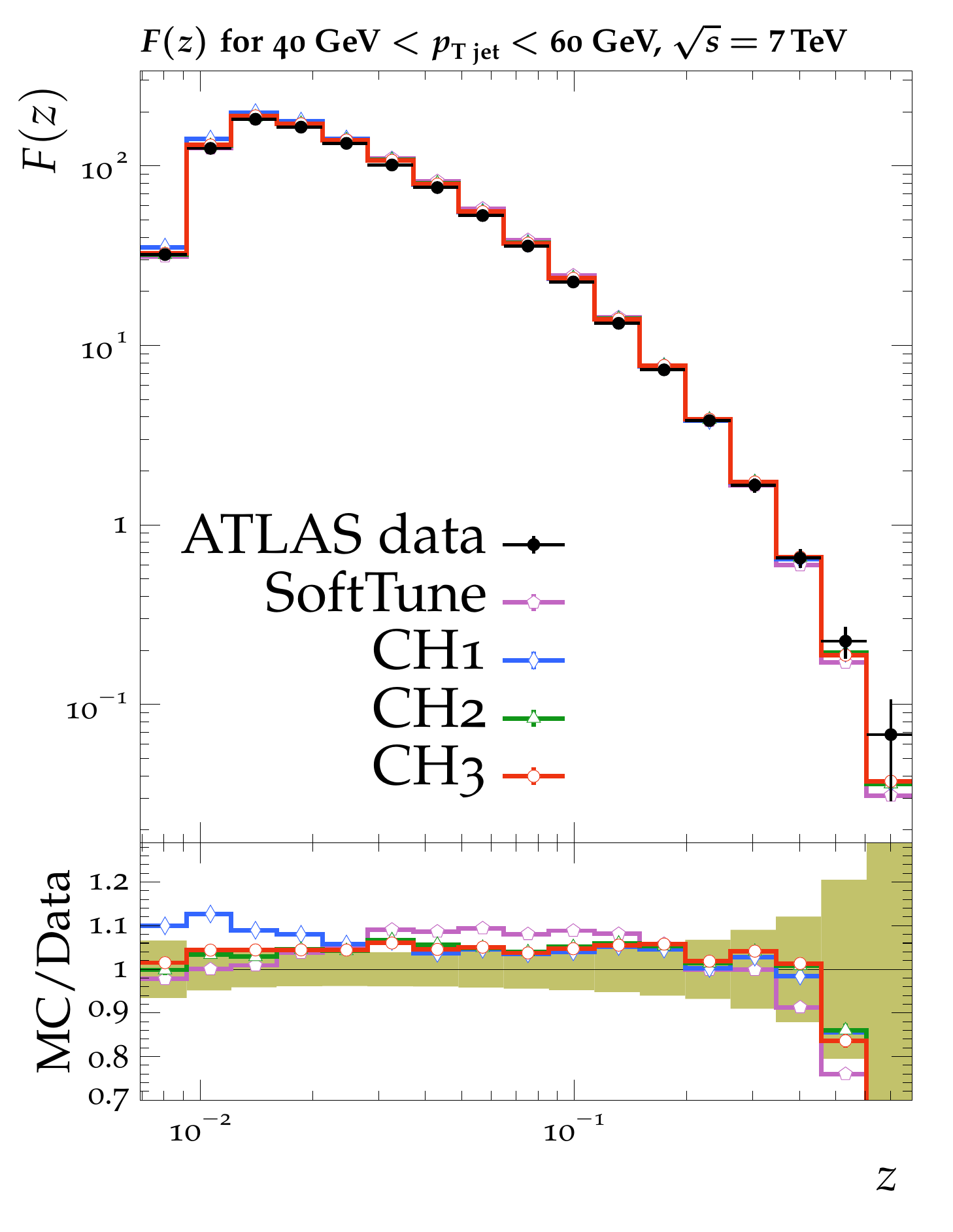}
  \includegraphics[width=0.49\textwidth]{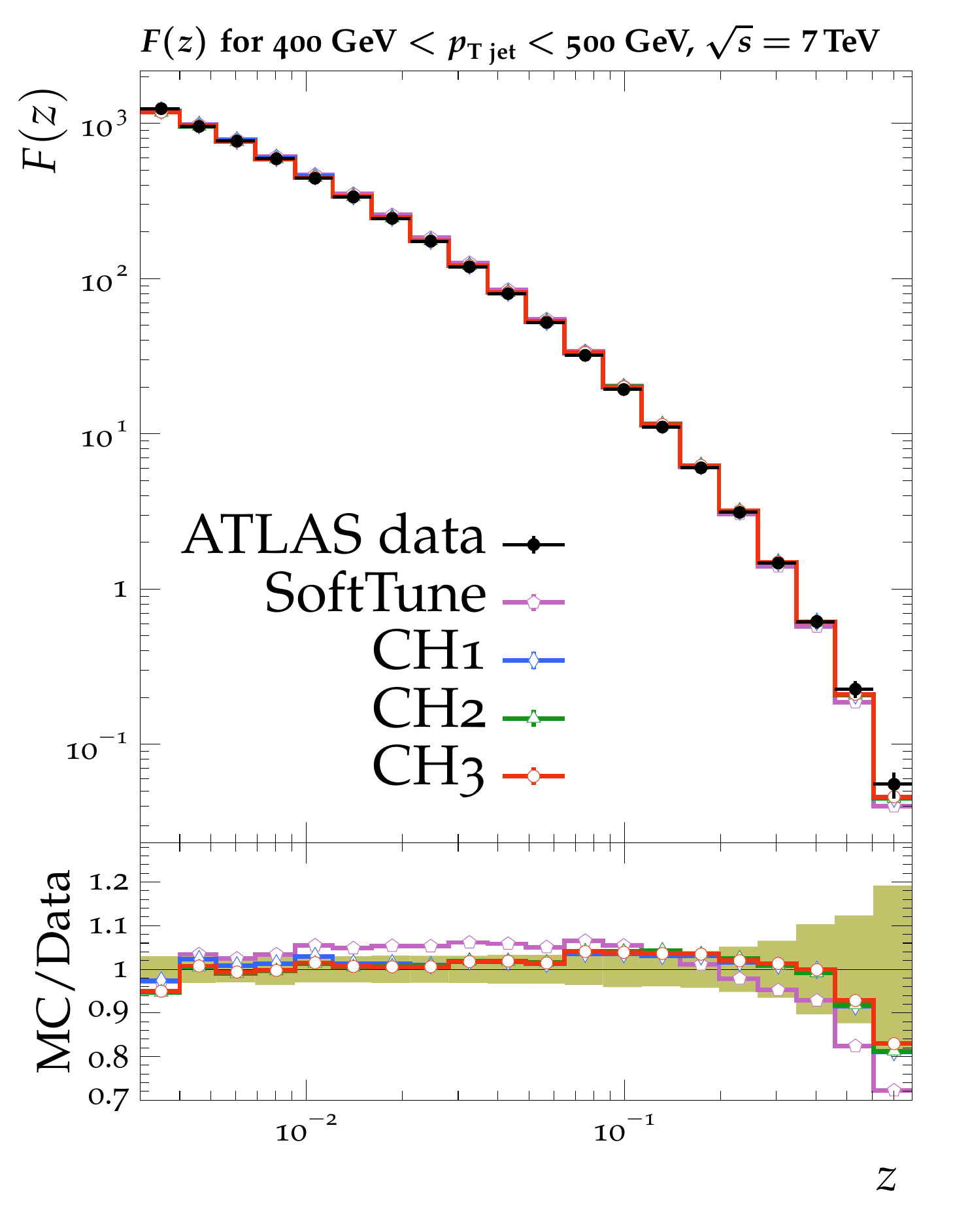} \\
  \includegraphics[width=0.49\textwidth]{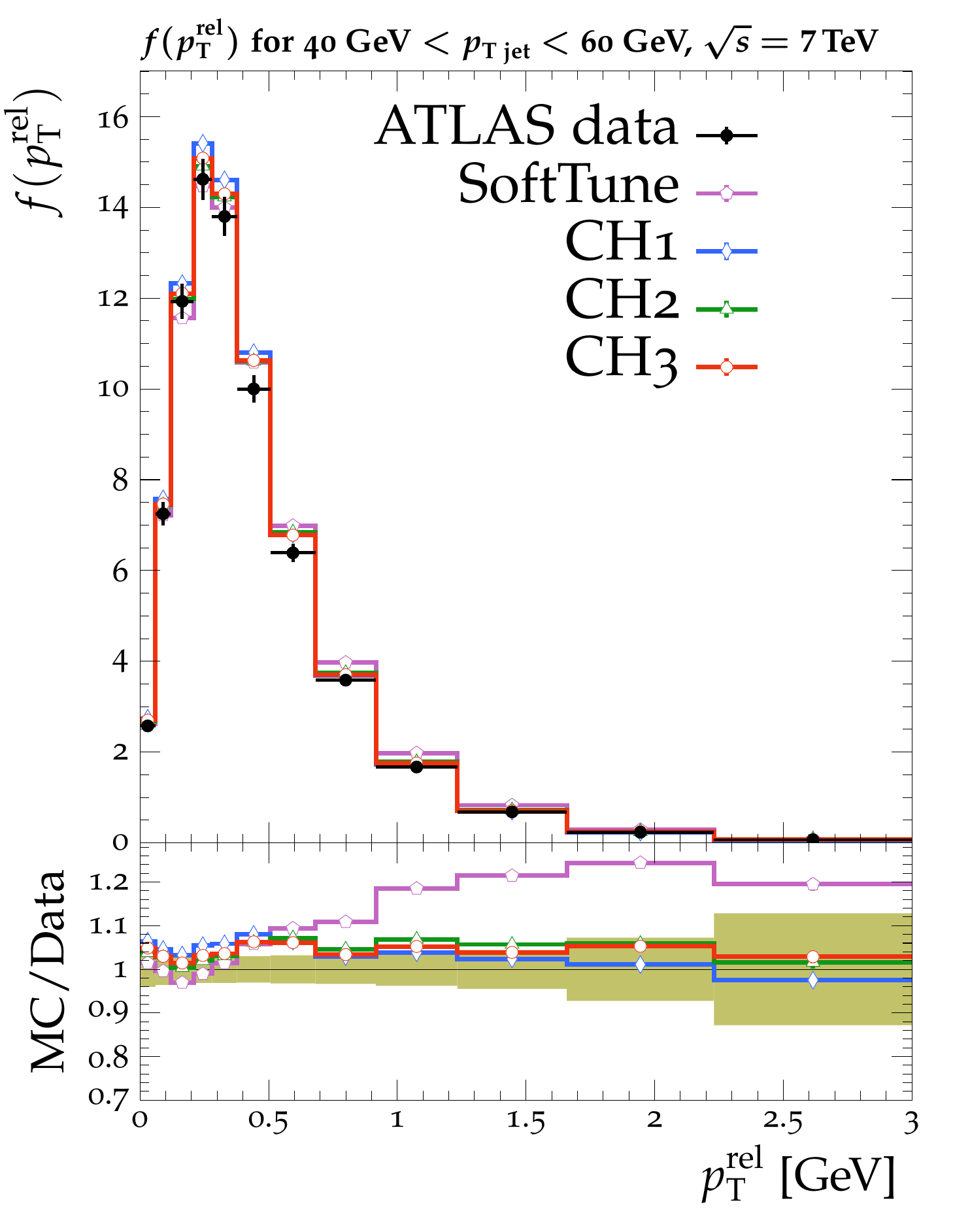}
  \includegraphics[width=0.49\textwidth]{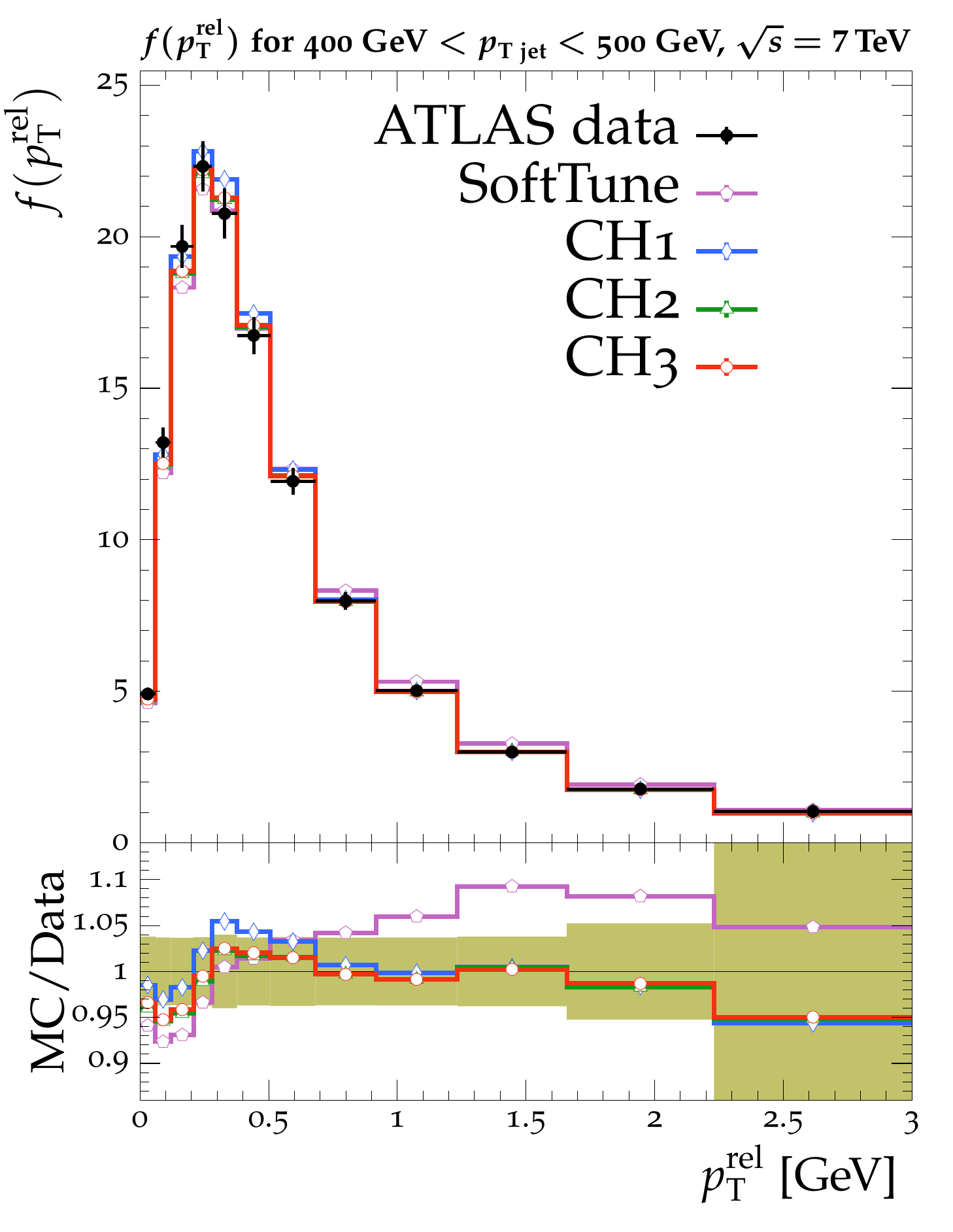}
  \caption{The ATLAS data at $\sqrts=7\TeV$ on the \Fz and \fptrel distributions~\cite{ATLASJetFrag}.  ATLAS inclusive jet data are compared with the predictions from \HerwigS, with the \SoftTune and \CH tunes.  \captionColouredShadedBand}
  \label{fig:ATLASJetFrag}
\end{figure*}

Figure~\ref{fig:ATLASJetFrag} shows the \Fz distribution as a function of \z, and the \fptrel distribution as a function of \ptrel, as measured by the ATLAS experiment, along with the \HerwigS predictions.  The former distribution is a differential measurement of the charged-particle multiplicity inside jets as a function of the fraction of the jet longitudinal momentum carried by the jet constituents, \z.  The latter distribution is the same data but as a function of the transverse momentum of the jet constituents, \ptrel, with respect to the jet axis.  The jets are clustered using the \antikt algorithm with a distance parameter of 0.6, and the distributions are shown for two ranges of jet \pt (\ptjet): $40 < \ptjet < 60 \GeV$ and $400 < \ptjet < 500 \GeV$.  For all distributions, \SoftTune provides the least consistent prediction of the data.  At low jet \pt, the \CHt and \CHth tunes provide the best description of the data, whereas the \CHo tune deviates somewhat from the data both at low \z and at low \fptrel.  At high jet \pt, only \SoftTune shows significant differences with respect to the data; however, these differences are smaller than those observed at low jet \pt.

\cleardoublepage \section{The CMS Collaboration \label{app:collab}}\begin{sloppypar}\hyphenpenalty=5000\widowpenalty=500\clubpenalty=5000\vskip\cmsinstskip
\textbf{Yerevan Physics Institute, Yerevan, Armenia}\\*[0pt]
A.M.~Sirunyan$^{\textrm{\dag}}$, A.~Tumasyan
\vskip\cmsinstskip
\textbf{Institut f\"{u}r Hochenergiephysik, Wien, Austria}\\*[0pt]
W.~Adam, F.~Ambrogi, T.~Bergauer, M.~Dragicevic, J.~Er\"{o}, A.~Escalante~Del~Valle, R.~Fr\"{u}hwirth\cmsAuthorMark{1}, M.~Jeitler\cmsAuthorMark{1}, N.~Krammer, L.~Lechner, D.~Liko, T.~Madlener, I.~Mikulec, F.M.~Pitters, N.~Rad, J.~Schieck\cmsAuthorMark{1}, R.~Sch\"{o}fbeck, M.~Spanring, S.~Templ, W.~Waltenberger, C.-E.~Wulz\cmsAuthorMark{1}, M.~Zarucki
% ...
\vskip\cmsinstskip
\dag: Deceased\\
1:  Also at Vienna University of Technology, Vienna, Austria\\
% ...
\end{sloppypar}
\end{document}